\documentclass[epj]{svjour}
\usepackage{graphicx,mathrsfs,amsmath,bm,amssymb}
\usepackage{physics}
\usepackage{hyperref}
\usepackage{multirow}
\usepackage{lscape}
\usepackage{color}
\usepackage{hyperref}
\newcommand{\gtorder}{\mathrel{\raise.3ex\hbox{$>$}\mkern-14mu
            \lower0.6ex\hbox{$\sim$}}}
\newcommand{\ltorder}{\mathrel{\raise.3ex\hbox{$<$}\mkern-14mu
            \lower0.6ex\hbox{$\sim$}}}
\begin{document}
\title{Bayesian analysis of multimessenger M-R data with interpolated hybrid EoS}

\author{
A.~Ayriyan\inst{1,2,3}
\and
D. Blaschke\inst{4,5,6}
\and
A.~G.~Grunfeld\inst{7,8} 
\and
D. Alvarez-Castillo\inst{5,9}
\and
H. Grigorian\inst{1,2,10}
\and
V. Abgaryan\inst{1,11,12}
}
%\thanks{\emph{On leave from:} 
%Instituto de F\'{i}sica, Universidad Aut\'{o}noma de San Luis Potos\'{i}, M\'{e}xico}
% etc

\institute{
Laboratory of Information Technologies, JINR, 6 Joliot-Curie Str., Dubna, 141980, Russian Federation % 1
\and
IT and Computing Division, A. Alikhanyan National Laboratory, 2 Alikhanian Brothers Str., Yerevan, 0036, Armenia % 2
\and
Dubna State University, 19 Universitetskaya Str., Dubna, 141980,Russia % 3
\and
Institute of Theoretical Physics, University of Wroclaw, 9 M. Borna Sq., Wroclaw, 50-204, Poland % 4
\and
Bogoliubov Laboratory of Theoretical Physics, JINR, 6 Joliot-Curie Str., Dubna, 141980, Russian Federation % 5
\and
National Research Nuclear University (MEPhI), 31 Kashirskoe Hwy, Moscow, 115409, Russian Federation % 6
\and
CONICET, Godoy Cruz 2290, Buenos Aires, Argentina % 7
\and
Departamento de F\'\i sica, Comisi\'on Nacional de Energ\'{\i}a At\'omica, Av. Libertador 8250, (1429) Buenos Aires, Argentina % 8
\and
Henryk Niewodnicza\'{n}ski Institute of Nuclear Physics, 152 Radzikowskiego Str, Cracow, 31-342, Poland % 9
\and
Department of Physics, Yerevan State University, 1 Alex Manoogian Str, Yerevan, 0025, Armenia % 10
\and
Theoretical Physics Division, A. Alikhanyan National Laboratory, 2 Alikhanian Brothers Str., Yerevan, 0036, Armenia % 11
\and
Peoples’ Friendship University of Russia (RUDN University), 6 Miklukho-Maklaya Str., Moscow, 117198, Russian Federation % 12
}
%\fi
%
\date{Received: date / Revised version: date}
\abstract{
We introduce a family of equations of state (EoS) for hybrid neutron star (NS) matter that is obtained by a two-zone parabolic interpolation between a soft hadronic EoS at low densities and a set of stiff quark matter EoS at high densities within a finite region of chemical potentials $\mu_H < \mu < \mu_Q$. Fixing the hadronic EoS as the APR one and choosing the color-superconducting, nonlocal NJL model with two free parameters for the quark phase, we perform  Bayesian analyses with this two-parameter family of hybrid EoS. Using three different sets of observational constraints that include the mass of PSR J0740+6620, the tidal deformability for GW170817, and the mass-radius relation for PSR J0030+0451 from NICER as obligatory (set 1), while set 2 uses the possible upper limit on the maximum mass from GW170817 as an additional constraint and set 3 instead of the possibility that the lighter object in the asymmetric binary merger GW190814 is a neutron star. We confirm that in any case, the quark matter phase has to be color superconducting with the dimensionless diquark coupling approximately fulfilling the Fierz relation $\eta_D=0.75$ and the most probable solutions exhibiting a proportionality between $\eta_D$ and $\eta_V$, the coupling of the repulsive vector interaction that is required for a sufficiently large maximum mass. We used the Bayesian analysis to investigate with the method of fictitious measurements the consequences of anticipating different radii for the massive $2~M_\odot$ PSR J0740+6220 for the most likely equation of state. With the actual outcome of the NICER radius measurement on PSR J0740+6220 we could conclude that for the most likely hybrid star EoS would not support a maximum mass as large as $2.5~M_\odot$ so that the event GW190814 was a binary black hole merger.
 % end of PACS codes
} %end of abstract
\maketitle
\section{Introduction}
\label{sec:intro}

The observation of the first binary neutron star merger GW170817 in gravitational waves \cite{TheLIGOScientific:2017qsa} 
and the subsequent electromagnetic signals from the gamma-ray burst to the light curve of the kilonova 
\cite{GBM:2017lvd} have opened the era of multi-messenger astronomy.
This extends the available mass range for neutron star observations up to $2.6~M_\odot$ for the companion star of the $23~M_\odot$ black hole in the binary merger GW190814 \cite{Abbott:2020khf}, if that object was indeed the heaviest neutron star and not the lightest black hole, which is a currently disputed question. 
The observation of gravitational waves from the inspiral phase of the merger GW170817 did allow to extract for the first time a new constraint on the equation of state (EoS) of dense matter, the tidal deformability, to be in the range of 
$70 < \Lambda_{1.4} < 580$ \cite{Abbott:2018exr}
for a neutron star with the mass of $1.4~M_\odot$.
From this measurement, together with other constraints, the authors of \cite{Capano:2019eae} could constrain the radius of a neutron star in that mass range to 
the rather narrow limits of $R_{1.4}=11.0^{+0.9}_{-0.6}$~km.
An open and controversially discussed question is the interior composition of neutron stars, in dependence of their mass.

It is very likely that the quark substructure of nucleons manifests itself at increasing densities first by a stiffening of the EoS due to quark Pauli blocking
in nuclear matter \cite{Blaschke:2020qrs}
and at still higher densities by a delocalization of the quark wave function and the occurrence of deconfined quark matter. 
For a recent discussion of soft delocalization vs. hard deconfinement in the transition from nuclear to quark matter,  see \cite{Fukushima:2020cmk}.
A crucial open question, to which the present work intends to contribute, concerns the onset mass of deconfinement and the character of the transition
\cite{Baym:2017whm}.

A standard approach to the hadron-to-quark-matter transition would start from separate EoS models for these two phases and obtain the phase transition from a Maxwell construction (for sufficiently large surface tension between these phases) or a Glendenning construction of a homogeneous mixed phase (for vanishing surface tension) \cite{Glendenning:1992vb}.
In-between these limiting cases, the more realistic scenario of the first-order phase transition would consider structures of finite size formed by the balance between Coulomb interactions and surface tension (pasta phases), see \cite{Maslov:2018ghi} and references therein.
This approach has been used recently for a Bayesian analysis with observational constraints for masses and radii of neutron stars \cite{Blaschke:2020qqj} which reaches the conclusion that very likely the phase transition onset occurs in the center of neutron stars with masses around $1~M_\odot$ and would then match the observed compactness \cite{Capano:2019eae} in this way.
For this scenario to work, it is customary to have a sufficient stiffness of nuclear matter at supersaturation densities so that the deconfinement transition is driven to relatively low densities. 

It is worth noticing that the approach \cite{Blaschke:2020qqj} is based on a strong first-order phase transition which entails the formation of hybrid stars as a third family of compact stars 
\cite{Gerlach:1968zz}
featuring the mass twin
\cite{Glendenning:1998ag}
phenomenon.

\begin{figure}[ht]
	\centering
	\includegraphics[width=\linewidth]{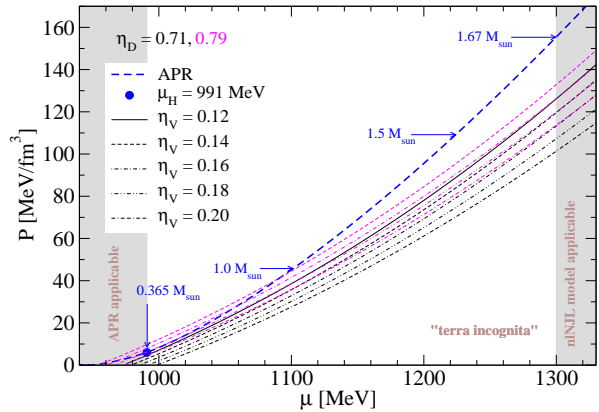}
	\caption{
	Pressure vs. chemical potential for the nlNJL EoS with different values of $\eta_V=0.12 (0.02) 0.20$ and the two limiting cases of $\eta_D=0.71$ and $\eta_D=0.79$ is compared to that of the 
	APR EoS. 
	The point  $\mu_H$ corresponds here to $n_H=1.5~n_0$.
	For orientation, a few selected values of central pressure and chemical potential in neutron stars are indicated by 
	blue arrows labelled with their mass.
	\label{fig:eos}}
\end{figure}

In Fig.~\ref{fig:eos} we illustrate this situation.
A soft hadronic EoS like that of Akmal, Pandharipande and Ravenhall (APR) \cite{Akmal:1998cf} has either no
crossing (Maxwell construction) with the color superconducting quark matter EoS (considering a nonlocal version of the NJL model, nlNJL, described below) for the weak diquark coupling strength ($\eta_D=0.71$) or, at slightly increased dimensionless diquark coupling  ($\eta_D=0.79$) an unrealistically early transition that is followed by a "reconfinement" or there is again no transition, depending on the value of the dimensionless vector meson coupling $\eta_V$.

\begin{figure}[ht]
	\centering
	\includegraphics[width=\linewidth]{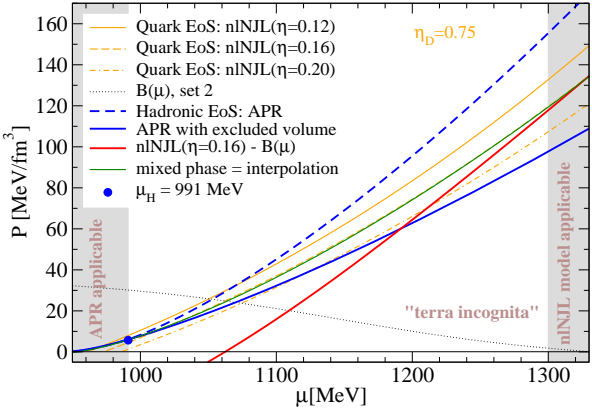}
	\caption{
	Pressure vs. chemical potential for the nlNJL EoS (orange lines) with $\eta_D=0.75$ and the three values $\eta_V=0.12 (0.04) 0.20$ compared 
	to that of the APR EoS (dashed blue line) shows that no reasonable Maxwell construction is possible. When a nucleonic excluded volume is applied to APR (solid blue line) and a density-dependent bag pressure $B(\mu)$ according to set 2 of Ref. \cite{Alvarez-Castillo:2018pve} (black dotted line) to the quark matter EoS, a Maxwell transition point is obtained and a mixed phase construction (green solid line) can be performed which would correspond to an interpolation between APR and nlNJL($\eta_V=0.16$).
	%The point  $\mu_H$ corresponds to $n_H=2.0~n_0$.
	\label{fig:eos2}}
\end{figure}

In Fig.~\ref{fig:eos2} we show how such pathologies of the phase transition construction (or its inapplicability) with too simple EoS which are not suitable for such a construction, could be cured.
A stiffening of the hadronic EoS, here realized by an excluded nucleon volume, leads already to a reasonable transition at not too low densities and to circumvention of the reconfinement problem of a second (unphysical) crossing of hadronic and quark matter EoS at higher densities\footnote{For a discussion of the reconfinement problem see, e.g., \cite{Zdunik:2012dj}, for the related masquerade problem, see \cite{Alford:2004pf} and for their solution see, e.g., \cite{Blaschke:2015uva}.}.  
The situation would still be improved towards a more realistic description when confining effects would be included to the quark matter description, e.g., by a (density-dependent) bag pressure that resembles the effect of a nonperturbative QCD vacuum surrounding color charges (quarks) and leads to their confinement 
in color singlet multiquark states (hadrons).
We note that without such a negative pressure (and/or a confining force), by the larger number of quark and gluon degrees of freedom on the one hand and the larger masses of hadrons in the spectrum on the other, the quark-gluon plasma phase would be favorable over the hadronic matter phase at low temperatures and densities, see Fig.~\ref{fig:eos} and \cite{Satz:2010wxs}.
As we already noted above, within a first-order phase transition, the formation of structures such a bubbles, droplets, rods and plates (pasta phases) is likely, with their sizes defined by an interplay of surface tension and Coulomb interaction effects. 

The resulting pressure (green curve in Fig.~\ref{fig:eos2} looks as if a direct interpolation between the hadronic and the quark matter EoS would have been performed and the underlying three main microphysical ingredients (quark Pauli blocking, quark confinement and pasta structures in the mixed phase) could be circumvented by a direct shortcut from the  nuclear matter phase just above saturation density to the quark matter phase which would then appear as a crossover-like EoS.

Such crossovers have been invoked on physical grounds by symmetry arguments as a quark-hadron continuity in Refs.~\cite{Schafer:1998ef,Schafer:1999pb,Wetterich:1999vd} and by the combined effects of chiral symmetry breaking and diquark condensation intertwined by the axial anomaly so that they result in a crossover at low temperatures which eventually entails a second critical endpoint in the QCD phase diagram 
\cite{Hatsuda:2006ps,Yamamoto:2007ah,Baym:2019iky}.
The crossover behaviour has subsequently been realized in effective interpolating constructions following 
\cite{Masuda:2012kf,Masuda:2012ed,Masuda:2015kha,Whittenbury:2015ziz} and further literature in this direction. 

While in interpolating constructions the information about the composition of the matter in the mixed phase is lost, they allow for conclusions on the likely properties of the pure quark matter EoS that is used as input, once this phase is reached in neutron star interiors.
In contrast, the dominant class of EoS used to extract EoS constraints from observations of masses and radii \cite{Hebeler:2013nza,Kurkela:2014vha,Fraga:2015xha}
as well as tidal deformabilities \cite{Annala:2017llu,Paschalidis:2017qmb} and in near future also the moment of inertia \cite{Greif:2020pju}
are the multi-polytrope extrapolations of the EoS at supersaturation densities. Here we referred only to some prominent examples.
These constraints have also been used to perform 
Bayesian analyses of the most likely EoS, see for instance the recent work of Miller et al. 
\cite{Miller:2019nzo}.
We note that analyses based on the multi-polytrope ansatz for the high-density EoS are agnostic of the composition of matter as well as constraints on the microphysics of dense quark matter.

In this paper, we will perform a Bayesian analysis study with modern mass and radius constraints, as well as fictitious radius measurements, on the basis of a new, two-zone interpolation construction for obtaining hybrid EoS that 
allows for conclusions on the most favorable parameter set for the lagrangian model of color superconducting quark matter with nonlocal chiral interactions of the Nambu--Jona-Lasinio type.
The EoS model, including the new interpolation procedure is described in the next section. 
The results for masses, radii and tidal deformabilities of hybrid stars are outlined in Sect. \ref{sec:masses} and astrophysical inputs for the Bayesian analyses are given in Section \ref{sec:bayes}. 
The results of the Bayesian analysis are presented in Section \ref{sec:results} and in Section \ref{sec:conclusion} we draw the conclusions from this study.

\section{New class of quark-hadron hybrid EoS by two-zone interpolation}
\label{sec:hybrid}
The idea is to interpolate between hadron and quark EoS models 
from the trustable region of the hadronic EoS at the nuclear saturation density ($n_H=1.0 ... 1.5~n_0$) 
to the trustable region for the quark matter model at 
$n_Q\gtrsim 3 n_0$ 
(see fig.~\ref{fig:eos}).
Before we outline in detail the new interpolation method in subsection
\ref{ssec:interpol}, we specify in the following two subsections the hadronic and the quark matter EoS that we employ in the present study to describe the pure phases outside the region of "terry incognita" 
indicated in Figs.~\ref{fig:eos} and \ref{fig:eos2}.

\subsection{Hadronic EoS }

Our choice of hadronic equation of state for this work is the well known APR model \cite{Akmal:1998cf}.  
It is a non-relativistic model derived by means of variational chain summation methods which included Urbana potentials of two and  three nucleon interactions and  features a pion condensate. Moreover, it exhibits a causality breach in neutron star matter for massive stars, a problem that shall not appear for the hybrid star models build in this work. The APR EoS version we have chosen is A$18+\delta$v+UIX$^{*}$ which is not extremely stiff, reaching the maximum neutron star mass right below 2M$_{\odot}$. 

In addition, in order to complete the description of the neutron star matter EoS, we adjoin a low density region EoS corresponding to the crust of neutron stars, namely the SLy4 model~\cite{Douchin:2000kad}.

\subsection{Quark matter EoS}

For the description of the quark matter phase we consider a nonlocal chiral quark model, as in Ref.~\cite{Alvarez-Castillo:2018pve}, which includes scalar
quark-antiquark interaction, anti-triplet scalar diquark
interactions and vector quark-antiquark interactions. The grand canonical thermodynamic potential per unit volume at zero temperature and finite density in the mean field approximation (MFA) reads
\begin{eqnarray}
\Omega^{\rm MFA}  &=&   \frac{ \bar
\sigma^2 }{2 G_S} + \frac{ {\bar \Delta}^2}{2 H} 
- \frac{\bar \omega^2}{2 G_V} \nonumber\\
& &-\frac{1}{2} \int \frac{d^4 p}{(2\pi)^4} \ \ln
\mbox{det} \left[ \ S^{-1}(\bar \sigma ,\bar \Delta, \bar \omega,
\mu_{fc}) \right] ~,
\label{mfaqmtp}
\end{eqnarray}
see  Ref.~\cite{Blaschke:2007ri} for details of the calculation.
The input parameters of the model are determined as to reproduce meson properties in the vacuum, at vanishing temperature and densities, then, $m_c$, $p_0$ (the cutoff) and $G_S$ can be determined under that conditions. 
The remaining coupling constants $G_D$ and $G_V$ are driving the terms that, after bosonization, give rise to the superconducting gap field and the vector field. Then, the ratios $\eta_D=G_D/G_S$ and $\eta_V=G_V/G_S$ are input parameters.  
For OGE interactions in the vacuum, Fierz transformation leads to $\eta_D =3/4$ and $\eta_V = 1/2$. 
As the microscopic interaction is not derived directly from QCD then, the above coupling ratios have in principle no strong phenomenological constraint except for the fact that $\eta_D$ values larger than 
$\eta_D^{*} = (3/2)m/(m - m_c)$ may lead to color symmetry breaking in the vacuum \cite{Zablocki:2009ds} (where $m$ stands for the dressed mass and $m_c$ for the current quark mass). 
In the present work we consider $\eta_D$ and $\eta_V$ as free parameters to be varied in reasonable limits, as it has been done 
for the NJL model case in Ref.~\cite{Klahn:2013kga}.
The mean field values $\bar \sigma$, $\bar \Delta$ and $\bar \omega$ satisfy the coupled equations
\begin{equation}
\frac{ d \Omega^{\rm MFA}}{d\bar \Delta} = 0, \quad 
\frac{ d \Omega^{\rm MFA}}{d\bar \sigma} = 0, \quad 
\frac{ d \Omega^{\rm MFA}}{d\bar \omega} = 0.
\label{gapeq}
\end{equation}
As we are focused on describing the behaviour of quark matter in the core of NSs, we have to impose: equilibrium under weak interactions, chemical equilibrium,  
and color and electric charge neutrality. Then, the six different chemical potentials $\mu_{fc}$ in Eqn. (\ref{mfaqmtp}) (depending on the two quark flavors $u$ and $d$ and quark colors $r,g$ and $b$), can be written in terms of three independent quantities: the baryonic chemical potential $\mu$, the electron chemical potential $\mu_e$ and a color chemical potential $\mu_8$. So basically, for each value of $\mu$ we self-consistently solve the gap equations (\ref{gapeq}),
complemented with  the  conditions  for $\beta$-equilibrium and electric charge and color charge 
neutrality (details of the calculation can be found in the Appendix of Ref.~\cite{Blaschke:2007ri}).

In the present work, we consider a Gaussian form factor $g(p)=\exp (-p^2/p_0^2)$ in Euclidean 4-momentum space. 
The input parameters of the quark model are fixed to $m_c=5.4869$ MeV, $p_0=782.16$ MeV and $G_S p_0^2 = 19.804$ so that the pion mass $m_\pi=139$ MeV, the pion decay constant $f_\pi=92.4$ MeV and the chiral condensate $-\langle \bar{q}q \rangle ^{1/3} = 244$ MeV are reproduced.% \cite{Blaschke:2007ri}. 
As mentioned above, we perform our analysis considering $\eta_D$ and $\eta_V$ as free parameters.

\subsection{
Two-zone interpolation method}
\label{ssec:interpol}
We introduce here a two-zone interpolation scheme that is inspired by the discussion in subsection V.D (Fig.~15) and 
V.G (Fig.~18) of Ref.~\cite{Baym:2017whm},
see also Fig.~\ref{fig:eos2} and its discussion in the 
previous section.
According to that discussion, the interpolated part of the 
hybrid EoS can be motivated as a result of hadronic interactions and many-body forces in the hadronic phase 
(leading to $P_H(\mu) \to P^*_H(\mu)$) and 
confining effects in the quark matter phase (resulting in 
$P_Q(\mu) \to P^*_Q(\mu)$). 
In general, one would have to expect that these effects lead
to a "normal" crossing of hadronic and quark matter $P(\mu)$ curves, but with a "kink" at the location $\mu_c$ of the  crossing point which results in a jump the density
$\Delta n(\mu_c) = dP^*_Q/d\mu|_{\mu_c}-dP^*_H/d\mu|_{\mu_c}>0$.
In this work we discuss the continuous interpolation as the limiting case for $\Delta n (\mu_c)\to 0$.
	\begin{figure}[!ht]
		\centering
		\includegraphics[width=1.0\linewidth]{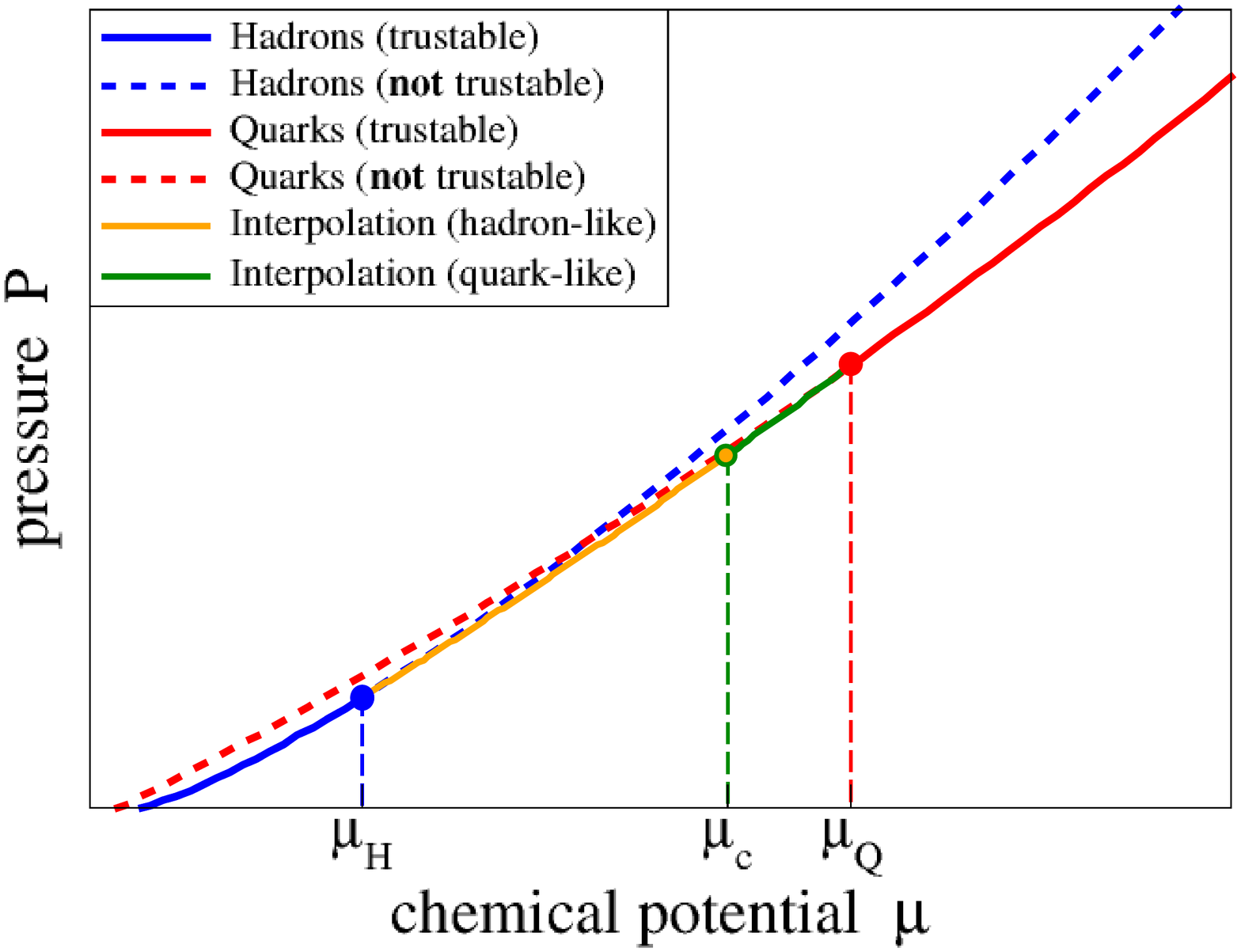}
		\includegraphics[width=1.0\linewidth]{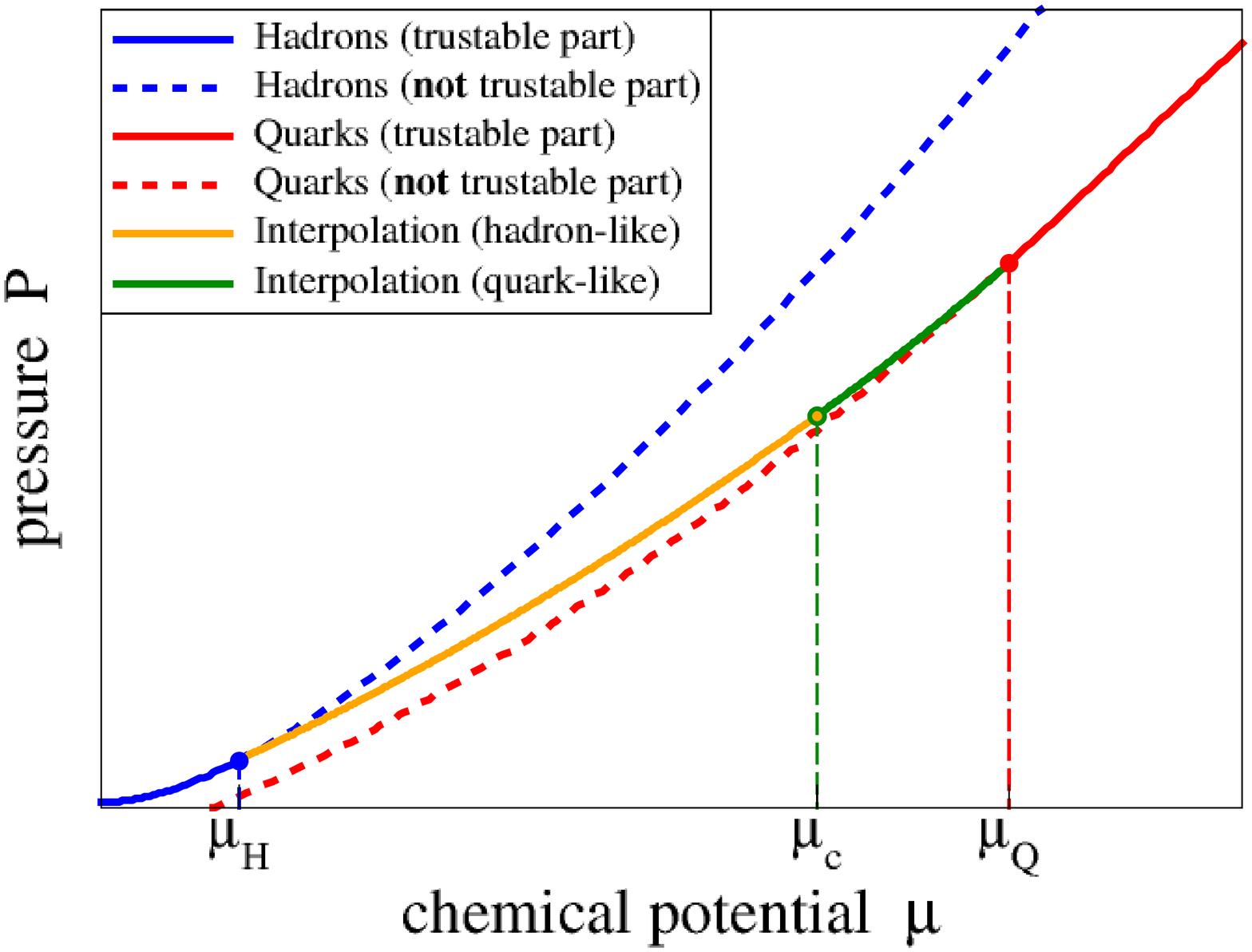}
		\caption{A hybrid equation of state that joins a soft nuclear equation of state with a stiffer quark matter equation of state by interpolation in the intermediate region between $\mu_H$ and $\mu_Q$ for 
		$n(\mu_H) = n_H\sim n_0$ and 
$n(\mu_Q)=n_Q \sim (2.5 \dots 5)n_0$. 
The dotted curves  indicate where the extrapolations of the nuclear and quark matter equations of state become unreliable. 
		We can treat the case where these extrapolations cross each other (upper panel) as well as the case when they do not cross (lower panel).
				\label{fig:interpol}
}
	\end{figure}
An advantage of the two-zone interpolation between $n_H$ and $n_Q$ over a simple single-zone interpolation is that one can obtain good results without employing higher-order polynomials
\cite{Ayriyan:2017nby,Abgaryan:2018gqp},
just by using two parabolic functions
\begin{equation}
\left\{
\begin{array}{lcll}
P_{\eta}(\mu) &=& a_{\eta}(\mu - \mu_H)^2 + b_{\eta}(\mu - \mu_H) + c_{\eta}, & \mu\le\mu_c\\[2mm]
P_{\rho}(\mu) &=& a_{\rho}(\mu - \mu_Q)^2 + b_{\rho}(\mu - \mu_Q) + c_{\rho}, & \mu\ge\mu_c
\end{array}
\right.
\label{eq:interp}
\end{equation}
where $\mu_H$ and $\mu_Q$ correspond to $n_H$ and $n_Q$ respectively, and $\mu_c$ is free parameter taking value between them: $\mu_H<\mu_c<\mu_Q$.

The four parameters $b_{\eta}$, $b_{\rho}$, $c_{\eta}$ and $c_{\rho}$ can be immediately defined from the 
boundary 
conditions
\begin{eqnarray}
P_{\eta}(\mu_H) \equiv  P_H(\mu_H) &=& c_{\eta}\\[2mm]
n_{\eta}(\mu_H) \equiv  n_H(\mu_H) &=& b_{\eta} \\[2mm]
P_{\rho}(\mu_Q) \equiv P_Q(\mu_Q) &=& c_{\rho}\\[2mm]
n_{\rho}(\mu_Q) \equiv n_Q(\mu_Q) &=& b_{\rho}
\label{eq:main_conds}
\end{eqnarray}
The remaining parameters $a_{\eta}$, $a_{\rho}$ are defined by the 
matching conditions at at the intermediate 
$\mu_c$ where both functions should be sewed 
\begin{equation}
\left\{
\begin{array}{lcl}
P_{\eta}(\mu_c) & = & P_{\rho}(\mu_c)\\[2mm]
n_{\eta}(\mu_c) & = & n_{\rho}(\mu_c) 
- \Delta n(\mu_c)~.
\end{array}
\right.
\label{eq:aa_conds}
\end{equation}
These conditions 
are equivalent to 
the following system of linear algebraic equations (SLAE)
\begin{equation}
\left\{
\begin{array}{lcl}
a_{\eta}(\mu_c - \mu_H)^2 - a_{\rho}(\mu_c - \mu_Q)^2 & = & \kappa_1 \\[4mm]
2a_{\eta}(\mu_c - \mu_H)  -  2a_{\rho}(\mu_c - \mu_Q)&  = & \kappa_2~,
\end{array}
\right.
\label{eq:slae}
\end{equation}
where
\begin{equation}
\left\{
\begin{array}{lcl}
\kappa_1 & = & n_Q(\mu_c - \mu_Q) - n_H(\mu_c - \mu_H) + P_Q  - P_H,   \\[4mm]
\kappa_2 & = & n_Q - n_H
- \Delta n(\mu_c)~.
\end{array}
\right.
\label{eq:kappas}
\end{equation}
The determinant of this SLAE 
\begin{equation}
\Delta = 2(\mu_c - \mu_Q)(\mu_c - \mu_H)(\mu_H - \mu_Q)
\label{eq:det}
\end{equation}
shows that the system has always a solution when $\mu_c \neq \mu_Q$, $\mu_c \neq \mu_H$ and $\mu_H \neq \mu_Q$.
The solution to the SLAE is
\begin{equation}
\left\{
\begin{array}{lcl}
a_{\eta} & = &  \displaystyle \frac{-2\kappa_1+\kappa_2(\mu_c-\mu_Q)}{2(\mu_c - \mu_H)(\mu_H - \mu_Q)}~, \\[8mm]
a_{\rho} &=& \displaystyle \frac{-2\kappa_1+\kappa_2(\mu_c-\mu_H)}{2(\mu_c - \mu_Q)(\mu_H - \mu_Q)} ~.
\end{array}
\right.
\label{eq:solution}
\end{equation}
We note, that this two-zone interpolation allows for a generalization to describe a first-order phase transition. 
This could be achieved by 
choosing a nonzero value for the jump in the density $\Delta n (\mu_c)$, which appears as additional parameter in the
second equation of the system~\eqref{eq:aa_conds}.

\subsection{Constant speed-of-sound representation}
For the nlNJL model EoS following from (\ref{mfaqmtp}),
a causality violation at high energy densities 
(which corresponds to very high chemical potentials) 
appears due to a backbending of the quark pressure as a function of the energy density. 
To circumvent that obstacle, we make use of the recently discovered fact that in the range of densities relevant for NS applications the nlNJL model can be represented with high accuracy by a constant-speed-of-sound (CSS) EoS \cite{Antic:2021zbn}.
This EoS is given by \cite{Alford:2013aca,Blaschke:2020vuy}
\begin{equation}
\left\{
\begin{array}{lcl}
P(\mu) & = & P_0 + P_1 \left(\displaystyle \frac{\mu}{\mu_x} \right)^{1+1/c_s^2} %& \mathrm{for~} \mu>\mu_x
\\[4mm]
\varepsilon(\mu) & = &-P_0 + P_1 \frac{1}{c_s^2} \left(\displaystyle \frac{\mu}{\mu_x} \right)^{1+1/c_s^2} 
%& \mathrm{for~} \mu>\mu_x
\\[4mm]
n_B(\mu) & = & P_1 \displaystyle \frac{1+1/c_s^2}{\mu_x} \left(\displaystyle \frac{\mu}{\mu_x} \right)^{1/c_s^2} 
%& \mathrm{for~} \mu>\mu_x
\end{array}
\right.
\label{eq:ccs}
\end{equation}
Here $\mu_x$ is a scale for the chemical potential 
which we set to $\mu_x=1$ GeV in correspondence with \cite{Antic:2021zbn,Blaschke:2020vuy}.
The parameters $P_0$, $P_1$ and $c_s^2$ are obtained from 
the parameters $\eta_D$ and $\eta_V$ of the nlNJL model 
by a functional mapping that has been defined in Ref.~\cite{Antic:2021zbn}.
We note that the values for the squared sound speed obtained 
by this fit are in the range $c_s^2 = 0.45 \dots 0.54$.

\iffalse
where the matching to the CSS extrapolation starts. 
The parameter $\beta$ is directly related to the squared speed of sound  
\begin{equation}
c_s^2=\frac{\partial P/ \partial \mu}{\partial \varepsilon / \partial \mu} = \frac{1}{\beta - 1}
\label{eq:css}
\end{equation}
so that
\begin{equation}
\beta  = 1 + \displaystyle \frac{1}{c_s^2} ~.
\label{eq:css_beta}
\end{equation}
It is obvious that fulfilling the causality constraint $c_s^2 \le 1$ entails that $\beta \ge 2$. 
The coefficients $P_0$ and $P_1$ in Eq. (\ref{eq:ccs}) are defined as
\begin{eqnarray}
P_0 & = & \left[(\beta - 1) P_x - \varepsilon_x \right]/\beta\\
P_1 & = & \left(P_x + \varepsilon_x \right) / \beta~,
\label{eq:css_pp}
\end{eqnarray}
where $P_x=P(\mu_x)$ and $\varepsilon_x=\varepsilon(\mu_x)$.

The results below have been obtained for the choice $\beta=2$, i.e. $c_s^2 = 1$, with $\mu_x$ being the chemical potential at which for first time the squared speed of sound of the quark matter EoS reached the causality limit, $c_s^2(\mu_x)=1$. 
For those EoS which have no causality violation the CSS extrapolation has been performed with $c_s^2$ corresponding to the value at end of the given table.
\fi

\subsection{Hybrid EoS from a two-zone interpolation}

\begin{figure}[ht]
	\centering
	\includegraphics[width=\linewidth]{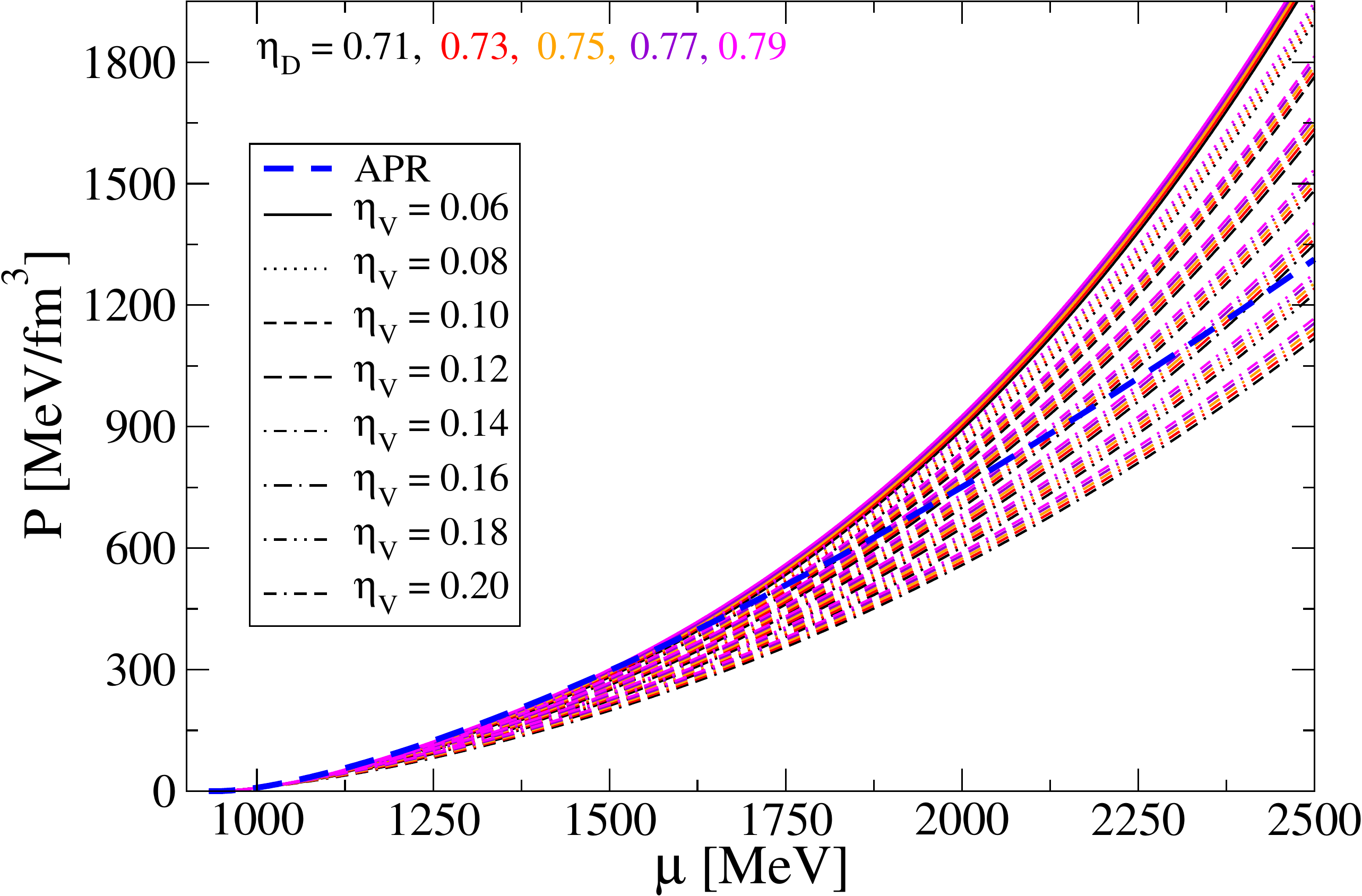}
	\caption{Two-zone interpolation construction between APR and nlNJL on $P$-$\mu$ plot.
	\label{fig:p_mu}}
\end{figure}

\begin{figure}[ht]
	\centering
	\includegraphics[width=\linewidth]{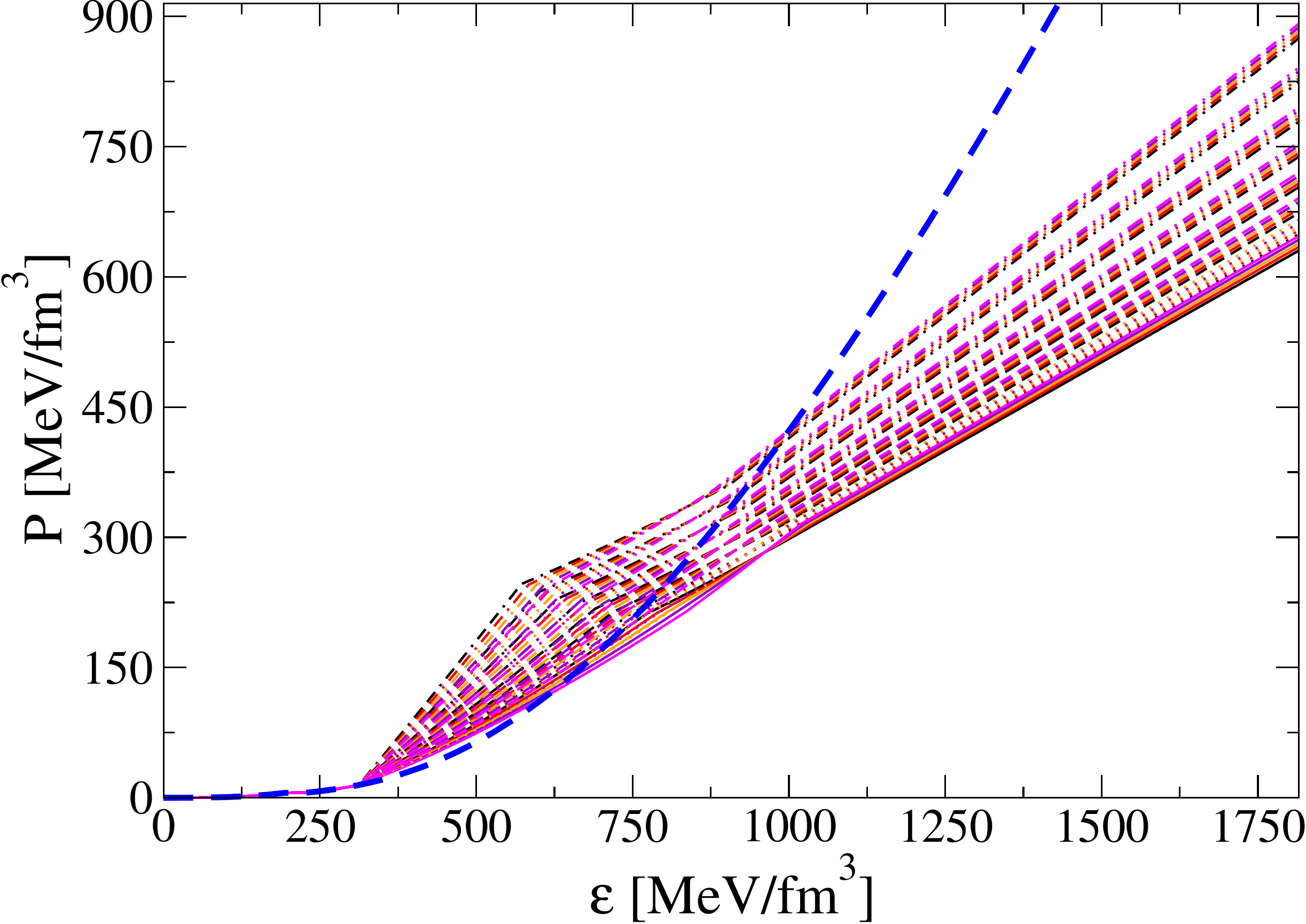}
	\caption{Two-zone interpolation construction between APR and nlNJL on $P$-$\varepsilon$ plot.
	Legend as in Fig.~\ref{fig:p_mu}.
	\label{fig:p_eps}}
\end{figure}

\begin{figure}[ht]
	\centering
	\includegraphics[width=\linewidth]{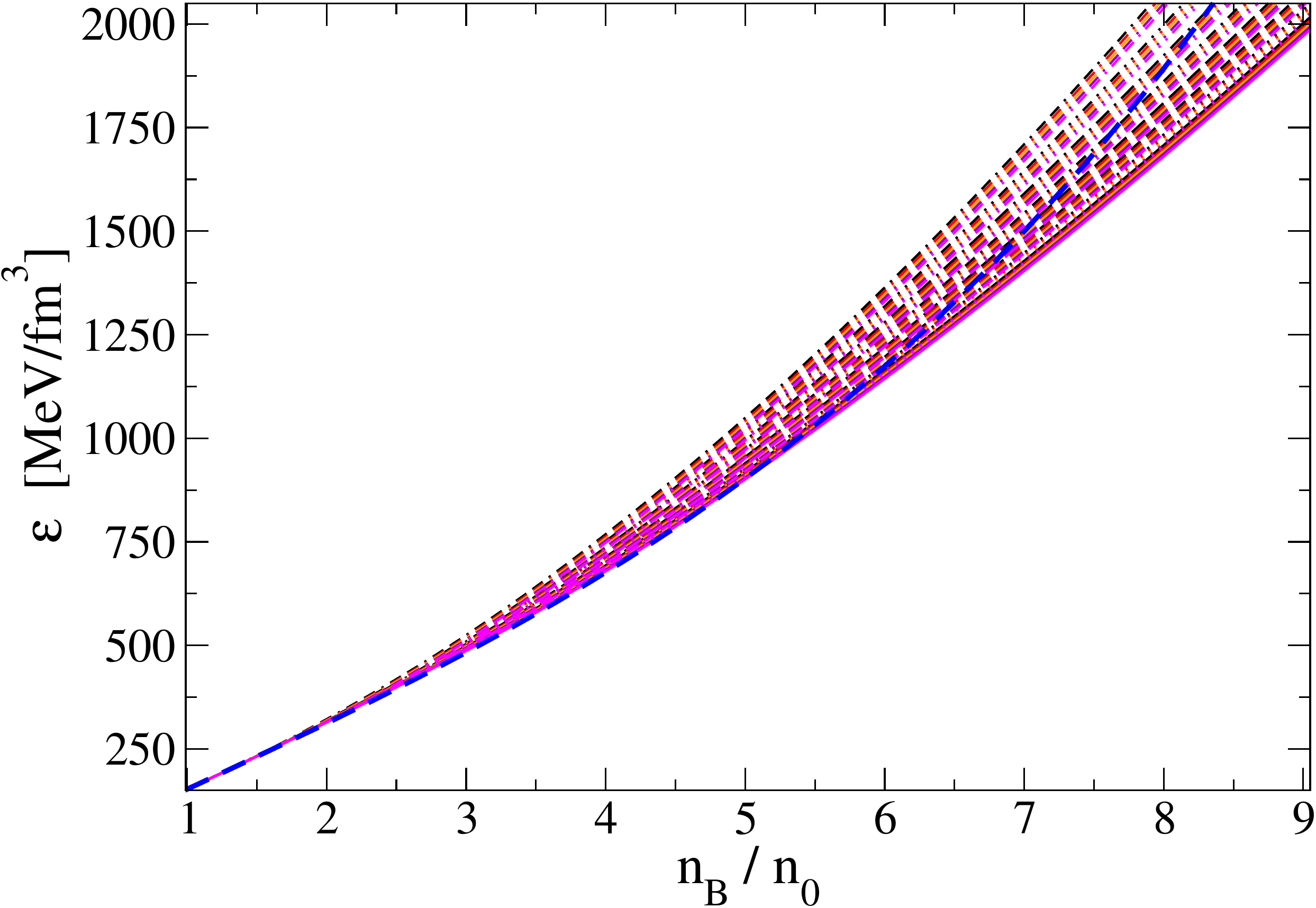}
	\caption{Energy density dependence on the baryon density for the EoS curves under consideration.
	Legend as in Fig.~\ref{fig:p_mu}.
	\label{fig:epsilon_n}
	}
\end{figure}

\begin{figure}[ht]
	\centering
	\includegraphics[width=\linewidth]{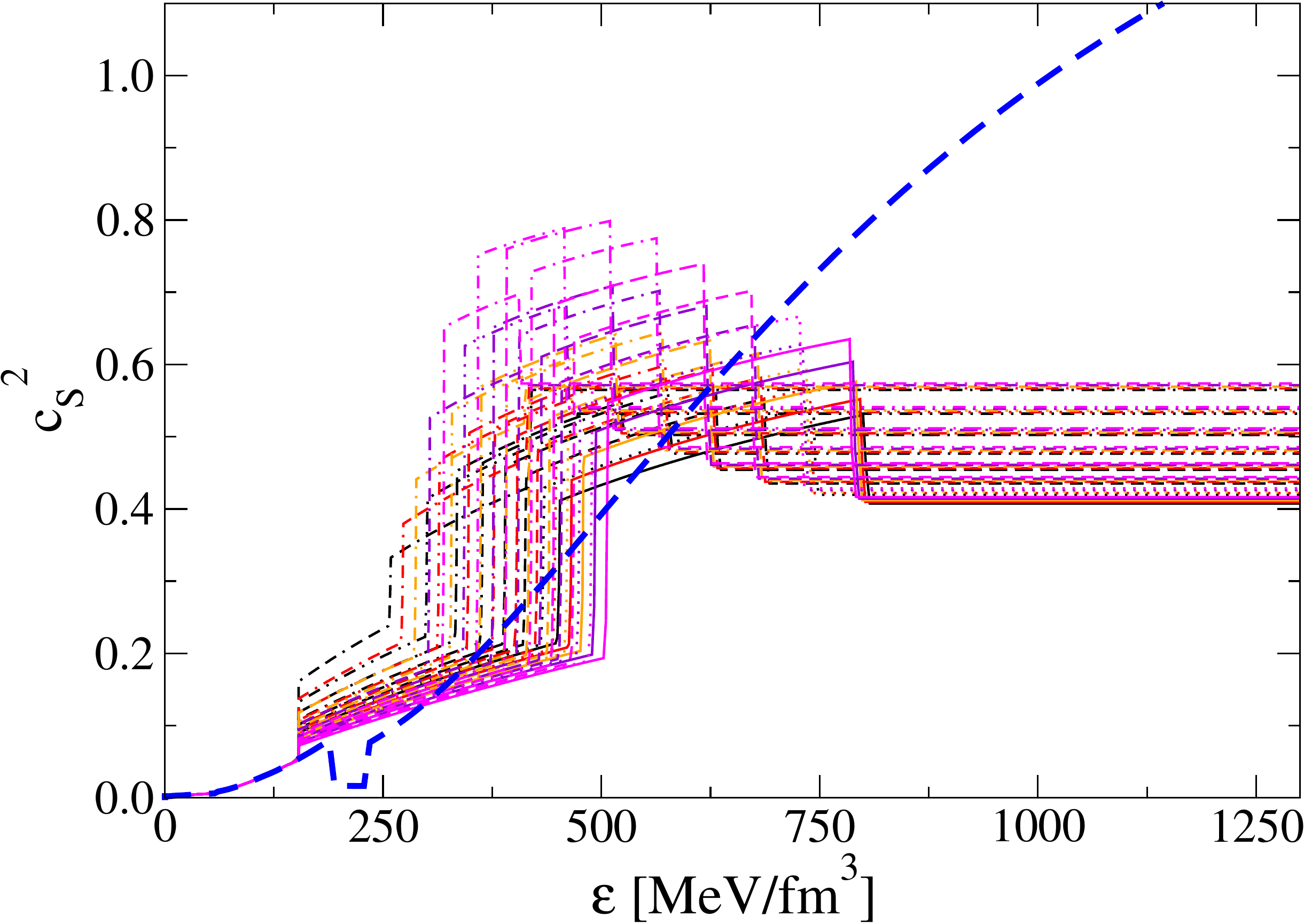}
	\caption{Speed of sound vs energy density for the two-zone interpolation construction between APR and nlNJL. 
	Legend as in Fig.~\ref{fig:p_mu}.
	\label{fig:cs2_eps}}
\end{figure}

\begin{figure}[ht]
	\centering
	\includegraphics[width=\linewidth]{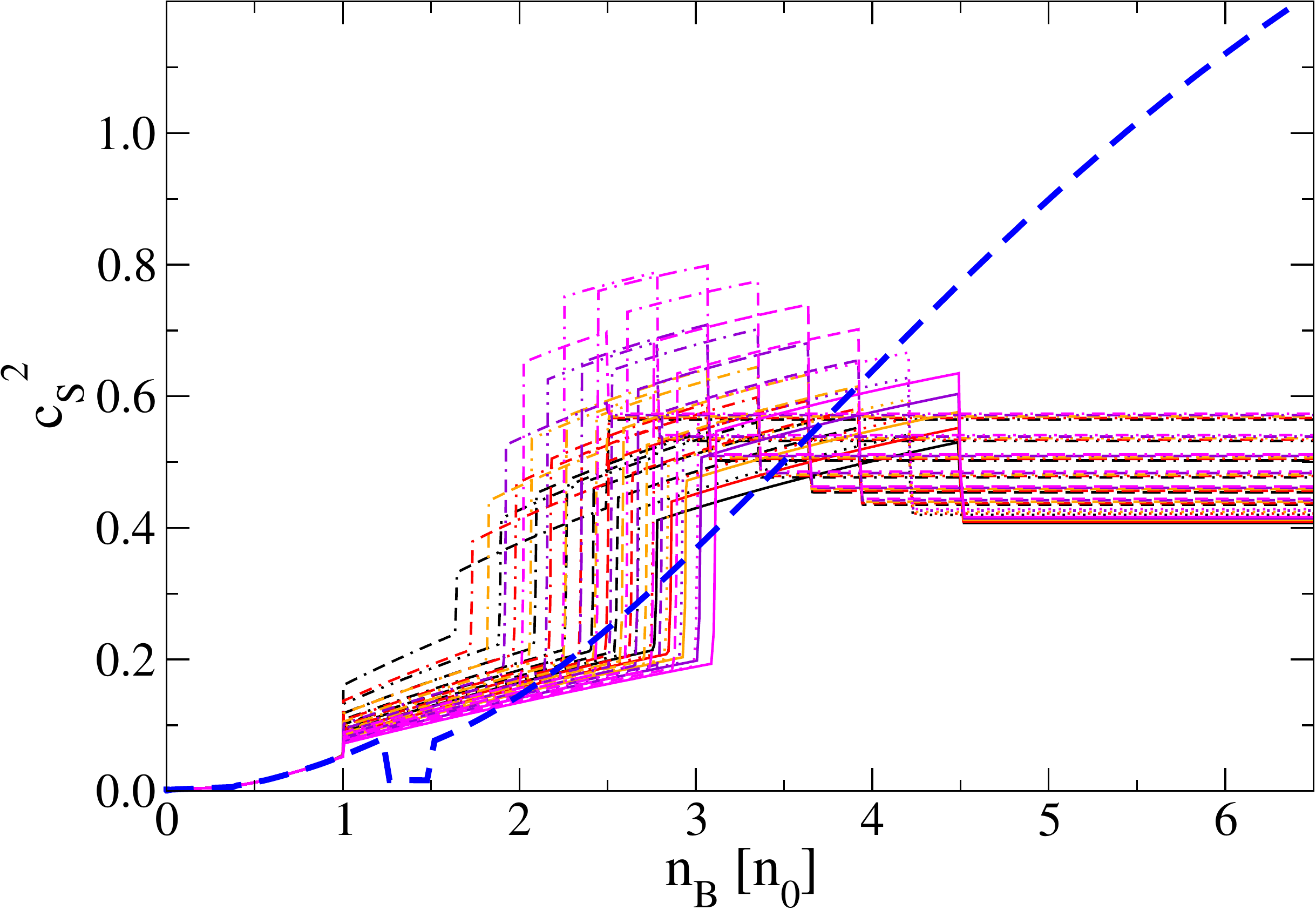}
	\caption{Speed of sound vs. density for the two-zone interpolation construction between APR and the set of nlNJL quark matter EoS from Fig.~\ref{fig:p_eps}.
	Legend as in Fig.~\ref{fig:p_mu}.
	\label{fig:cs2_n}}
\end{figure}

The interpolation method has been implemented for APR and NJL models with different  $\eta_D = 0.71 (0.02) 0.79$ and 
$\eta_V = 0.06 (0.02) 0.20$ (see~Figs.~\ref{fig:p_mu}--\ref{fig:cs2_n}). The onset density for the interpolation has been set to 
$n_H=n_0$, 
and the density where the interpolation matches the quark matter EoS has been varied depending on $\eta_V$ as 
$n_Q = 4.5 \dots 2.5\,n_0$ 
while simultaneously $\eta_V$ was incremented. The value of $\mu_c$ has been fixed as $\mu_c = \mu_H+0.75(\mu_Q-\mu_H)$.

It is interesting to observe in Figs.~\ref{fig:cs2_eps} and \ref{fig:cs2_n} the similarity in the behaviour of the squared speed of sound with that of recent models of quarkyonic compact star matter \cite{McLerran:2018hbz}.
A fast rise (stiffening of nuclear matter) is followed by a dip (hadron-quark mixed phase) and then saturates at an asymptotic value. This value in our case is the value obtained for the CSS fit of the nlNJL model and ranges between $0.45$ and $0.54$. 
Since the present neutron star phenomenology requires a stiffness of the EoS that results in a maximum mass of at least $2~M_\odot$, the corresponding energy densities that are probed by neutron star interiors do not exceed about 1 GeV/fm$^3$, see Fig.~\ref{MvsE}.
This means that with the present setting of the two-zone interpolation construction, the pure quark matter phase for $n_B > n_Q$ is barely reached in stable neutron star configurations, if at all.

\section{Masses, radii and tidal deformabilities for the hybrid EoS}
\label{sec:masses}

\begin{figure}[ht]
	\centering
	\includegraphics[width=\linewidth]{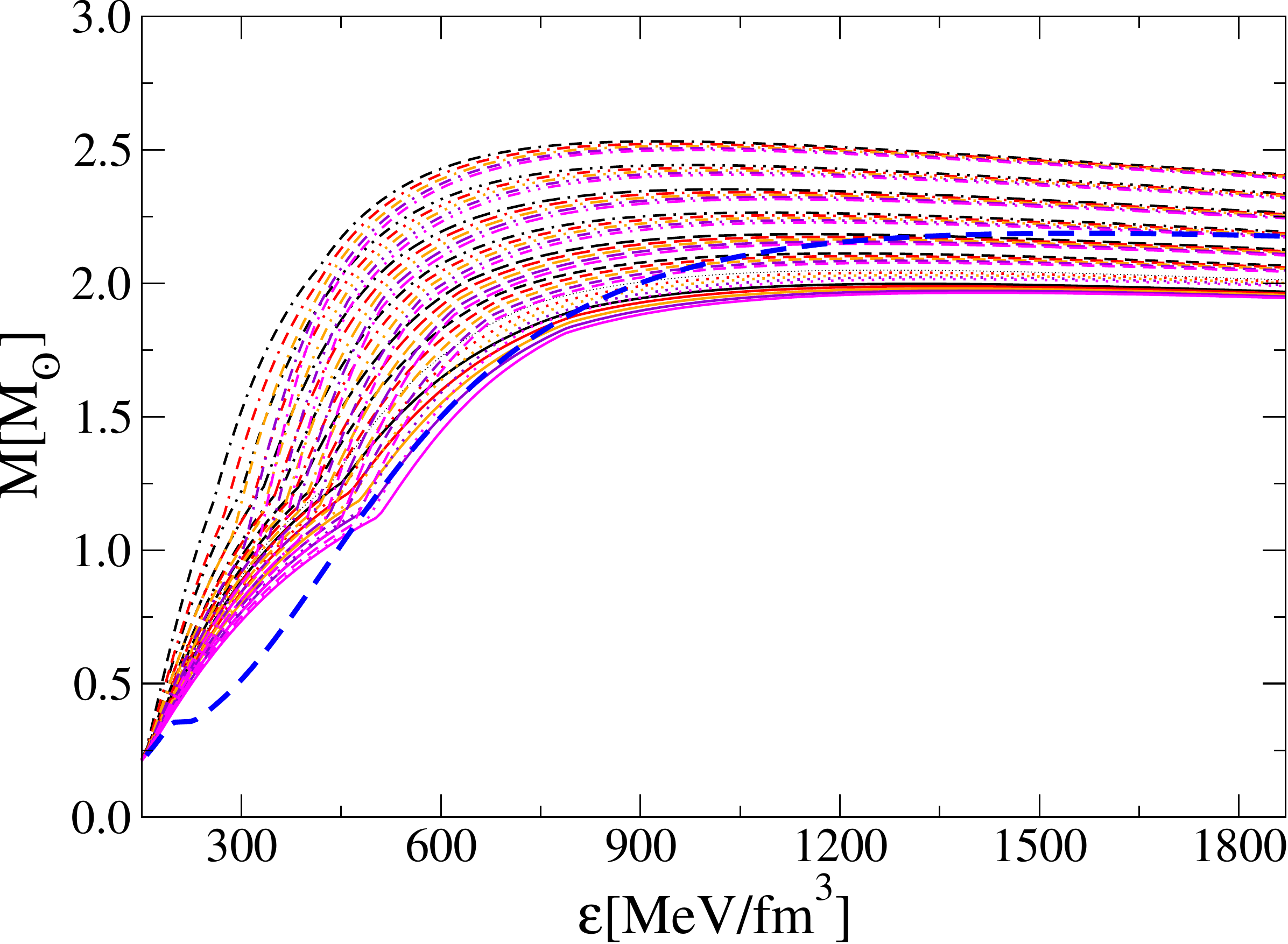}
	\caption{Mass vs. central energy density for the two-zone interpolation construction between APR and the set of nlNJL quark matter EoS.
	Legend as in Fig.~\ref{fig:p_mu}.
	}
	\label{MvsE}
\end{figure}

\begin{figure}[ht]
	\centering
	\includegraphics[width=\linewidth]{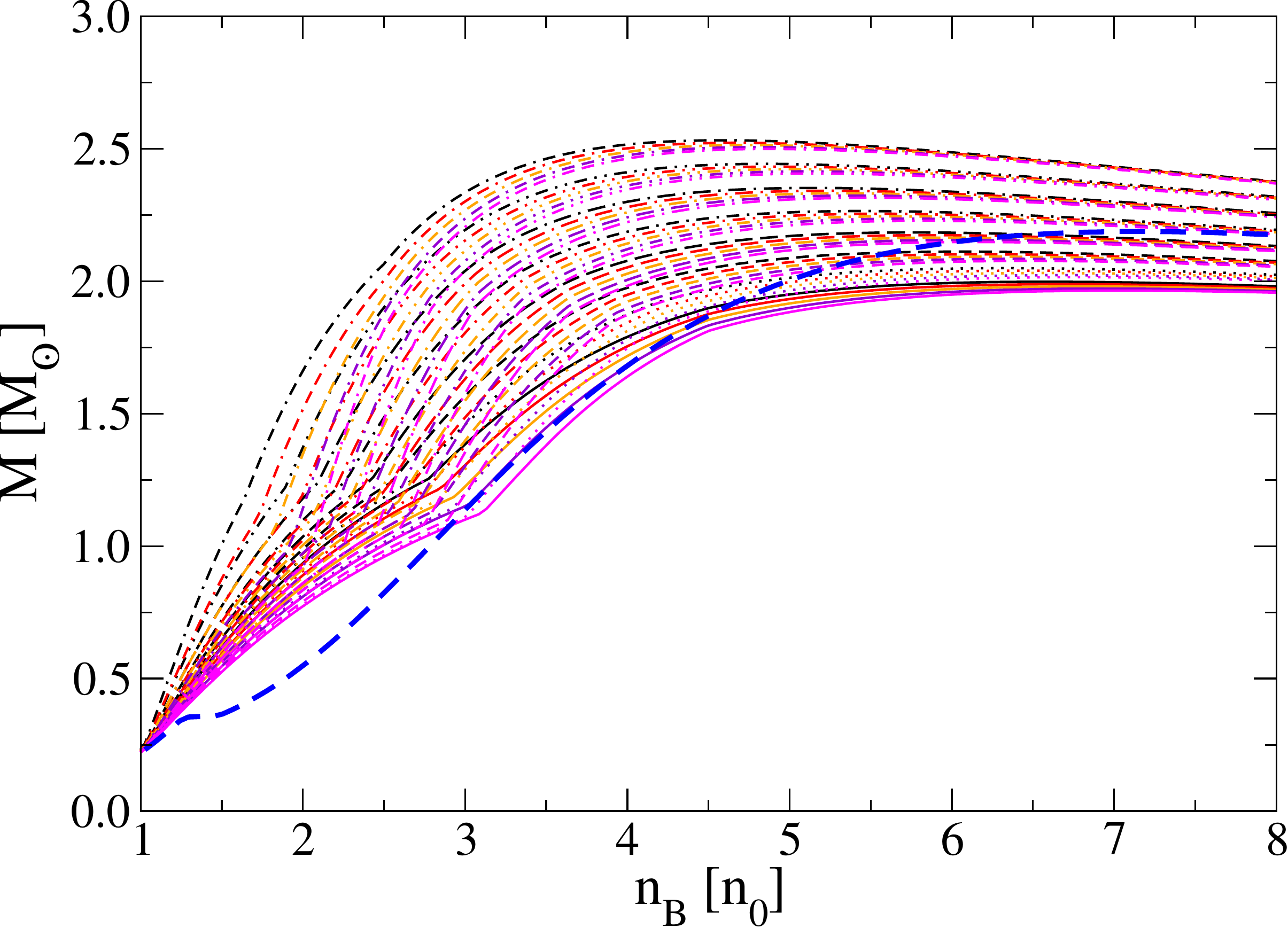}
	\caption{Mass vs. central baryon density for the two-zone interpolation construction between APR and the set of nlNJL quark matter EoS.
	Legend as in Fig.~\ref{fig:p_mu}.
	}
	\label{MvsnB}
\end{figure}

\begin{figure}[ht]
	\centering
	\includegraphics[width=\linewidth]{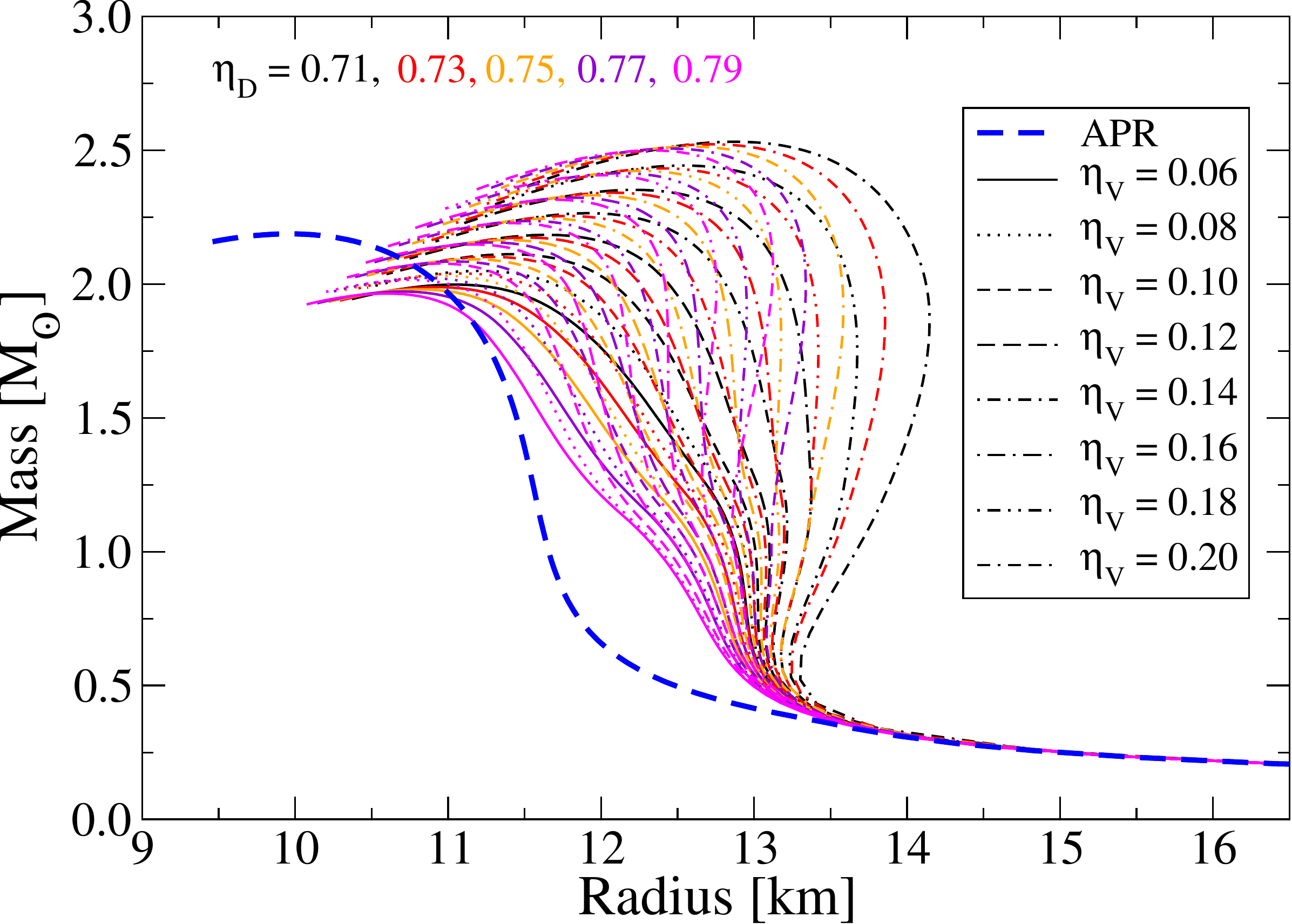}
	\caption{Mass-radius relations for the two-zone interpolation construction between APR and nlNJL.
	\label{fig:M-R}}
\end{figure}

Neutron stars are computed within the framework of general relativity by using the corresponding EoS in the form of $p(\varepsilon)$ to solve the Tolman-Oppenheimer-Volkoff (TOV) equations~\cite{Tolman:1939jz,Oppenheimer:1939ne}
\begin{eqnarray}
 \label{TOV}
\frac{\dd p( r)}{\dd r}&=& %\frac{-G }{r}
-\frac{\left(\varepsilon( r)+p( r)\right)
\left(m( r)+ 4\pi r^3 p( r)\right)}{r\left(r- 2m( r)\right)},\\
\frac{\dd m( r)}{\dd r}&=& 4\pi r^2 \varepsilon( r).
\label{eq:TOVb}
 \end{eqnarray}
which describe a static, spherical star. Radial mass $m( r)$, energy density $\varepsilon( r)$ and pressure $p( r)$ stellar internal profiles help to determined the mass $M$ and radius $R$ of a star with central density $\varepsilon_c$ with boundary conditions $m(r=0)=0$ and $p(r=R)=0$. 
In Fig. \ref{MvsE} and Fig. \ref{MvsnB} we show the compact star mass as a function of the central energy density and baryon density, respectively, for the two-zone interpolation construction.  The dashed blue line corresponds to the hadronic APR EoS and the remaining curves show the hybrid EoS for a range of input parameters for the quark matter phase (same patters/colors as in previous figures).
The compact star mass-radius sequence is a benchmark for every EoS model commonly presented in a mass-radius diagram that includes pulsar measurement regions as well as excluded ones by other astrophysical observations, see for instance Fig.~\ref{fig:M-R}. The EoS sequences displayed in there are obtained by a systematic integration of the the TOV equations for increasing $p_c$ for each single star up to the value of the maximum mass for which the condition $\partial M/ \partial \varepsilon_c>0$ holds.

In addition, deformation of the compact star is a feature of the EoS that is closely related to the last moments of the inspiral phase of compact star mergers. It is quantified by computing the tidal deformability $\Lambda$, for which estimated regions were derived by the observation of the GW170817 event~\cite{TheLIGOScientific:2017qsa,Abbott:2018exr}, also displayed in Fig.~\ref{fig:M-R}. The corresponding equations are derived from perturbations of the spherical metric of the compact star supplemented with the stellar internal profiles for physical quantities derived from the TOV equations. The equation
\begin{equation} 
\Lambda = \frac{2}{3} \frac{ R^{5}}{M^{5}} k_2
\end{equation}
relates the dimensionless tidal deformability $\Lambda$ with the tidal Love number $k_2$ and the total mass and radius of the star. Details on the derivation of the above formula can be found in~\cite{Hinderer:2007mb,Damour:2009vw,Binnington:2009bb,Yagi:2013awa,Hinderer:2009ca}. 
\begin{figure}[ht]
	\centering
	\includegraphics[width=1.0\linewidth]{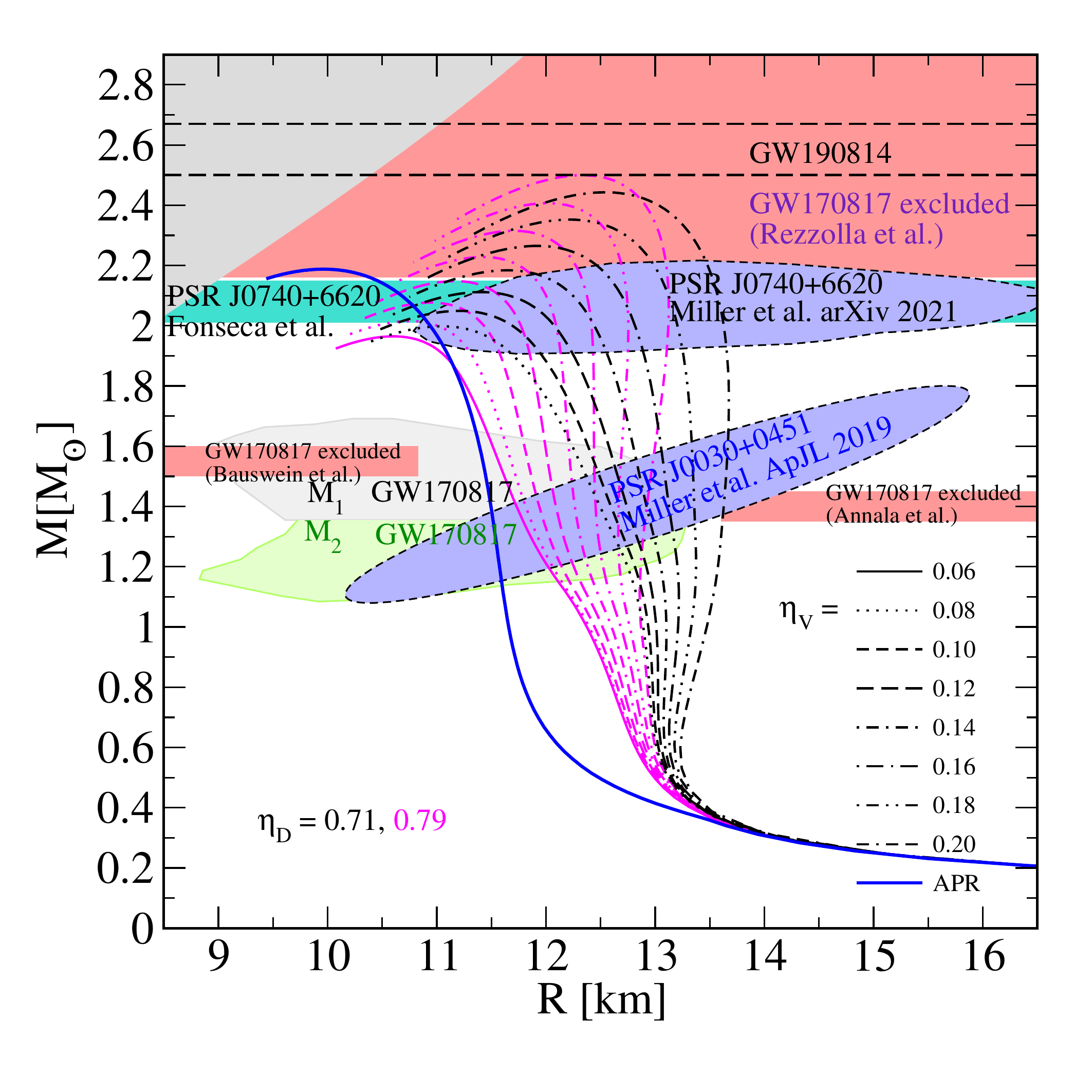}
	\caption{Mass-radius relations for EoS models featuring an interpolation scheme~\cite{Baym:2017whm} between the low density APR model for hadronic matter and high density nlNJL quark matter. A few compact star sequences are displayed for two fixed quark matter parameter values of $\eta_{D}$ with varying $\eta_{V}$ together with the hadronic APR EoS. Different color regions correspond to either pulsar measurements or forbidden regions that serve as  constraints for the compact star EoS. The green band region above 2M$_{\odot}$ corresponds to the updated mass measurement of PSR J0740+6620~\cite{Fonseca:2021wxt},
	which was recently upgraded to a mass-radius measurement 
	by NICER, shown by the blue ellipsoidal region for the result of the Maryland-Illinois team~\cite{Miller:2021qha}.
	The other blue ellipse corresponds to the mass and radius measurement of PSR J0030+0451 by NICER~\cite{Miller:2019cac}
	whereas
	the grey and light green regions correspond to the estimates of the components of the binary system labeled as $M_1$ and $M_2$ of the GW170817 merger~\cite{Abbott:2018exr}. Red bands correspond to excluded regions derived from GW170817 observations by  Bauswein et al.~\cite{Bauswein:2017vtn}, Annala et al.~\cite{Annala:2017llu} and Rezzolla et al.~\cite{Rezzolla:2017aly}. The black dashed horizontal lines are the upper and lower limit for the mass $2.59^{+0.08}_{-0.09}~M_{\odot}$ of the lighter component in the binary merger event GW190814 \cite{Abbott:2020khf}.
	\label{fig:M-R_final}}
\end{figure}

\begin{figure}[ht]
	\centering
	\includegraphics[width=\linewidth]{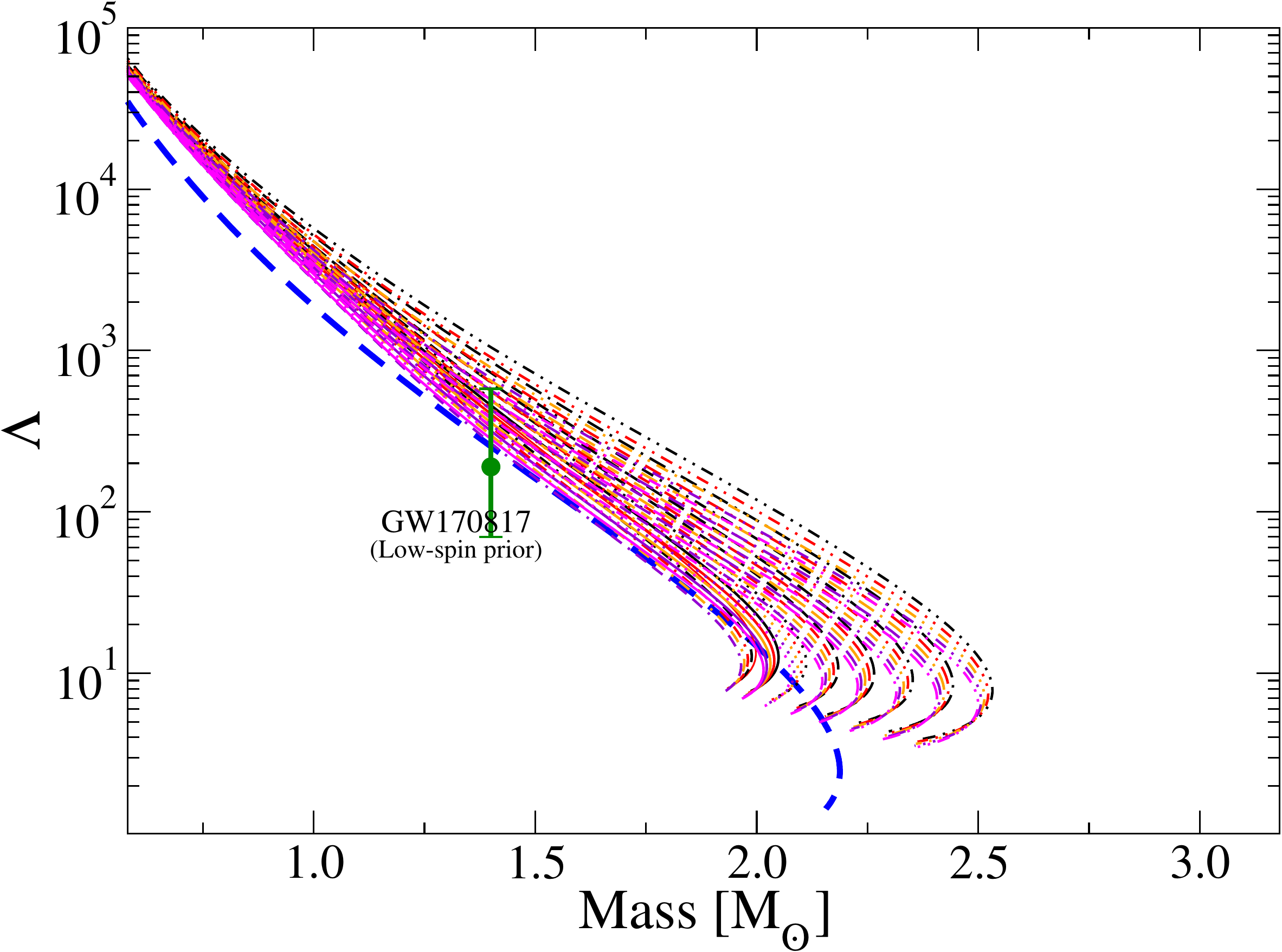}
	\caption{Dimensionless tidal deformabilities of hybrid compact stars together with the corresponding measurement from GW170817~\cite{Abbott:2018exr}.
	\label{fig:L-M}}
\end{figure}

\begin{figure}[ht]
	\centering
	\includegraphics[width=\linewidth]{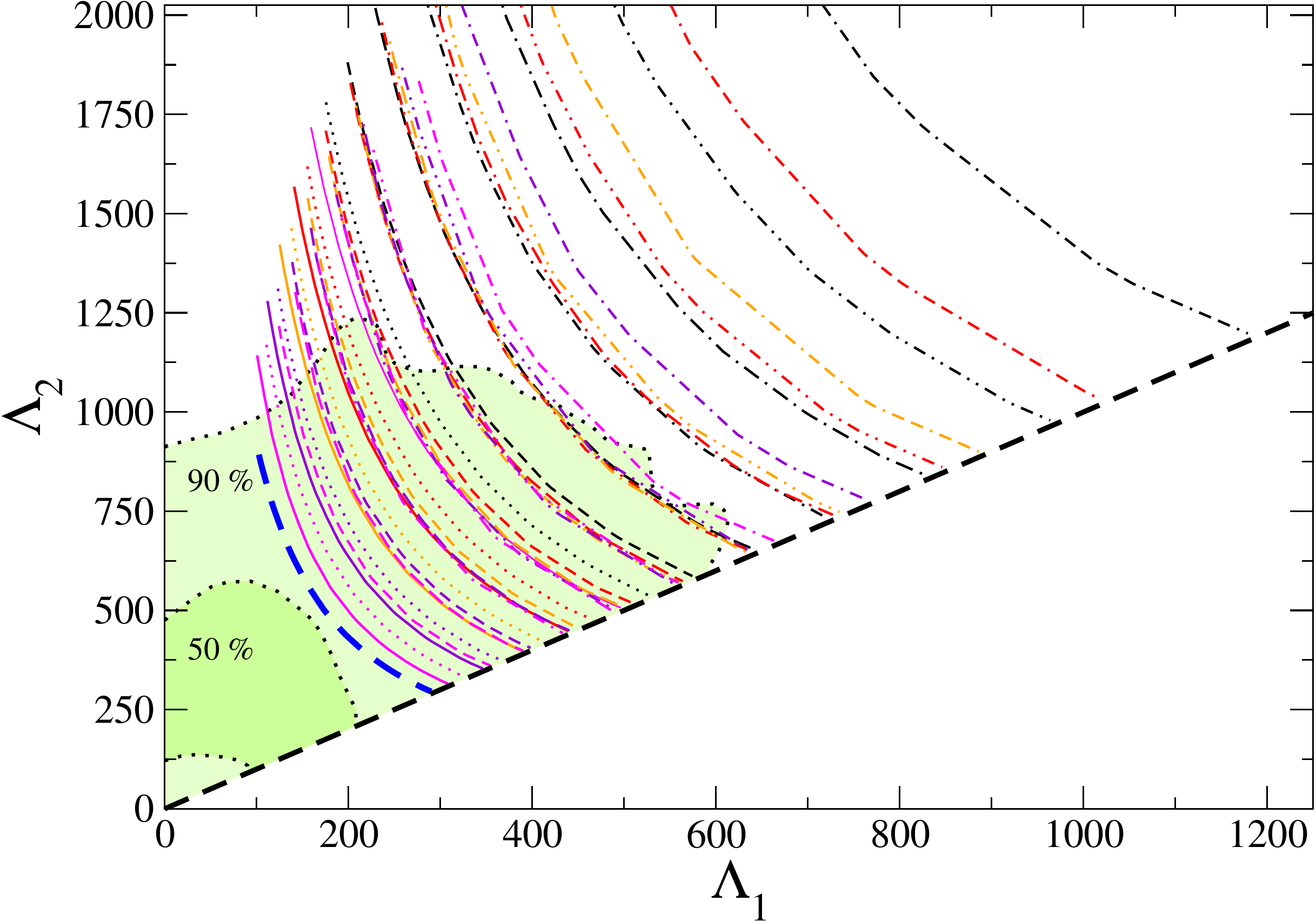}
	\caption{Tidal deformabilities diagram. $\Lambda_1$ and $\Lambda_2$ correspond to the dimensionless tidal deformability for each of the components of the binary in GW170817. Light and dark green regions correspond to the 50\% and 90\% credibility regions for the posteriors used in the LIGO-Virgo analysis~\cite{Abbott:2018exr}. All the EoS in this work result in curves that fall inside the lighter region as well as the APR EoS which is displayed in blue.
	\label{fig:L1-L2}}
\end{figure}

\section{Bayesian inference}
\label{sec:bayes}

In this section we introduce the Bayesian methodology and its application to the set of considered EoS characterized by the parameters $\eta_D$ and $\eta_V$ in order to find their best values that fulfill observational data.

The \textit{a\,posteriori} probability $P\left(\pi_{q}\vert E\right)$ is a conditional probability of the given vector of parameters $\bm{\pi}_q$ (introduced below), where $q$ denotes the indexes the values of parameters for each alliterative representation of the model of EoS. 
The condition $E$ is the set of the observational data (events), its likelihood for the given model is represented as a product
\begin{equation}
	\label{eq:p_event}
	P\left(E\vert\bm{\pi}_{q}\right)= \prod_{\alpha} P\left(E_{\alpha}\vert \bm{\pi}_{q}\right),
\end{equation}
which is the the conjunction of all events $E_\alpha$ (where $\alpha$ is an index of an event).
The \textit{a\,posteriori} probabilities, likelihoods and  \textit{a\,priori} probabilities are connected to each other via the Bayes formula
%$P(\pi_q\vert E_\alpha)$
\begin{equation}
	\label{eq:bayes}
	P\left(\pi_{q}\vert E\right)=\frac{P\left(E\vert\bm{\pi}_{q}\right)P\left(\bm{\pi}_{q}\right)}{\sum\limits _{p=0}^{N-1}P\left(E\vert \bm{\pi}_{p}\right)P\left(\bm{\pi}_{p}\right)},
\end{equation}
where the factor $P\left(\bm{\pi}_{q}\right)$ is the prior of a given model.

First, we define the set of values of the parameters $\eta_D$ and $\eta_V$ as 
\begin{equation*}
H_D = \lbrace0.71, 0.73, 0.75, 0.77, 0.79\rbrace
\end{equation*}
and  
\begin{equation*}
H_V = \lbrace0.06, 0.08, 0.10, 0.12, 0.14, 0.16, 0.18, 0.20\rbrace .
\end{equation*}

Then, the vector of parameters $\bm{\pi}_q $ will be an element of the set $H_D \times H_V $,
\begin{equation}
\label{pi_vec}
\bm{\pi}_q \in \lbrace {\eta_D}_{(i)}, {\eta_V}_{(j)}\,\,\vert\,\,i=0..N_D-1, j=0..N_V-1\rbrace,
\end{equation}
where $q = i N_V + j$ and $N_V = 8$, $N_D = 5$. Therefore,  $q=0..N-1$ and $N = 40$ is the full number of model representations. 
For the choice of the uniform distribution of the 
prior we have
\begin{equation*}
P\left(\bm{\pi}_{q}\right)=1/N.     
\end{equation*}
In the next section, we discuss the specific astrophysical constraints that we will employ in our Bayesian analysis (BA).

\section{Astrophysical constraints}
\label{constr}

\subsection{Lower limit of maximum mass}

Recently, the mass of the PSR\,J0740+6620 was obtained recently by a Shapiro delay based measurement combining data from the North American Nanohertz Observatory for Gravitational Waves (NANOGrav) and data from the orbital-phase-specific observations using the Green Bank Telescope. 
The 68.3\% of the credibility interval was given as $2.14^{+0.10}_{-0.09}\,M_{\odot}$ in~\cite{Cromartie:2019kug},
but it has recently been updated to $2.08\pm 0.07\,M_{\odot}$ in~\cite{Fonseca:2021wxt}. This value has been chosen as~the~\textit{lower limit of maximum mass} ($M_{\mathrm{max}}^{\mathrm{low}}$).

The likelihood for the lower limit of maximum mass constraint is given by
\begin{equation}
\label{eq:lhoodLowMass}
P\left(E_{M}\left|\pi_q\right.\right) = \Phi(M_q, \mu_l, \sigma_l)~,
\end{equation}
where $M_q$ is the maximum mass of the sequence of neutron star configurations for the given $\pi_q$, and $\Phi(M, \mu, \sigma)$ is the {\it cumulative distribution function} (CDF) of the standard normal distribution. 
And $\mu_l$ and $\sigma_l$ are the parameters of the uncertainty of a low limit maximum mass estimation.

Additionally, the assumption that one of the component of the binary merger GW190814 is a neutron star gives an estimation of the lower limit of the maximum mass as $2.59^{+0.08}_{-0.09}\,M_{\odot}$~\cite{Abbott:2020khf}. This value has been used in the Bayesian Inference as an alternative scenario for the lower limit.
In Fig. \ref{fig:M-R_final} we display the mass-radius relations for our hybrid configurations together with the APR model as the low density baseline for hadronic matter that becomes invalid at higher densities where it crosses over to the nlNJL  quark  matter model.  The different colored regions correspond to either mass and/or radius measurements or to forbidden regions following from GW170817 phenomenology that serve as constraints for the compact star EoS.
The horizontal black dashed lines show the mass range for the lighter object in the binary merger GW190814, that we employ as a possible lower limit on the maximum mass, in the case that this object was a neutron star.
\subsection{Upper limit of maximum mass}
There is an estimation of the upper limit of maximum mass of neutron star in the literature~\cite{Rezzolla:2017aly}. It was estimated with combination of the observation of gravitational waves (GW170817) and drawing from basic arguments on kilonova modeling of GRB 170817A, together with the quasi-universal relation between the maximum masses of static neutron stars and the fastest stable star under uniform rotation~\cite{Breu:2016ufb}. The upper limit of the maximum mass is  
$2.16^{+0.17}_{-0.15}\,M_{\odot}$ 
as shown in Fig. \ref{fig:M-R_final}.

The likelihood for the upper limit of maximum mass constraint is given by
\begin{equation}
\label{eq:lhoodUpperMass}
P\left(E_{M}\left|\pi_q\right.\right) = 1-\Phi(M_q, \mu_u, \sigma_u)~,
\end{equation}
where $M_q$ is the maximum mass of the sequence of neutron star configurations for the given $\pi_q$, and $\mu_u=2.16\,M_{\odot}$ and $\sigma_u=0.17\,M_{\odot}$.

However, in Ref.~\cite{Bauswein:2020aag}
a relationship between the onset mass of prompt collapse to a black hole, the tidal deformability at half this mass and the maximum TOV mass has been derived, according to which the fact that the merger GW170817 did not promptly collapse to a black hole implies a {\it lower} limit on the maximum TOV mass. 
Therefore, we will include the disputable maximum mass constraint of Ref.~\cite{Rezzolla:2017aly} only in one of the sets of our Bayesian analysis.

\subsection{Gravitational wave constraint}
The observation of the gravitational waves from binary NS-NS merger GW170817 allows to calculate relation between tidal deformabilities of the primary and secondary components~\cite{TheLIGOScientific:2017qsa,Abbott:2018exr}.
In order to implement the tidal deformability constraint to Bayesian Inference the Gaussian kernel density estimation has been used to recover the probability distribution function with use of the data on the $\Lambda_1-\Lambda_2$ publicly shared by LIGO collaboration~\footnote{https://dcc.ligo.org/LIGO-P1800115/public}.

The likelihood of the gravitational wave constraint is introduced as 
\begin{equation}
\label{eq:lhoodLL}
P\left(E_{GW}\left|\pi_q\right.\right) = \int_{l} \beta\left(\Lambda_1(n_c), \Lambda_2(n_c)\right)\dd{n_c}~,
\end{equation}
where $l$ is the length of the line on the $\Lambda_1$--$\Lambda_2$ diagram produced by $\overrightarrow{\pi_q}$, and $n_c$ is the central baryon density of a star.  $\beta(\Lambda_1, \Lambda_2)$ is the {\it probability distribution function} (PDF) that has been reconstructed (as previously done in~\cite{Alvarez-Castillo:2020fyn,Ayriyan:2018blj}).
In Fig.~\ref{fig:L-M} we show the dimensionless tidal deformabilities of hybrid com-pact stars configurations. The line colors/patters are the same as in previous figures. We display as well the corresponding measurement from GW170817. \\
Fig.~\ref{fig:L1-L2} shows our results for the tidal deformabilities $\Lambda_2$ as a function of $\Lambda_1$. Dark and light green 
regions correspond to the 50$\%$ and 90$\%$ credibility for the posteriors used 
in the LIGO-Virgo Collaboration analysis, respectively. All the hybrid EoS in this work result in curves that fall 
inside the 90$\%$ region as the APR EoS which is displayed as dashed blue line. \\

\subsection{Mass-radius constraint}
A simultaneous measurement of mass and radius has been performed use data collected by the Neutron Star Interior Composition Explorer (NICER) space observatory for the pulsar {PSR\,J0030+0451}. The results of the observation have been reported in a collection of publications\footnote{\href{https://iopscience.iop.org/journal/2041-8205/page/Focus_on_NICER_Constraints_on_the_Dense_Matter_Equation_of_State}{Z.\,Arzoumanian \& K.\,C.\,Gendreau. \textit{Focus on NICER Constraints on the Dense Matter Equation of State}, ApJ 887, 2019}}. %, see for instance Ref.~\cite{Miller:2019cac,Riley:2019yda}. 

There were two independent analysis of the mass and equatorial radius based on mutually exclusive assumptions about the uniform-temperature emitting spots. 
The first result for radius and mass is  $M_1={1.44}_{-0.14}^{+0.15}\,{M}_{\odot }$ and  $R_1={13.02}_{-1.06}^{+1.24}\,\mathrm{km}$~\cite{Miller:2019cac} whereas the second one is $M_2=1.34_{-0.16}^{+0.15}\,{M}_{\odot }$ and $R_2={12.71}_{-1.19}^{+1.14}\,\mathrm{km}$~\cite{Riley:2019yda}. A~bivariate probability distribution function $\alpha (M,R)$ has been reconstructed by the method of the Gaussian kernel density estimation using the data~\cite{Miller:2019zenodo}. 
A likelihood is formulated as
\begin{equation}
\label{eq:lhoodMR}
P\left(E_{GW}\left|\pi_q\right.\right) = \int_{l} \alpha\left(M(n_c), R(n_c)\right)\dd{n_c}~,
\end{equation}
where $M(n_c)$ and $R(n_c)$ are mass and radius of sequence of neutron star for a given $q^{\mathrm{th}}$ equation of state, and $n_c$ is the central baryon density.

Another simultaneous measurement of the mass and the equatorial radius has recently been published based on NICER observations combined with XMM Newton data of PSR\,J0740+6620. The two independent analysis teams reported $R=13.7_{-1.5}^{+2.6}$\,km~\cite{Miller:2021qha} and $R=12.39_{-0.98}^{+1.3}$\,km~\cite{Riley:2021pdl}. 
To implement these results in a Bayesian analysis, the function  $\alpha (M,R)$ has been reconstructed with a kernel density estimation using mass-radius data from~\cite{Miller:2021zenodo}.
The above constraints are shown in Fig. \ref{fig:M-R_final}.

\subsection{Fictitious radius measurements}
Simultaneous measurements of masses and radii like the ones provided recently by NICER are very important for deriving constraints on the EoS. However, due to limited observation time, the difficult decision has to be made for which object with a known mass should a precise radius measurement (and thus a long-term observation campaign) bear the largest discovery potential. A BA can support such decisions by using fictitious results of a radius measurement on the object under scrutiny. 
An essential input for the BA would be precise radius measurements for pulsars with known masses.
One example is the high-mass millisecond pulsar {PSR\,J0740+6620}, for which the mass is known from Ref.~\cite{Cromartie:2019kug} with the recent update by 
\cite{Fonseca:2021wxt}, but for which by the time this paper was written the results of the NICER radius measurement were not yet available.
Now, after their publication in Refs.~\cite{Miller:2021qha,Riley:2021pdl}, we can compare the these real results with the predictions of the fictitious radii and suggest the method of fictitious radii for supporting decisions on the selection of targets for future measurements.
%We use the BA to analyse the most probable EoS that would result from certain fictitious radius measurements. 
We consider here three different values for the radius
of {PSR\,J0740+6620}, namely $R=11$, $12$ and $13$\,km with 
the the same design uncertainty $\sigma=0.5$\,km of the NICER experiment. The likelihood for fictitious measurements is the same as introduced above~\eqref{eq:lhoodMR}.

\subsection{Sets of constraints}
We suggest three sets of constraints:
\begin{enumerate}
    \item[{\underline{\bf 1:}}] 
    the mass measurement for PSR J0740+6620 %\cite{Cromartie:2019kug} 
    \cite{Fonseca:2021wxt}
    as the lower limit for the maximum mass, the tidal deformability from GW170817 \cite{Abbott:2018exr} and the mass-radius constraint from PSR J0030+0451 
    \cite{Miller:2019cac} (set 1);
    \item[{\underline{\bf 2:}}] in addition to the constraints of set 1, the constraint on the upper limit of the maximum mass from Ref.~\cite{Rezzolla:2017aly} is included;
    \item[{\underline{\bf 3:}}] as for set 1, but assuming that the lower mass companion of the black hole in the asymmetric binary merger GW190814~\cite{Abbott:2020khf} was a neutron star,
    the lower limit for the maximum mass is replaced by the lower limit on its mass $M_{190814}=2.59_{-0.09}^{+0.08}~M_\odot$.
\end{enumerate}

\begin{table*}[htpb!]
%\small
\resizebox{\textwidth}{!}{ 
	\begin{tabular}{|c|c|c|c|c|c|c|c|c|c|}
	\hline
		\multirow{4}{*}{Set} & \multicolumn{3}{c|}{Maximum mass [M$_{\odot}$]} & \multicolumn{1}{c|}{$M$-$R$} & $\Lambda_1$-$\Lambda_2$ & \multicolumn{3}{c|}{\multirow{3}{*}{
		Fictitious radius measurement }} & \multirow{4}{*}{Fig.} \\ \cline{2-6}
		    & \multicolumn{2}{c|}{Lower limit}                   & Upper limit    &               &              & \multicolumn{3}{c|}{}                          &    \\ \cline{2-4} %\cline{7-9}
		    & PSR J0740+6620& GW190814 & GW170817& J0030+0451~\cite{Miller:2019cac}& GW170817 ~\cite{Abbott:2018exr}&  \multicolumn{3}{c|}{on PSR J0740+6620 [km] } 
		    & \\ \cline{2-4} \cline{7-9}
		    & $2.08\pm0.07$~\cite{Fonseca:2021wxt}  & $2.59\pm0.09$~\cite{Abbott:2020khf} & $2.16\pm0.17$~\cite{Rezzolla:2017aly} &               &              & $11\pm0.5$       & $12\pm0.5$ & $13\pm0.5$ &  \\ \hline
		1.0                   & \checkmark      &                &               & \multicolumn{1}{c|}{\checkmark}   & \checkmark   &                 &                 &              & 15(l) \\ \hline
		1.1            &                          &             &               & \multicolumn{1}{c|}{\checkmark}   & \checkmark   & \checkmark      &                 &              &  16(l,u)  \\ \hline
		1.2            &                         &                &               & \multicolumn{1}{c|}{\checkmark}   & \checkmark   &                 & \checkmark      &              & 16(c,u) \\ \hline
		1.3  &                   &                &               & \multicolumn{1}{c|}{\checkmark}   & \checkmark   &                 &                 & \checkmark &   16(r,u) \\ \hline
		2.0                   & \checkmark       &                & \checkmark    & \multicolumn{1}{c|}{\checkmark}   & \checkmark   &                 &                 &            &  15(c)  \\ \hline
		2.1                   &                          &                & \checkmark    & \multicolumn{1}{c|}{\checkmark}   & \checkmark   & \checkmark      &                 &             &  16(l,m) \\ \hline
		2.2                   &                          &                & \checkmark    & \multicolumn{1}{c|}{\checkmark}   & \checkmark   &                 & \checkmark      &             &  16(c,m) \\ \hline
		2.3                   &                          &                & \checkmark    & \multicolumn{1}{c|}{\checkmark}   & \checkmark   &                 &                 & \checkmark &  16(r,m)  \\ \hline
		3.0                   &  \checkmark                & \checkmark     &               & \multicolumn{1}{c|}{\checkmark}   & \checkmark   &                 &                 &            &   15(r) \\ \hline
		3.1                   &                  & \checkmark     &               & \multicolumn{1}{c|}{\checkmark}   & \checkmark   & \checkmark      &                 &             &  16(l,b) \\ \hline
		3.2                   &                  & \checkmark     &               & \multicolumn{1}{c|}{\checkmark}   & \checkmark   &                 & \checkmark      &              & 16(c,b) \\ \hline
		3.3                   &                  & \checkmark     &               & \multicolumn{1}{c|}{\checkmark}   & \checkmark   &                 &                 & \checkmark &  16(r,b)  \\ \hline
	\end{tabular}
	}
	 \caption{Overview on Bayesian analysis constraints employed in the present work. The rightmost column refers to the set associated figure number in the text and posterior probability LEGO plot position in it according to the following convention: columns are indicated by left (l), center (c), and right (r), while rows are labelled upper (u), middle (m), and bottom (b).}
         \label{tab:1}
\end{table*} 

Besides these "pure" sets of constraints, we investigate for each of them the possibility of an additional mass-radius constraint for the pulsar PSR J0740+6620, for which the mass 
%$2.14^{+0.10}_{-0.09}~M_\odot$
$2.08\pm 0.07~M_\odot$
is rather precisely measured and a result for the radius measurement by the NICER experiment is anticipated to yield  the values of $11$, $12$ and $13$ km. 
We denote these subsets with the numbers $1$, $2$ and $3$, respectively. 
The entirety of Bayesian constraint sets in this work is synoptically summarized in Tab.~\ref{tab:1}.
A separate BA is performed using the recent measurement of the combined mass-radius constraint for PSR J0740+6620 on the basis of NICER and XMM Newton data, and compared to the predictions using fictitious radii.

\begin{figure*}[!ht]
	\centering
	\includegraphics[width=0.32\linewidth]{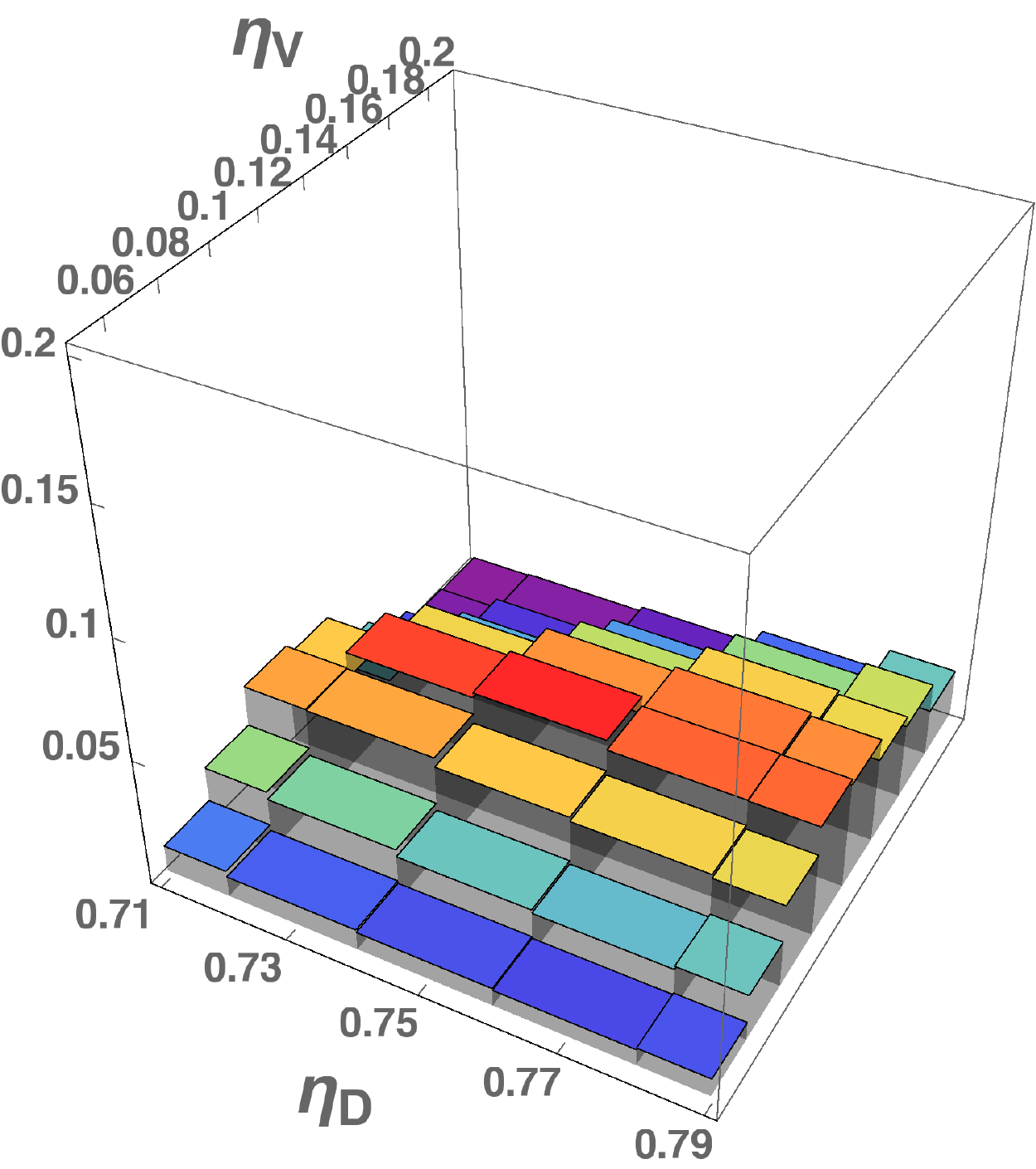}
	\includegraphics[width=0.32\linewidth]{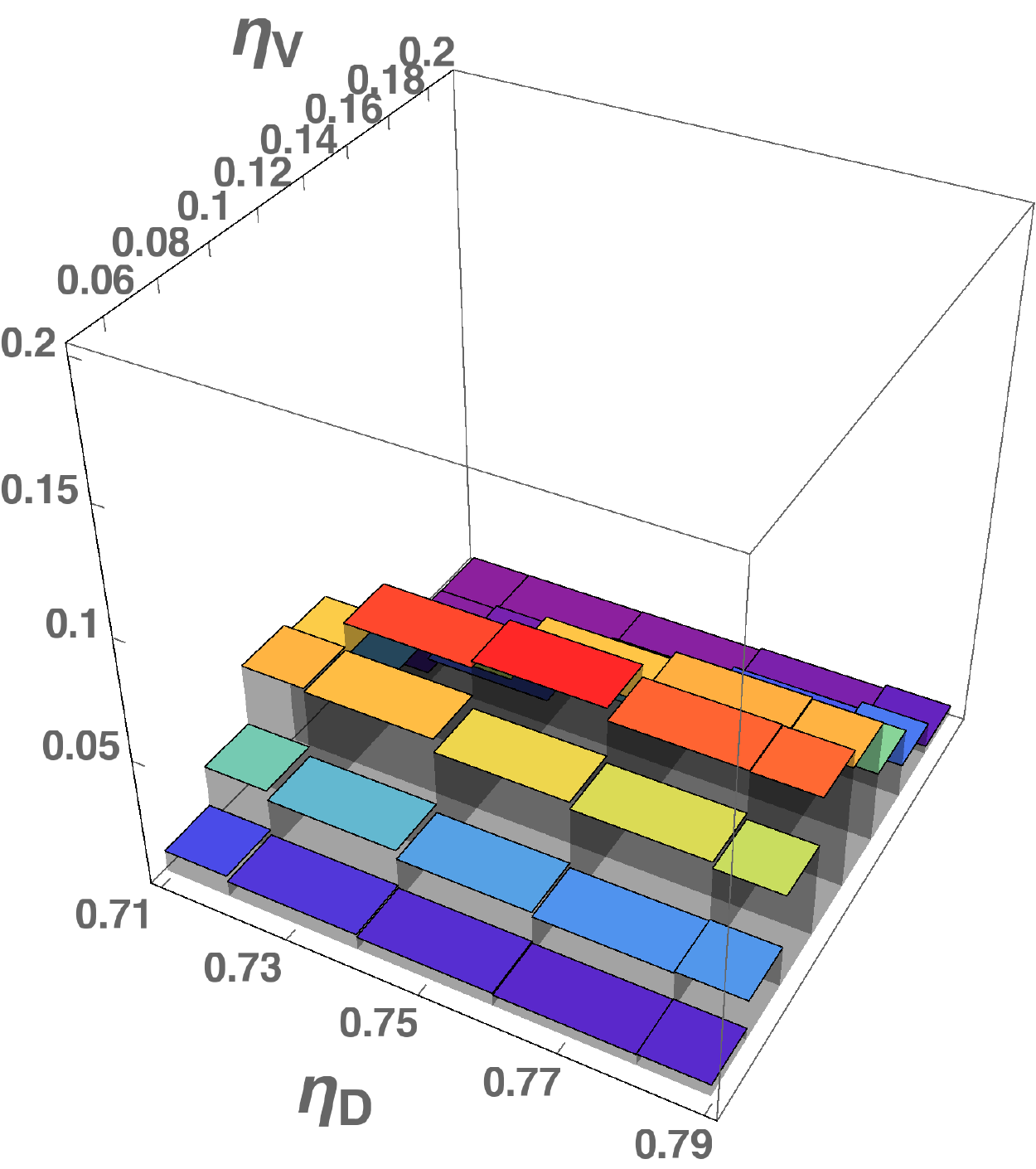}
	\includegraphics[width=0.32\linewidth]{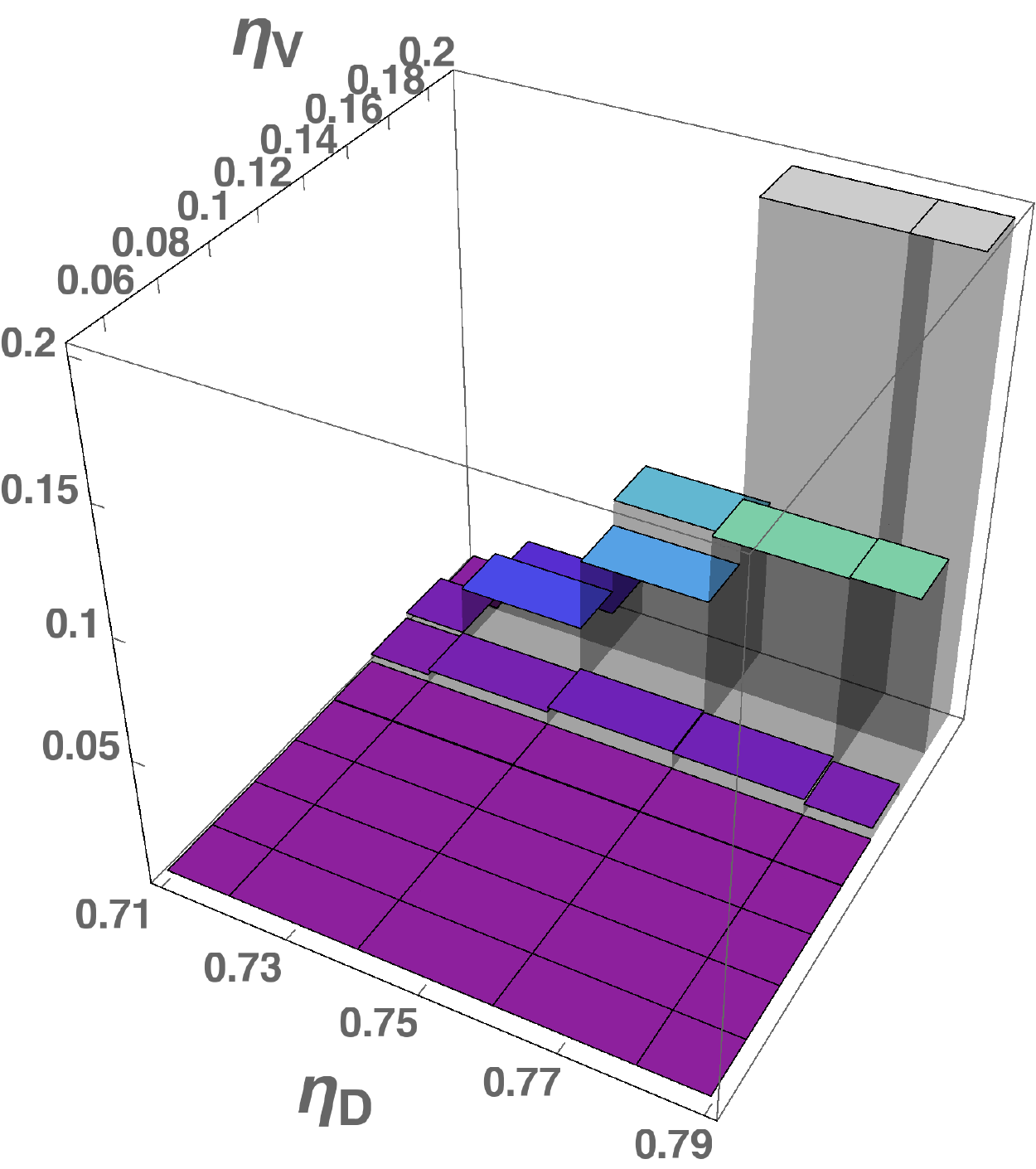}	\label{fig:LEGO1-3}
	\caption{Bayesian analysis using the mass measurement for PSR J0740+6620 \cite{Fonseca:2021wxt} as the
	lower limit for the maximum mass, the tidal deformability from GW170817 \cite{Abbott:2018exr} and the mass-radius constraint from PSR J0030+0451 for the class of hybrid EoS obtained with a two-zone interpolation between APR and nlNJL in the two-dimensional EoS parameter plane spanned by $\eta_V$ and $\eta_D$ (left panel; set 1). 
	In the middle panel the constraint on the upper limit of the maximum mass from Ref.~\cite{Rezzolla:2017aly} has been added (set 2) and in the right panel this limit has been lifted again in favor of the new lower limit on the maximum mass from lower-mass companion of the black hole in the asymmetric binary merger GW190814~\cite{Abbott:2020khf} that replaces the one from \cite{Fonseca:2021wxt} 
	(set 3), if the former object would not be a black hole.
	}
	\label{fig:BA}
\end{figure*}

\section{Results of the Bayesian analysis}
\label{sec:results}
In this section we show and discuss our results for the BA of the hybrid EoS for compact stars under the constraints defined in the previous section and summarized in Tab. \ref{tab:1}. 
The details of the considered families of hybrid EoS, the interpolation method, the constant speed of sound 
representation, the BA and astrophysical constraints were already presented in previous sections. 

Figure~\ref{fig:BA} shows the results of our BA for the $\eta_V$ and $\eta_D$ values of our EoS models under consideration. 
The difference between the three LEGO 
plots is that for their derivation the values of the maximum mass constraint have been changed. 
These three different cases comprise 
1) the mass measurement of the object PSR J0740+6620, \cite{Fonseca:2021wxt}, 
2) the previous measurement plus the upper limit of the maximum mass from Ref.~\cite{Rezzolla:2017aly}, 
3) the first measurement plus the mass for the lighter object in the binary merger GW190814~\cite{Abbott:2020khf} assuming that it was a 
neutron star. 
We can see that whereas for the first to cases the posterior probability distributions peak at intermediate values of the $\eta_V$ parameter of quark matter, the third case favours its highest values. 
The probability distribution in 
$\eta_D$ direction remains flat for the three cases.

%%%%%% New figure 14 %%%%%% 
\begin{figure*}[!th]
\begin{center}
%\begin{array}
\begin{tabular}{l|c|c|c}
\hline
\qquad $R$ $\to$ &&&\\[-2mm]
&11 km& 12 km& 13 km\\[-2mm]
$ \downarrow $ Constraints&&&\\
\hline
&&&\\[-2mm] 
\begin{minipage}{3cm}
\underline{\bf set 1}:\\[5mm]
inf\{$M_{\rm max}$\} \cite{Fonseca:2021wxt},\\
$\Lambda_{1.4}$ \cite{Abbott:2018exr},\\
$(M,R)_{\rm J0030+0451}$ \cite{Miller:2019cac}\\
\end{minipage}
&
\includegraphics[width=0.25\textwidth]{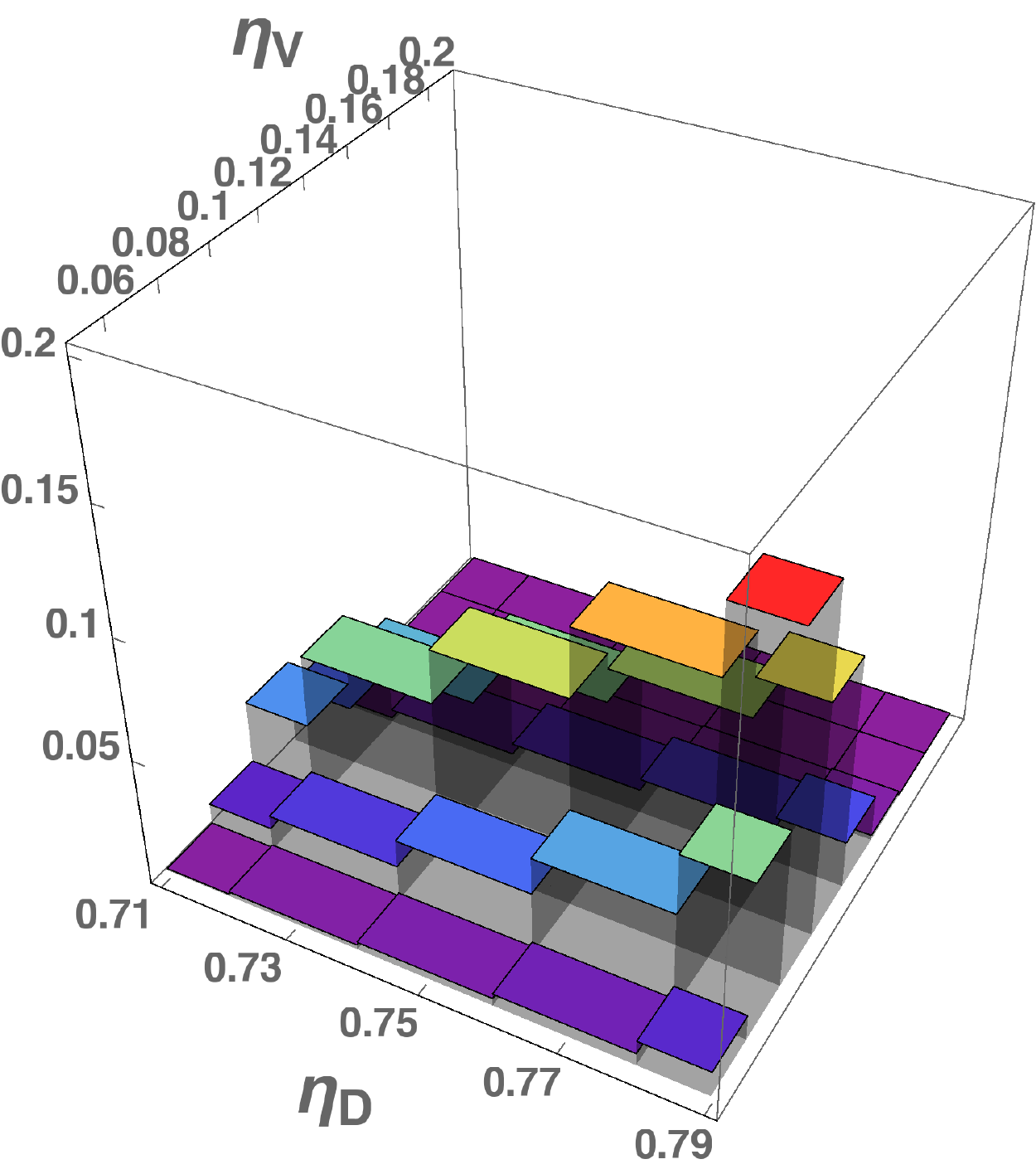} & 
\includegraphics[width=0.25\textwidth]{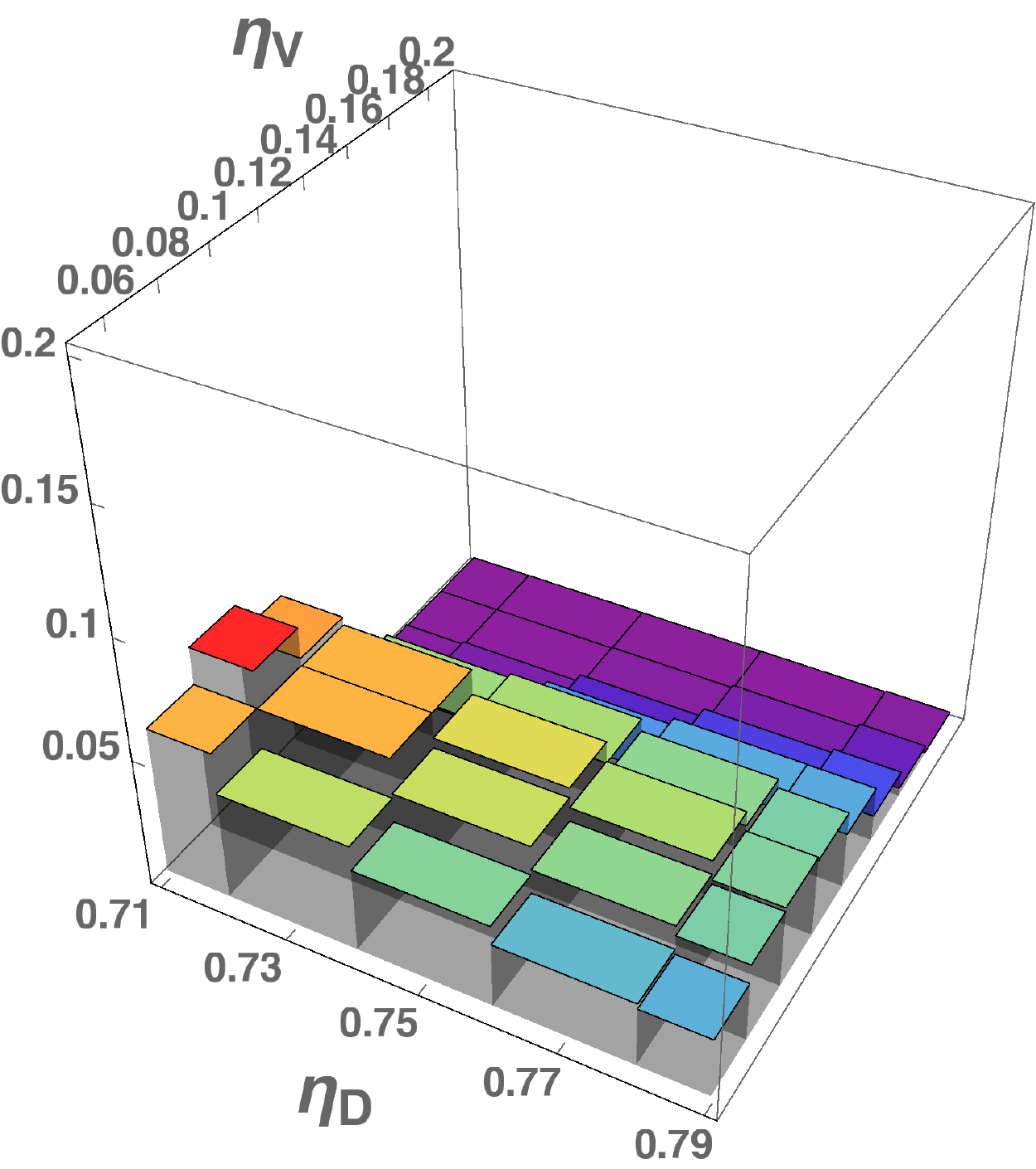} & 
\includegraphics[width=0.25\textwidth]{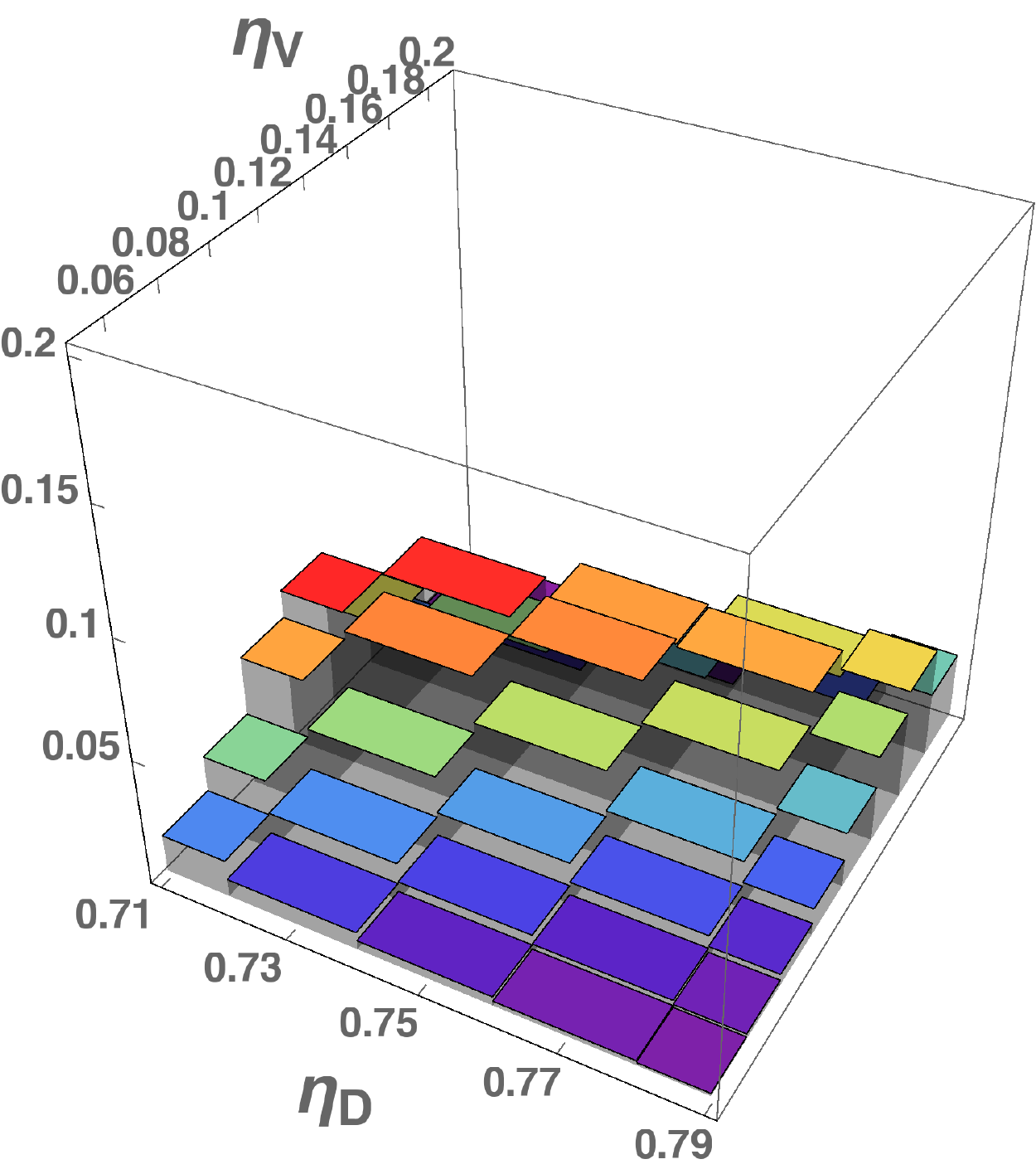} \\
&&&\\[-2mm]
\hline
&&&\\[-2mm] 
\begin{minipage}{3cm}
\underline{\bf set 2}:\\[5mm]
inf\{$M_{\rm max}$\} \cite{Fonseca:2021wxt},\\
$\Lambda_{1.4}$ \cite{Abbott:2018exr},\\
$(M,R)_{\rm J0030+0451}$ \cite{Miller:2019cac}\\{}
{sup}\{$M_{\rm max}$\} \cite{Rezzolla:2017aly}
\end{minipage}
&
\includegraphics[width=0.25\textwidth]{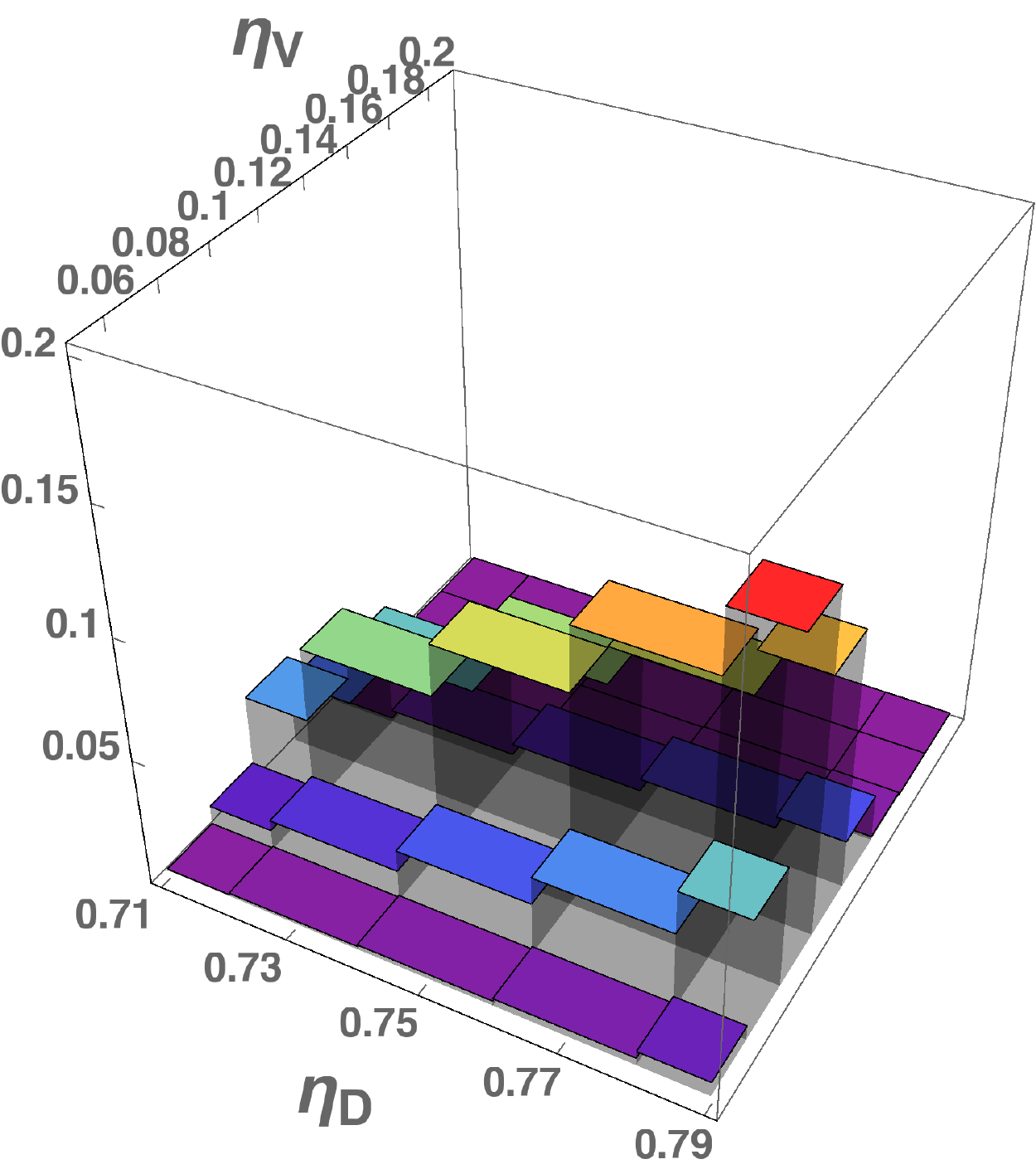} & 
\includegraphics[width=0.25\textwidth]{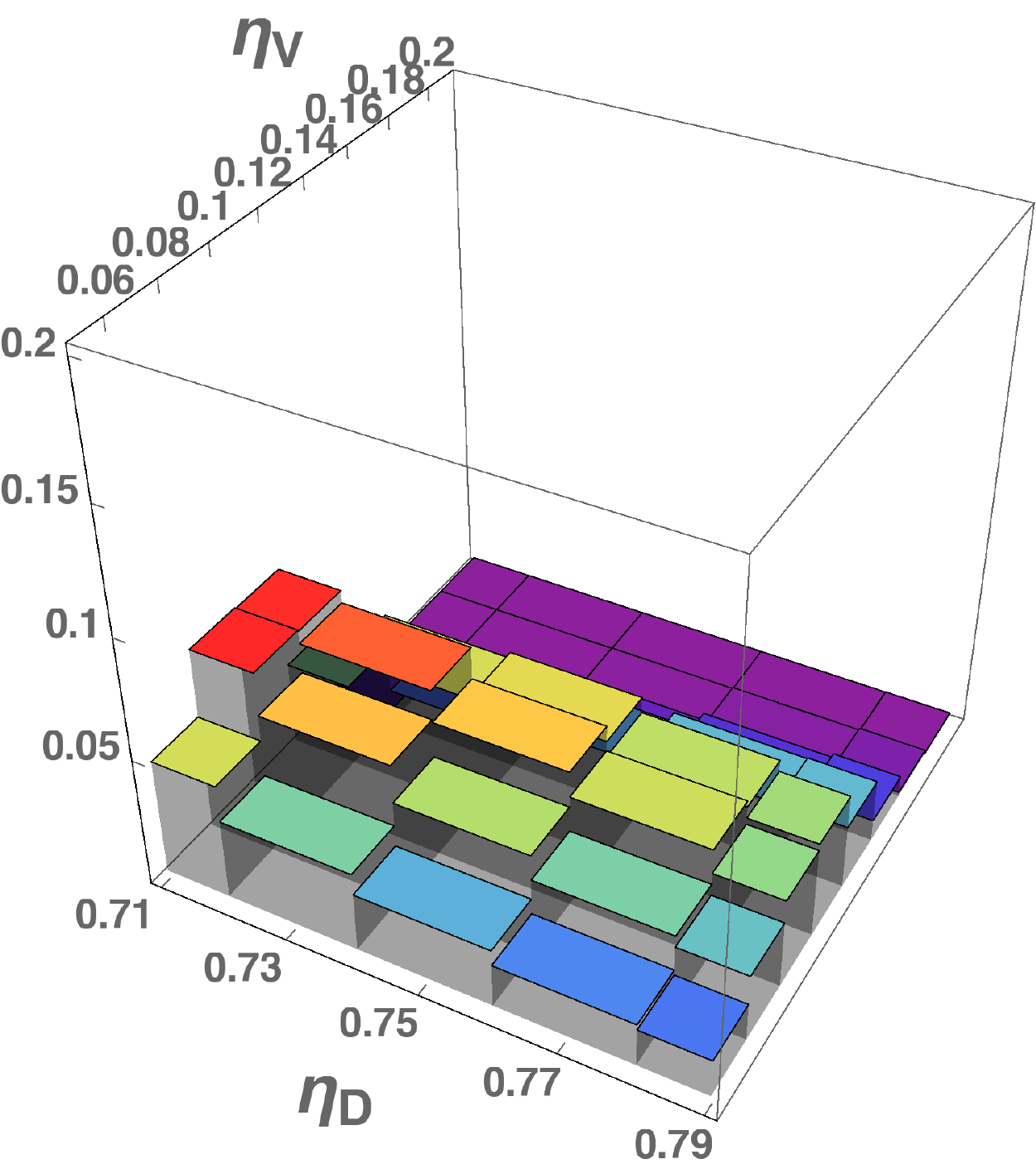} & 
\includegraphics[width=0.25\textwidth]{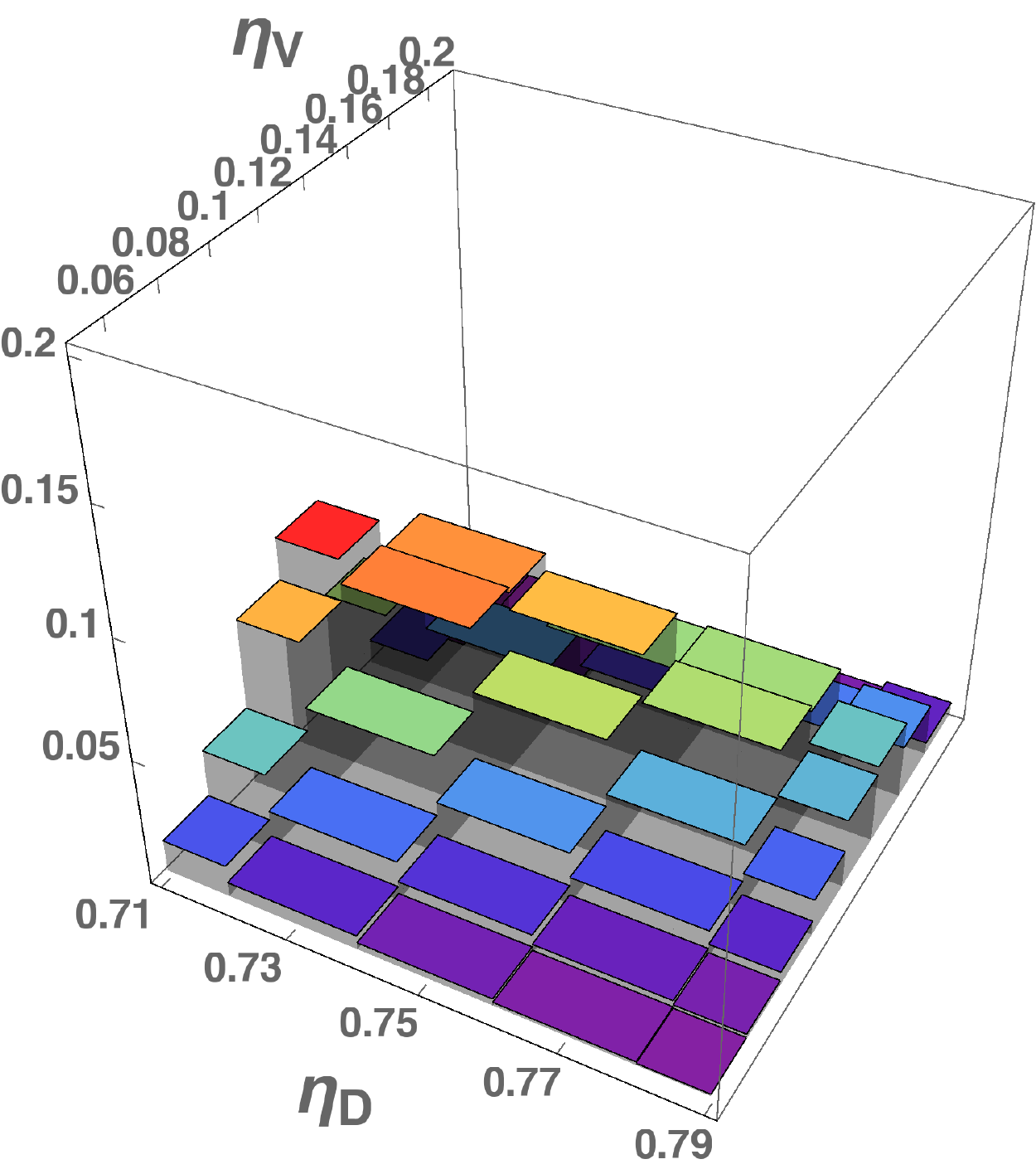} \\
&&&\\[-2mm]
\hline
&&&\\[-2mm] 
\begin{minipage}{3cm}
\underline{\bf set 3}:\\[5mm]
inf\{$M_{\rm max}$\} \cite{Abbott:2020khf},\\
$\Lambda_{1.4}$ \cite{Abbott:2018exr},\\
$(M,R)_{\rm J0030+0451}$ \cite{Miller:2019cac}
%\\{}{sup}\{$M_{\rm max}$\} %\cite{Rezzolla:2017aly}
\end{minipage}
&
\includegraphics[width=0.25\textwidth]{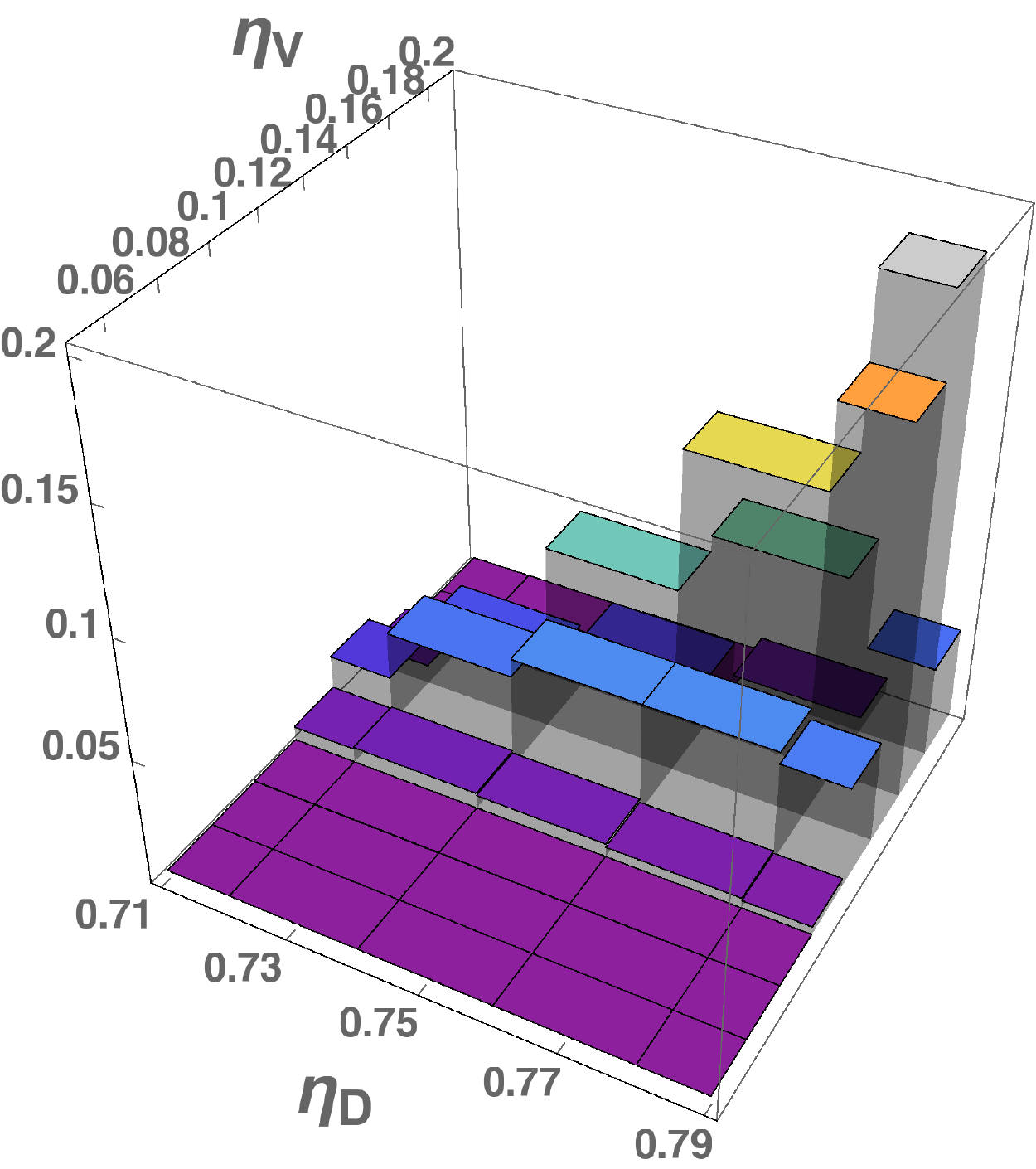} & 
\includegraphics[width=0.25\textwidth]{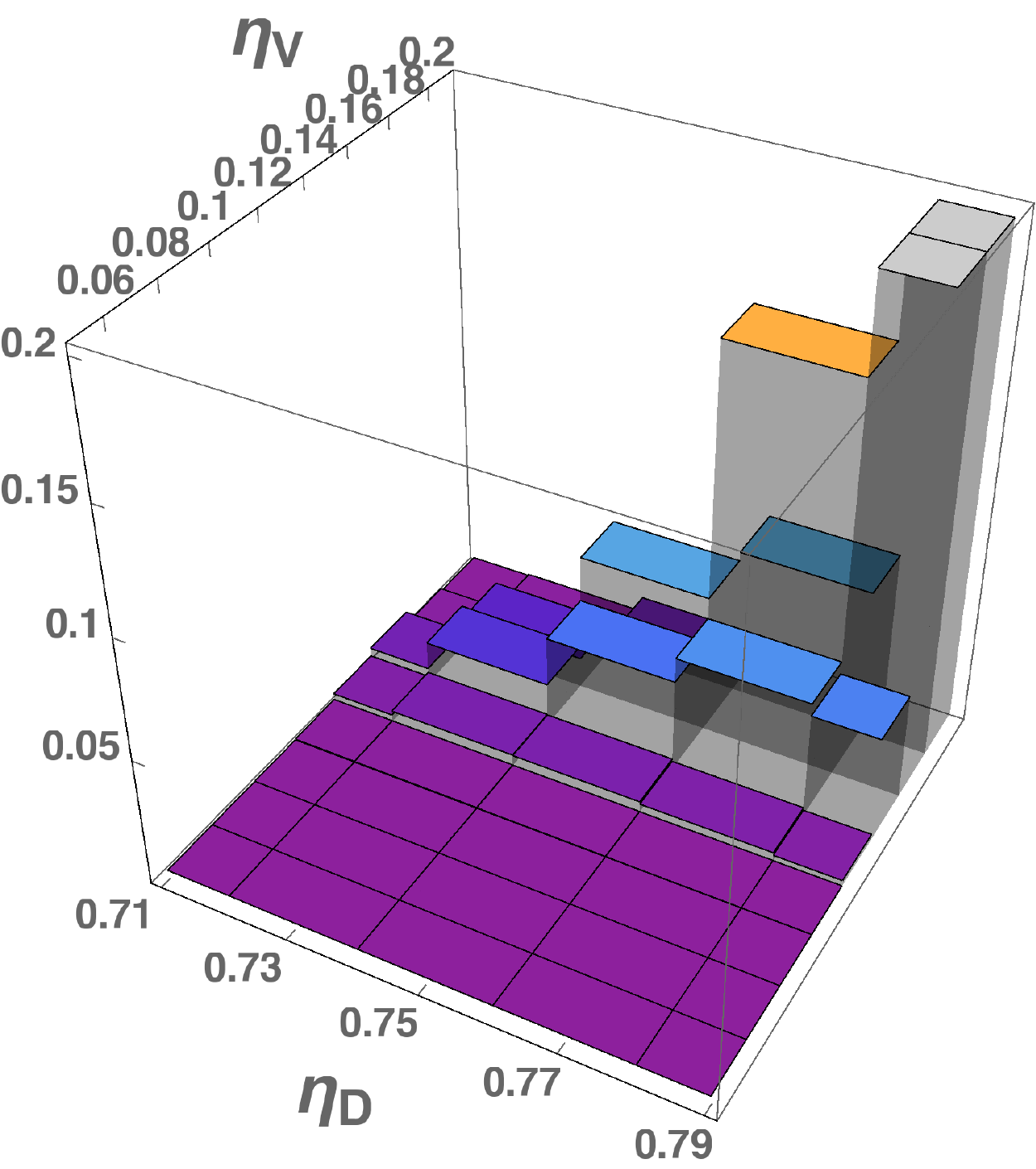} & 
\includegraphics[width=0.25\textwidth]{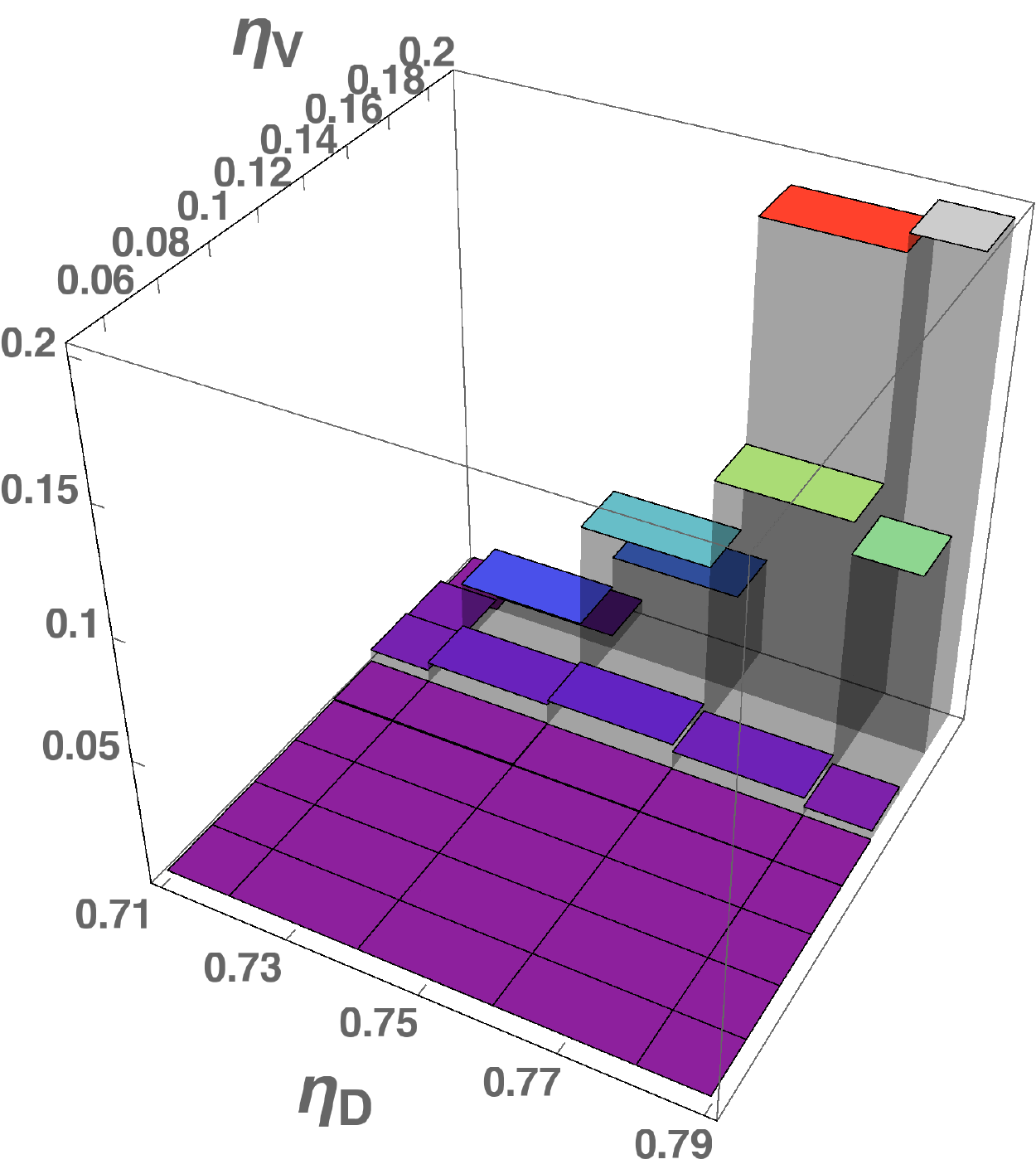} \\
&&&\\[-2mm]
\hline 
%\end{array}$
\end{tabular}
\end{center}
\caption{Probabilities when an additional measurement of the radius $R$ of PSR J0740+6220 (as provided recently by the NICER experiment \cite{Miller:2021qha}) is taken into account with anticipated values of $R=11$ km (second column), 12 km (third column) or 13 km (fourth column) and with a standard deviation of $\sigma_R =0.5$ km.
The mass is taken from the measurement by 
%Cromartie et al.~\cite{Cromartie:2019kug}.
Fonseca et al.~\cite{Fonseca:2021wxt}.
The three rows correspond to set 1, set 2 and set 3, respectively.
\label{fig:BA-fict}}
\end{figure*}

%\newpage

%%%%%%%%%%%%%%% Figure 17

\begin{figure*}[!ht]
\begin{center}
%\begin{array}
\begin{tabular}{l|c|c|c}
\hline
&&&\\[-2mm]
set 1 &set 1 + $R=11$ km & set 1 + $R=12$ km & set 1 + $R=13$ km\\%[-2mm]
&&&\\[-2mm]
\hline
&&&\\[-2mm]
\includegraphics[width=0.2\textwidth]{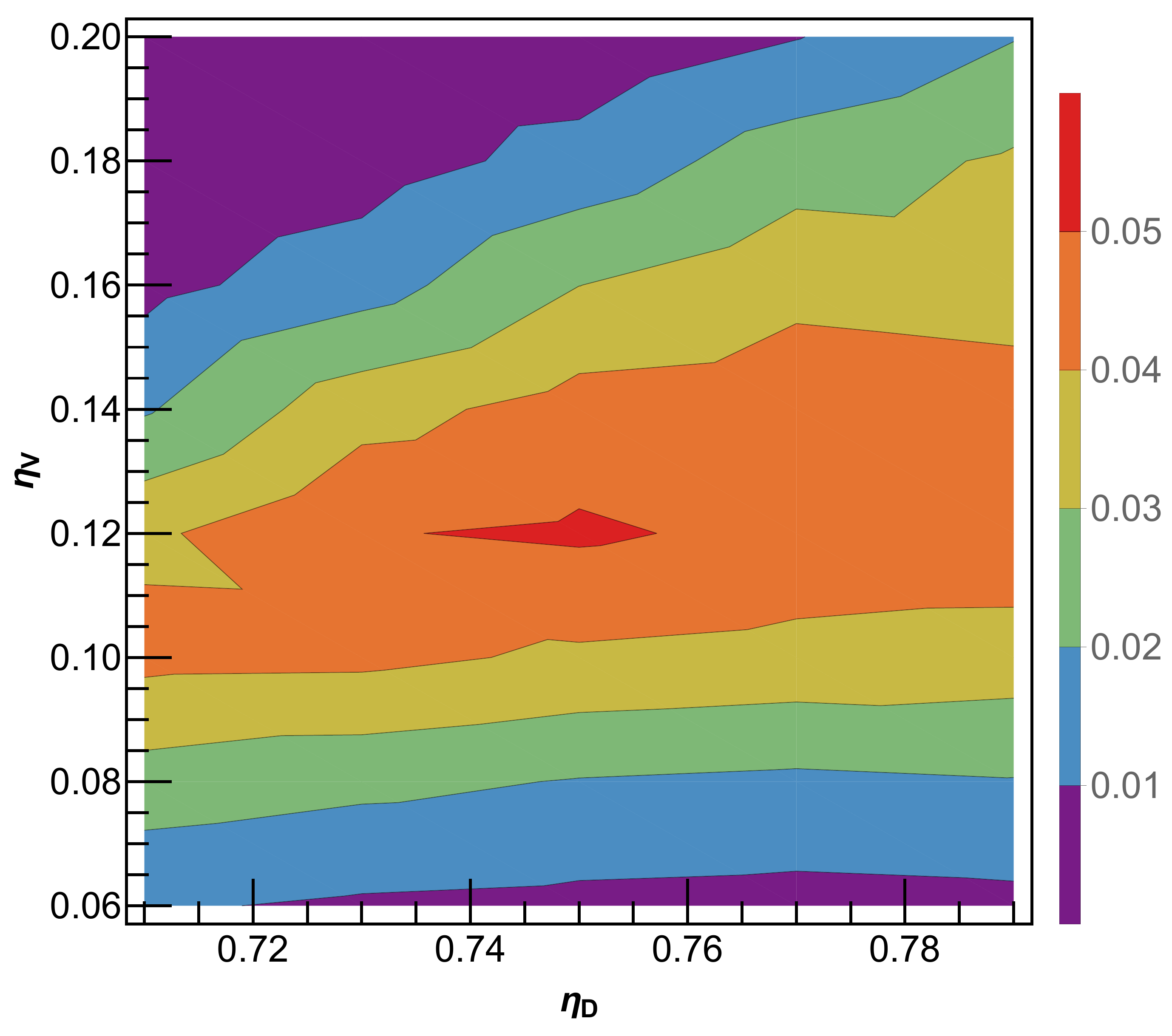} & 
\includegraphics[width=0.2\textwidth]{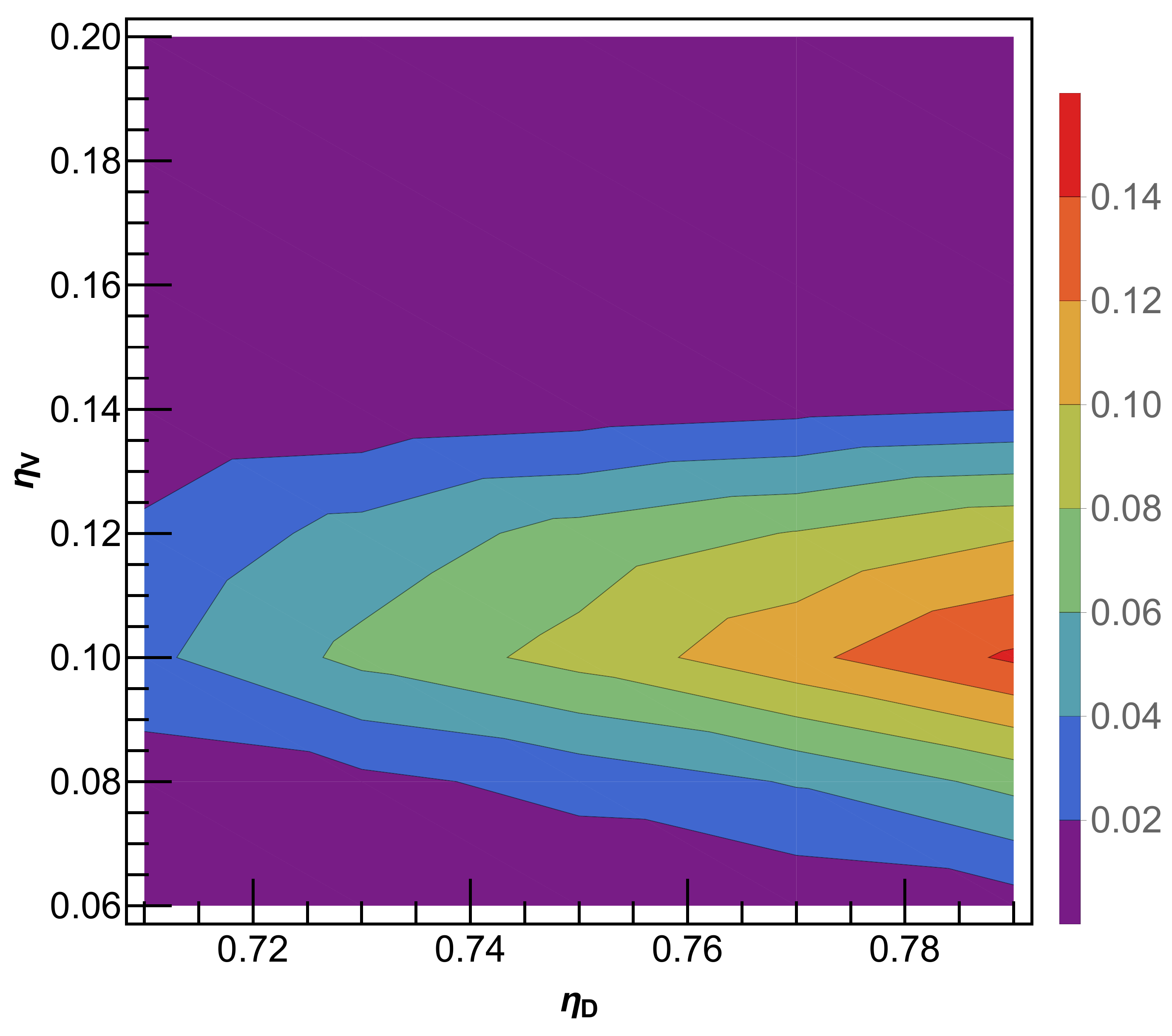} & 
\includegraphics[width=0.2\textwidth]{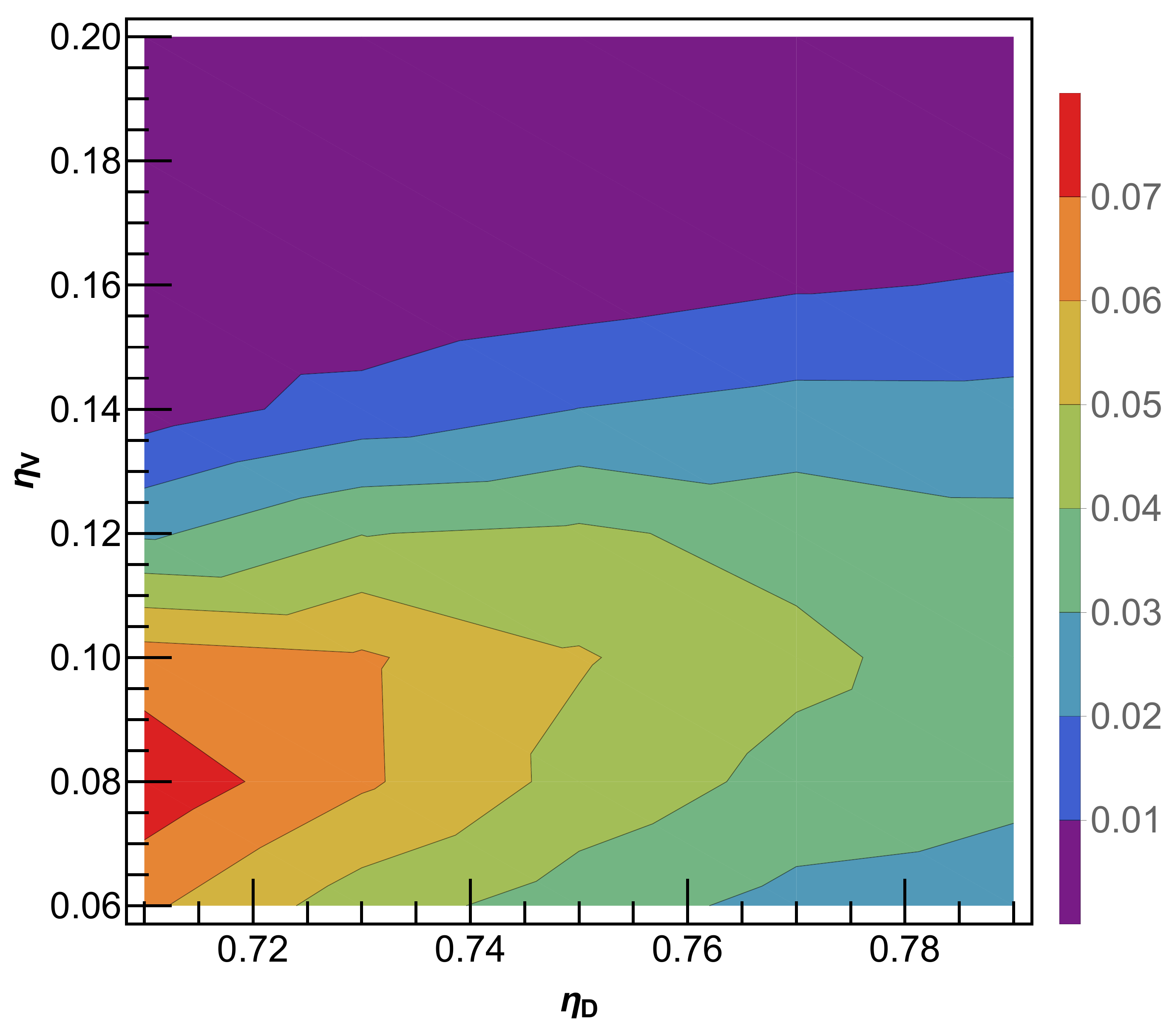} & 
\includegraphics[width=0.2\textwidth]{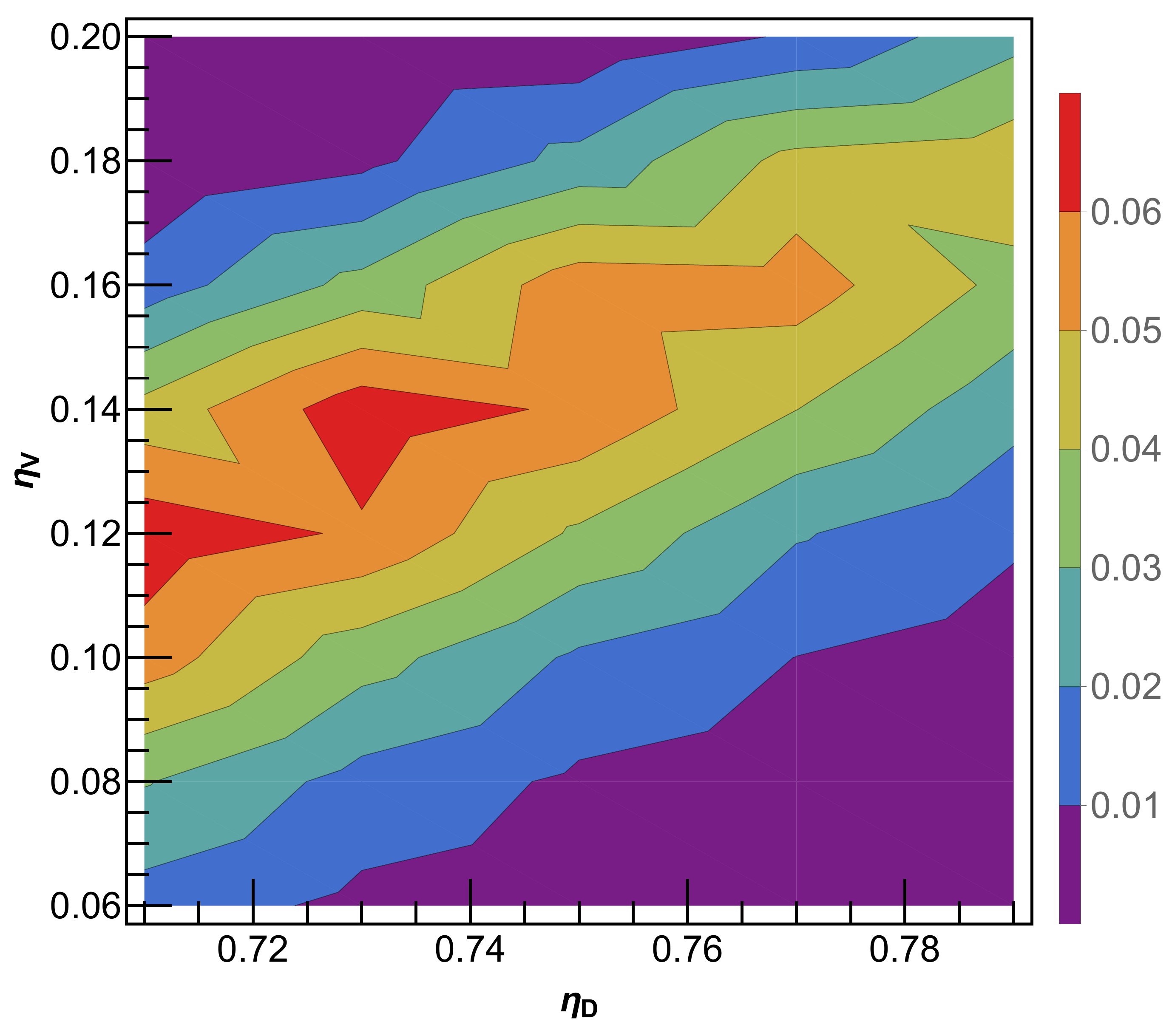} 
\\
%&&&\\[-2mm]
\hline
&&&\\[-2mm]
\includegraphics[width=0.2\textwidth]{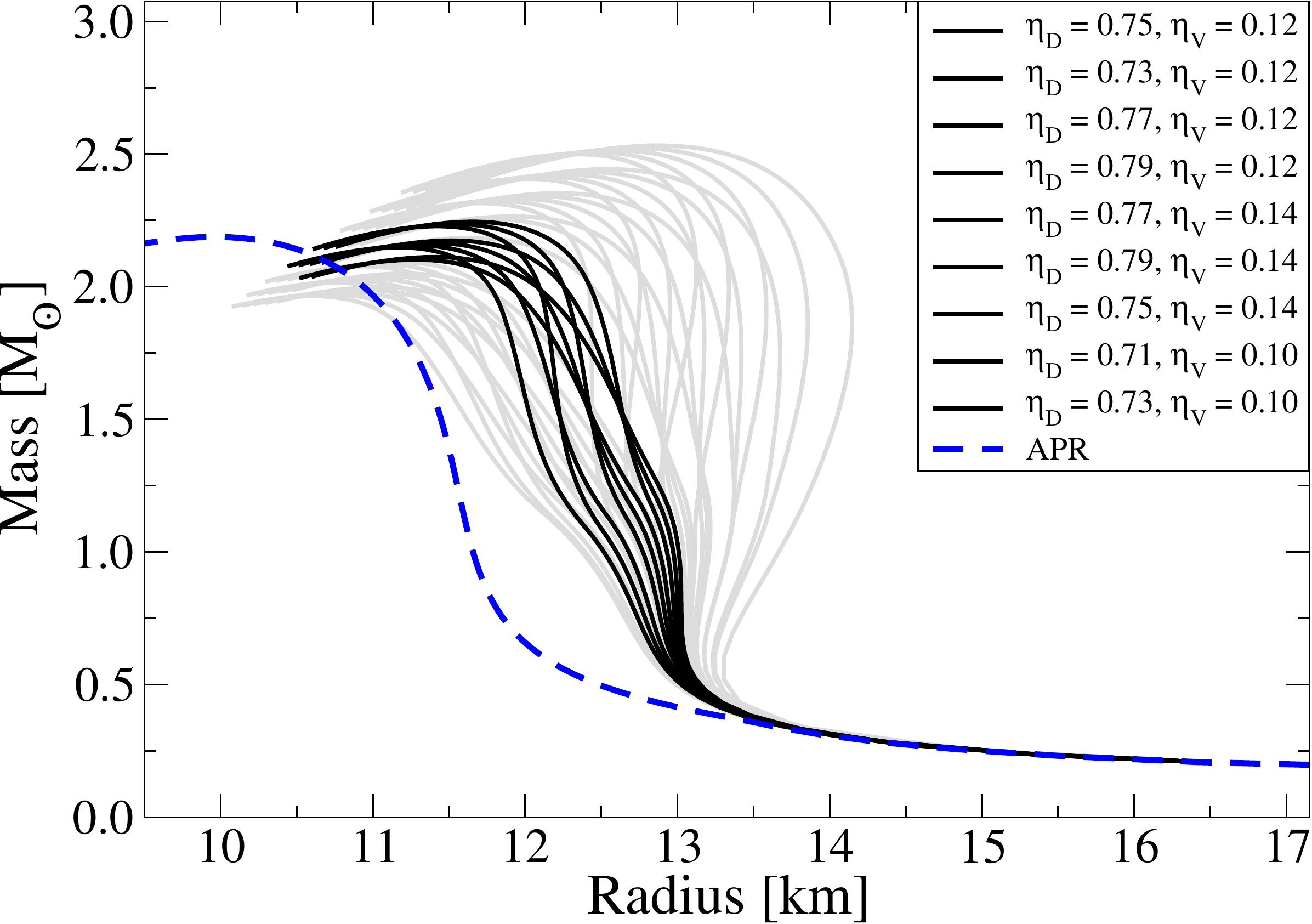}& 
\includegraphics[width=0.2\textwidth]{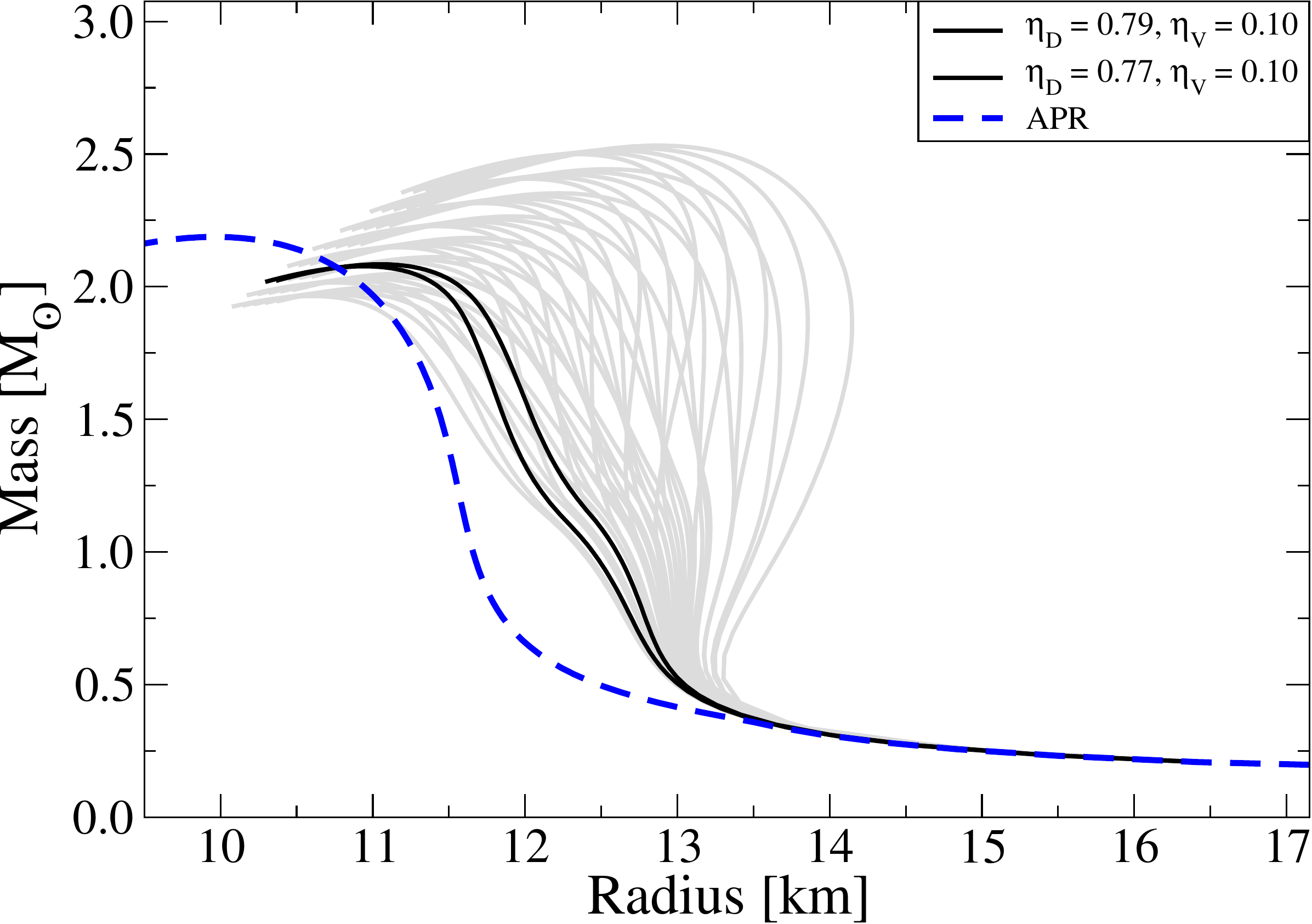}& 
\includegraphics[width=0.2\textwidth]{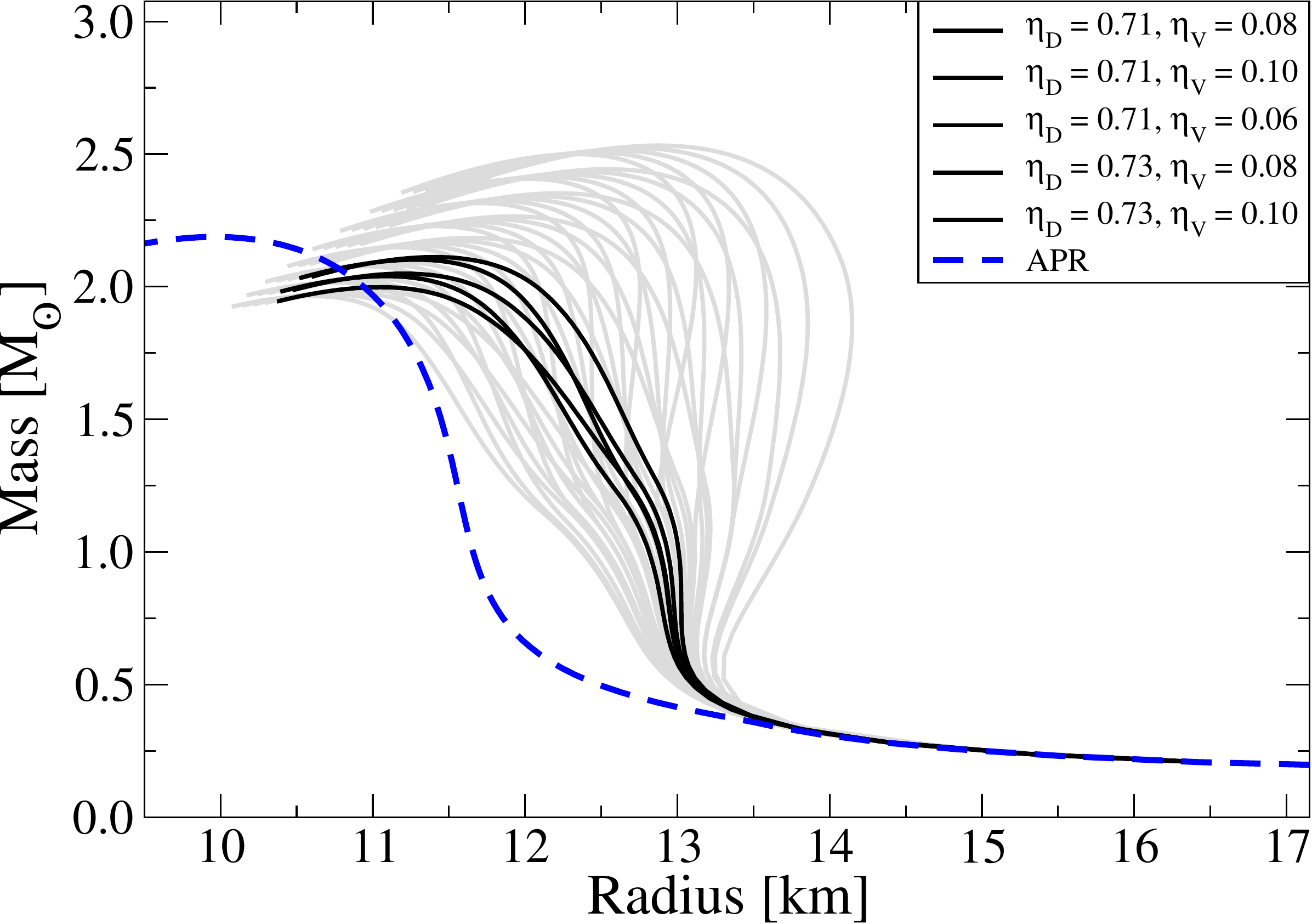}& 
\includegraphics[width=0.2\textwidth]{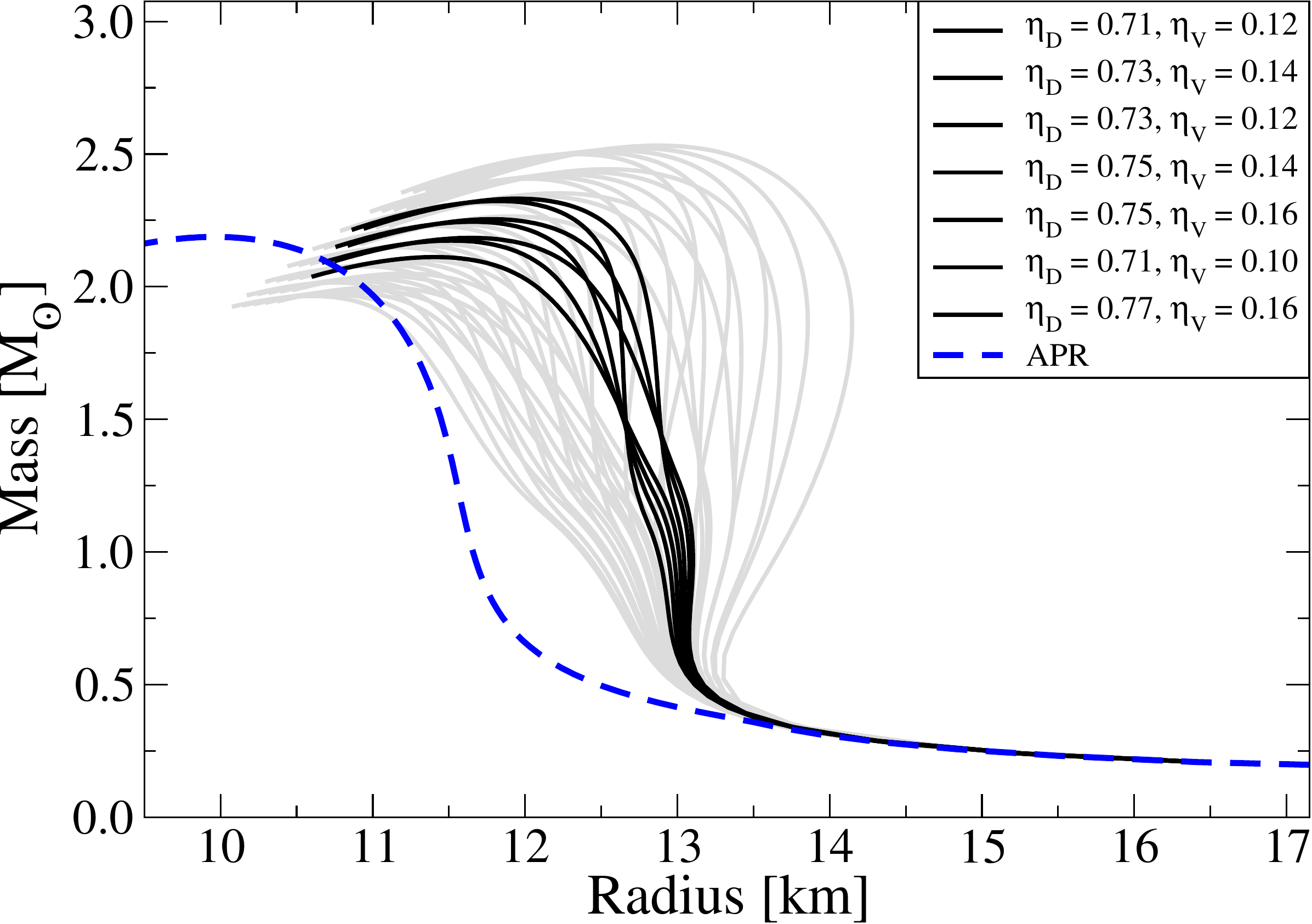}
\\
%&&&\\[-2mm]
\hline
&&&\\[-2mm]
\includegraphics[width=0.2\textwidth]{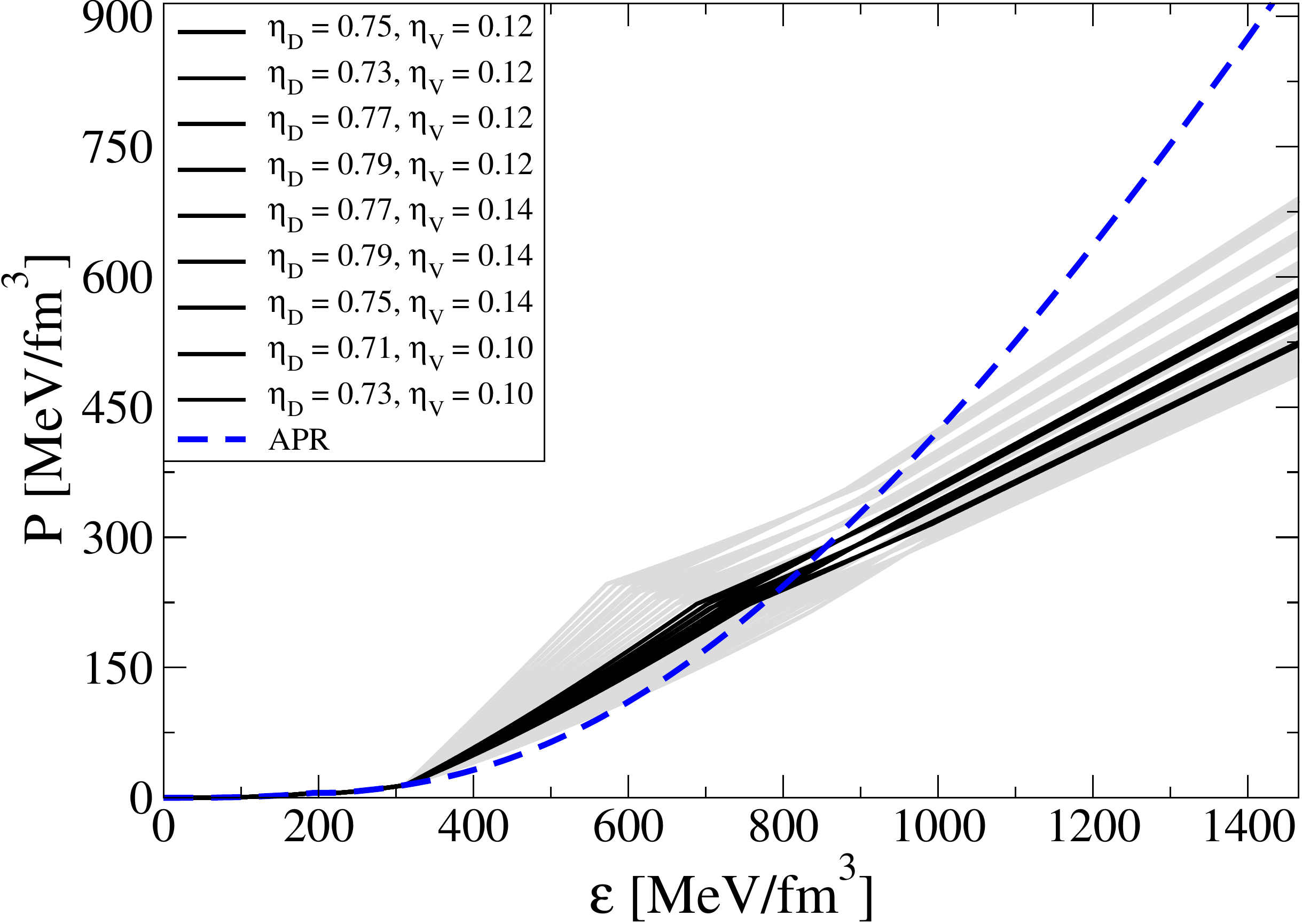}&
\includegraphics[width=0.2\textwidth]{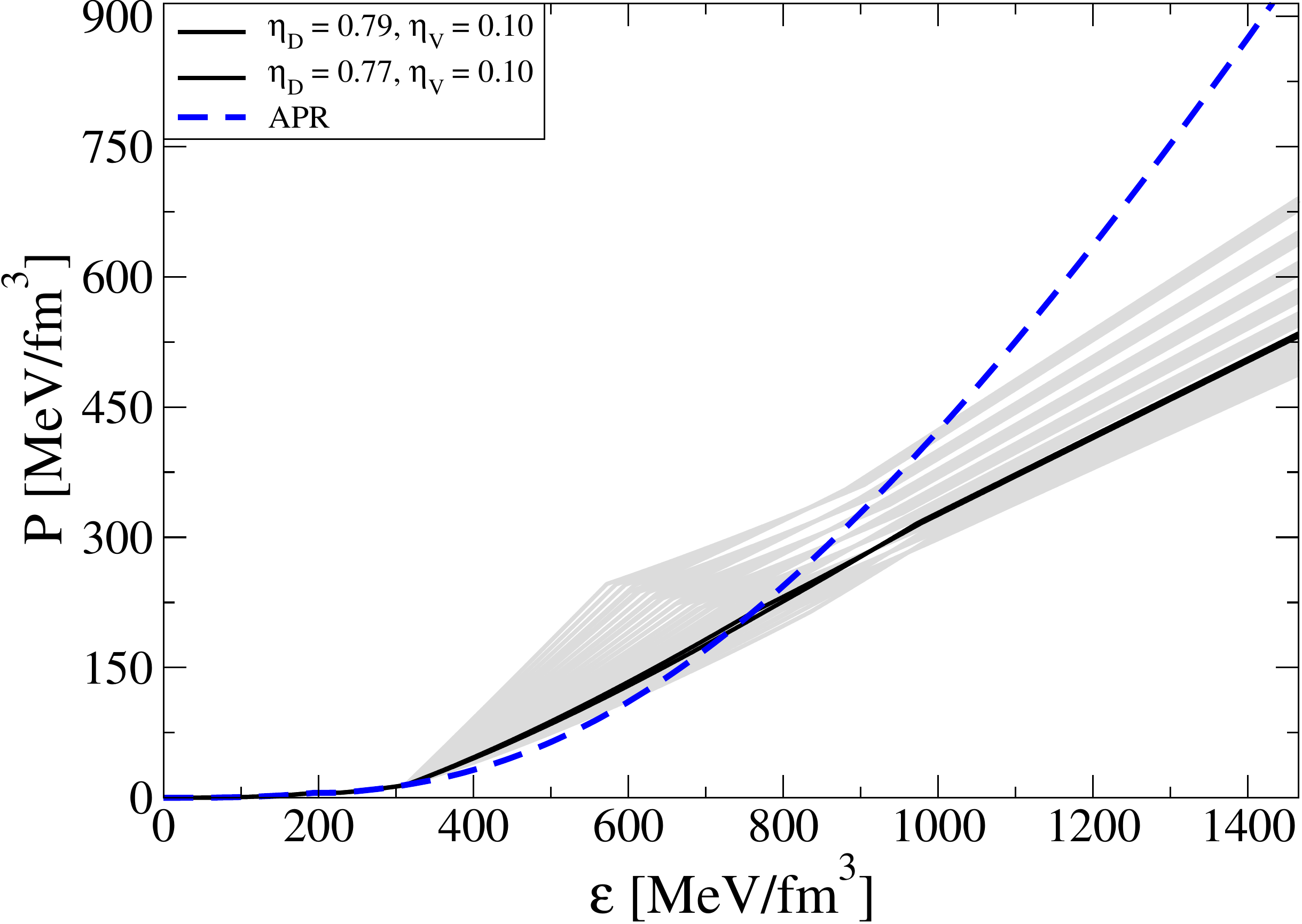}&
\includegraphics[width=0.2\textwidth]{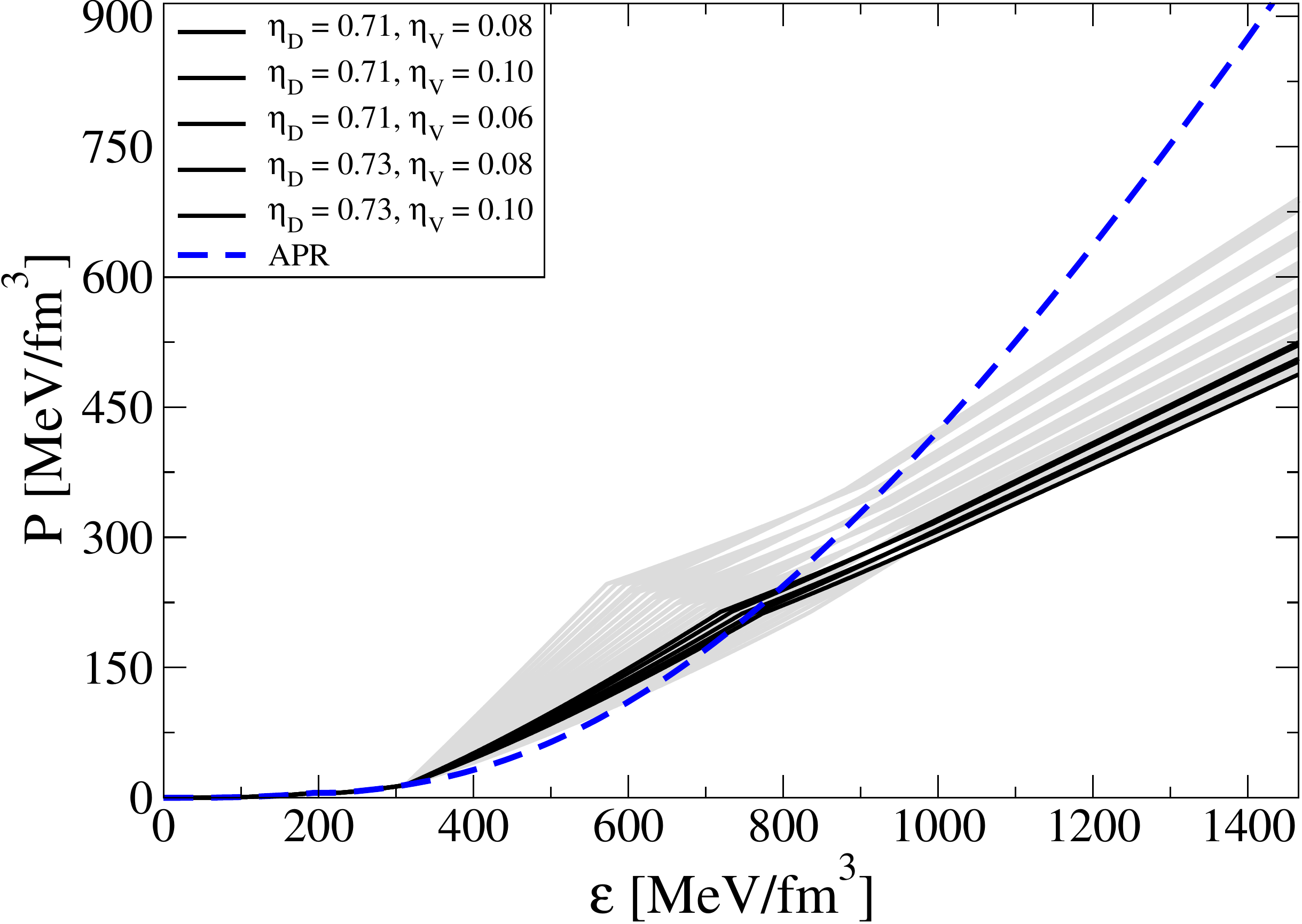}&
\includegraphics[width=0.2\textwidth]{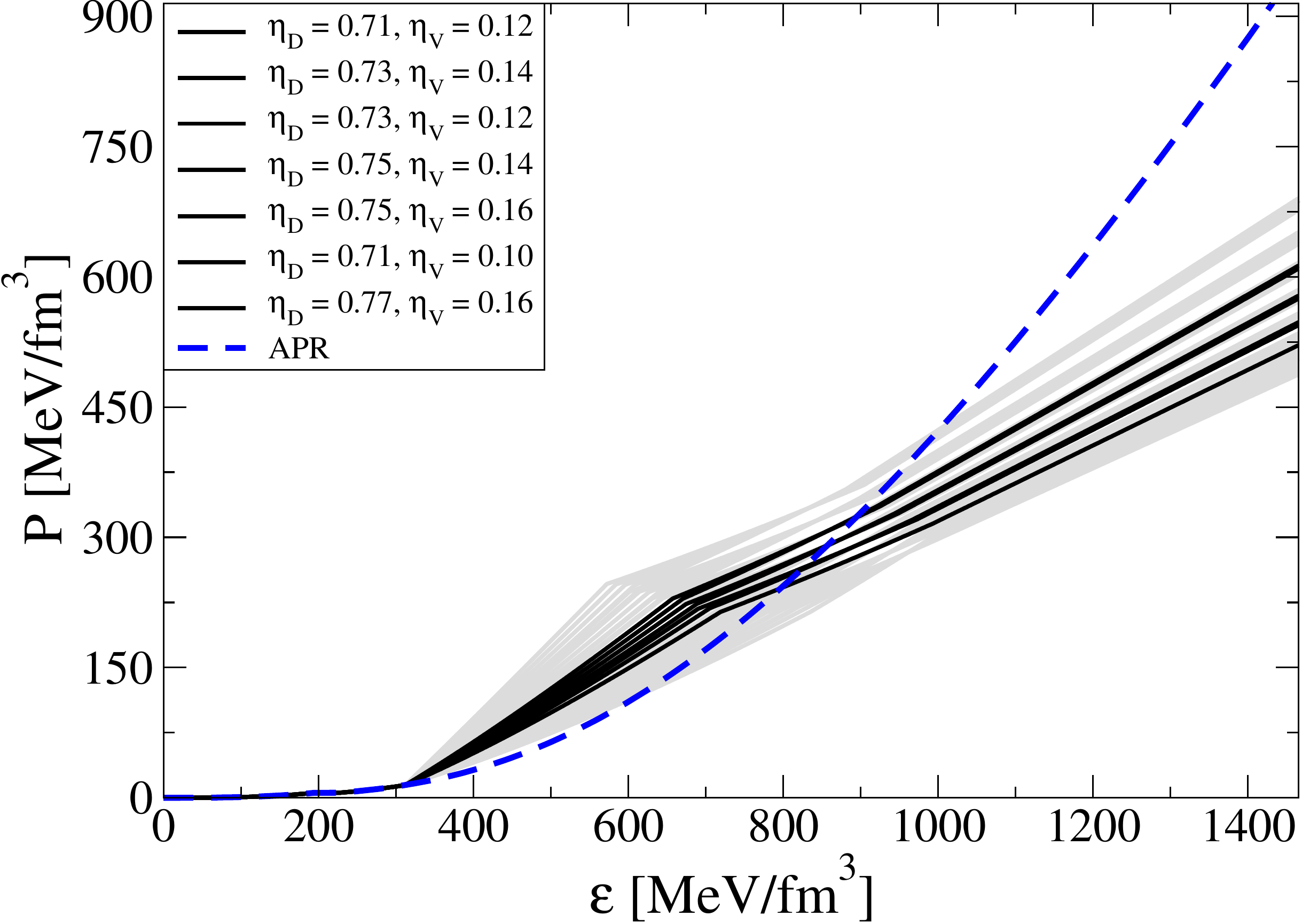}
\\
%&&&\\[-2mm]
\hline
&&&\\[-2mm]
\includegraphics[width=0.2\textwidth]{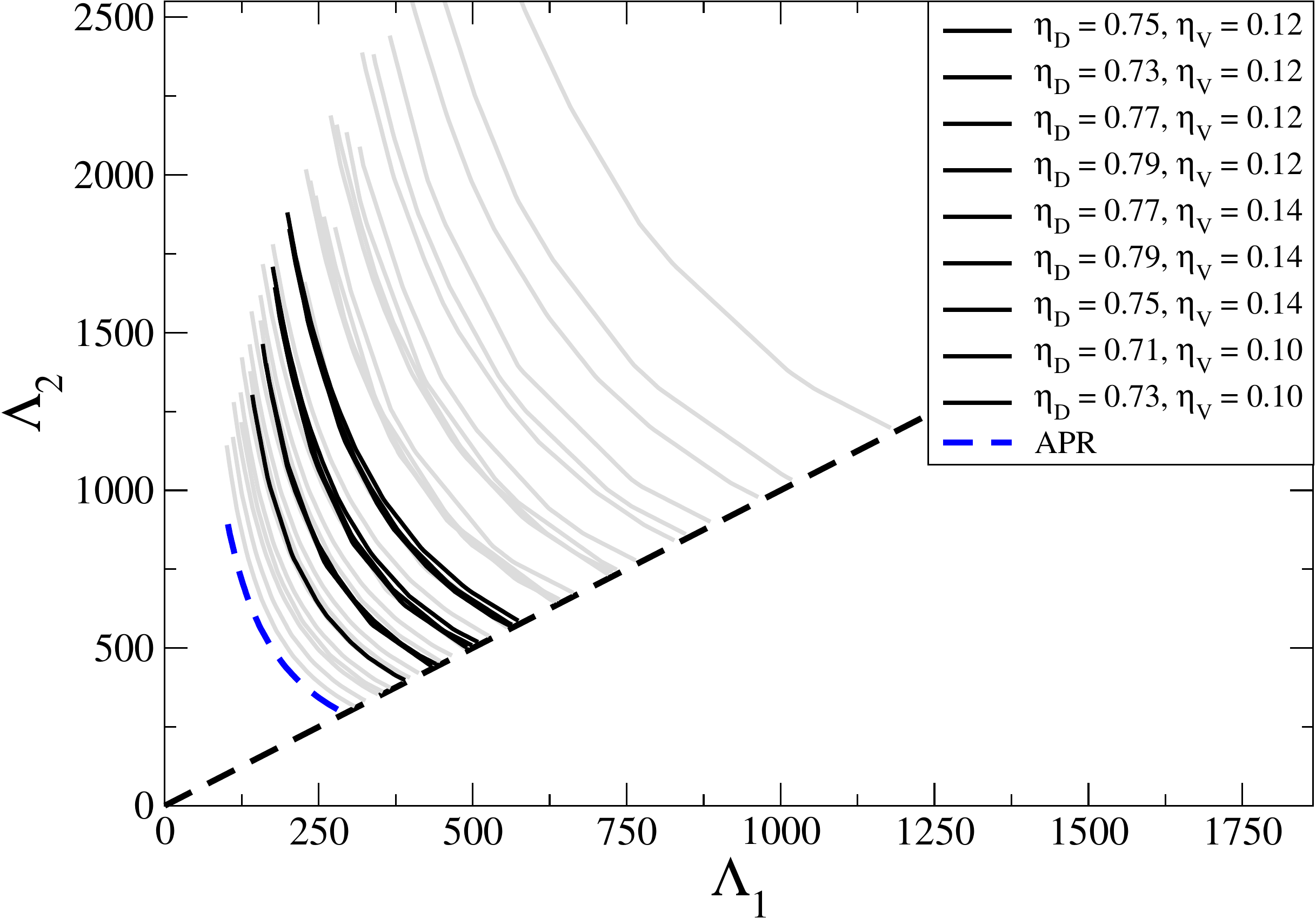}&
\includegraphics[width=0.2\textwidth]{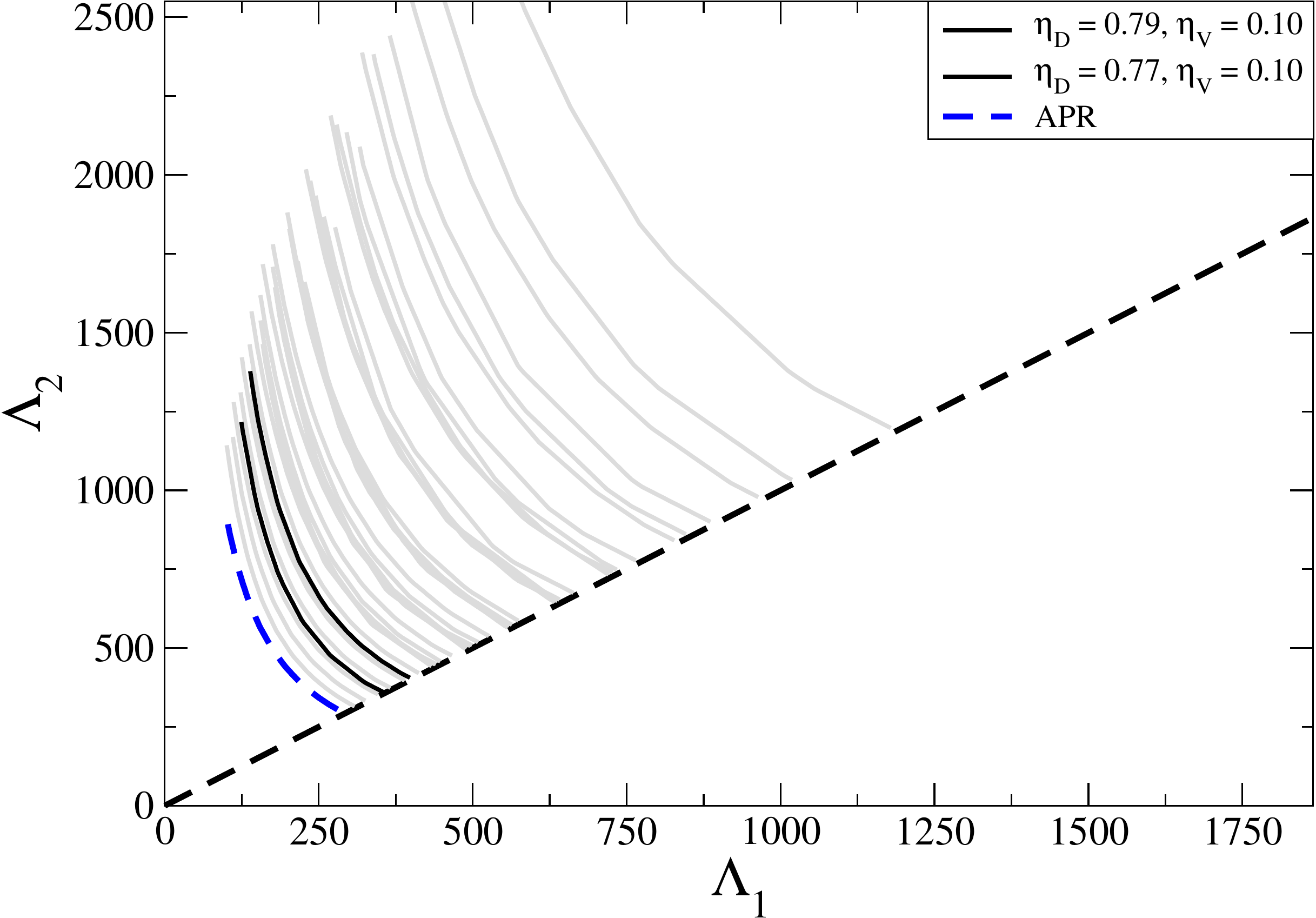}&
\includegraphics[width=0.2\textwidth]{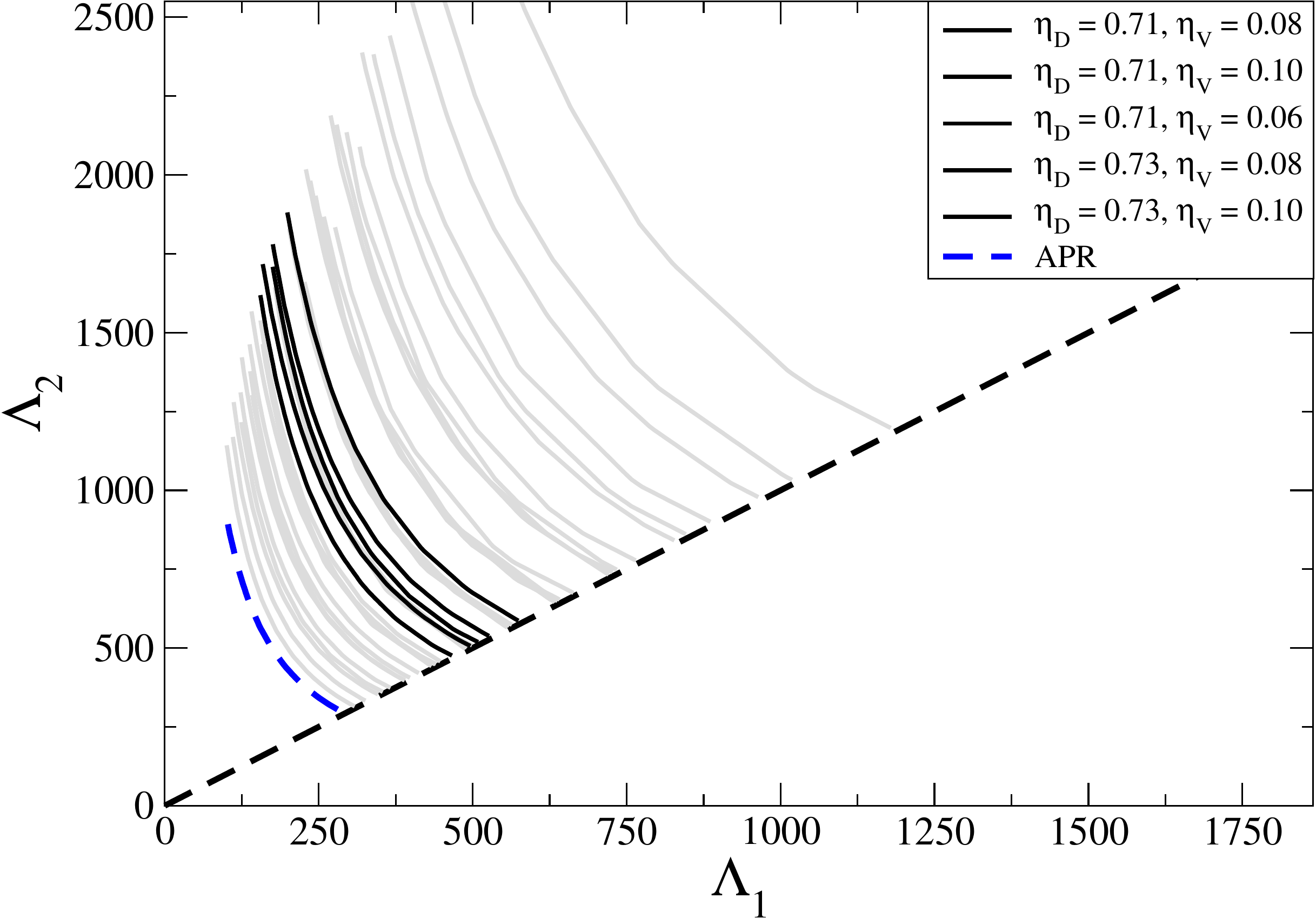}&
\includegraphics[width=0.2\textwidth]{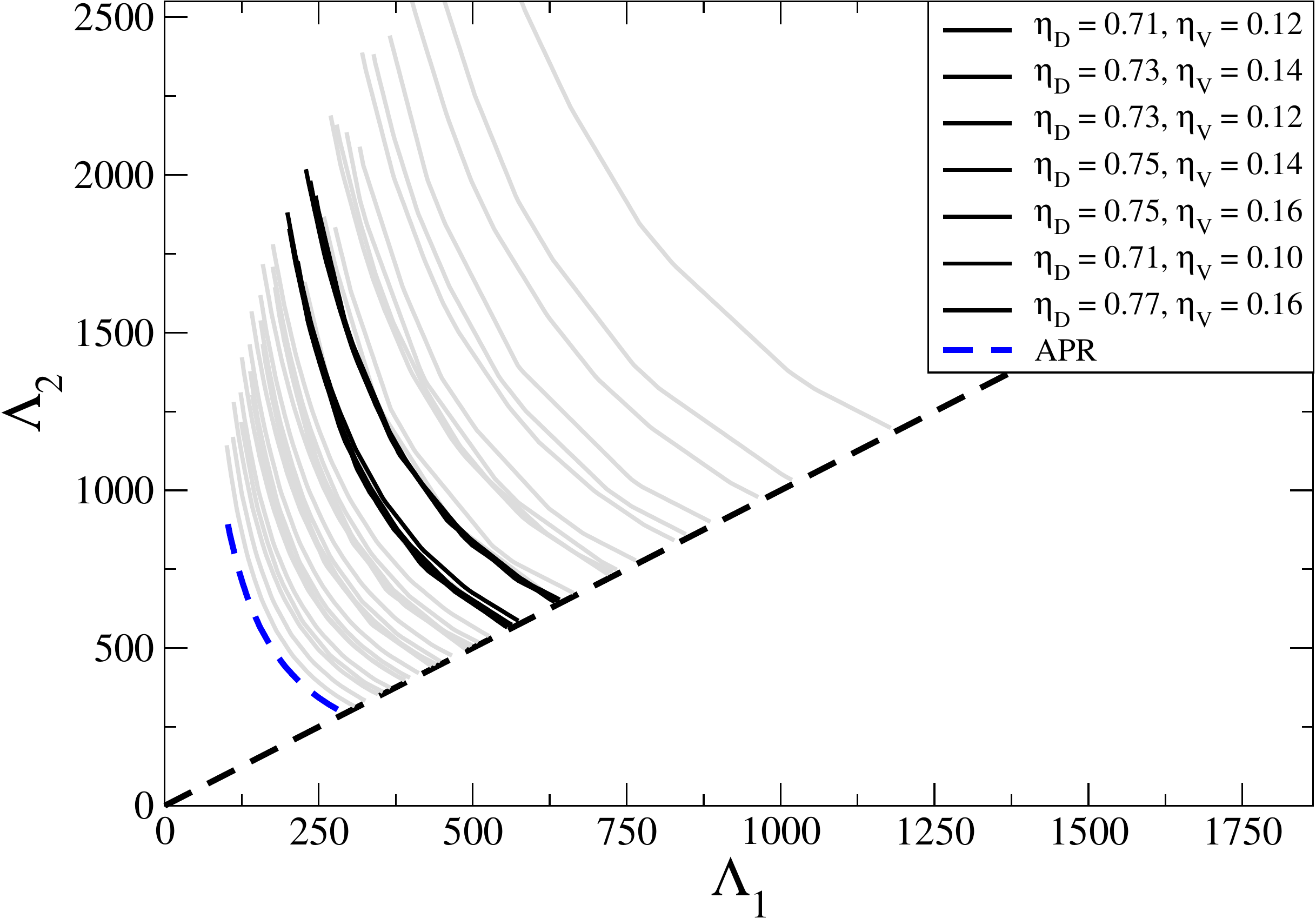}
\\
\hline 
%\end{array}$
\end{tabular}
\end{center}
\caption{Results of the BA for set 1 which includes the constraints (inf\{$M_{\rm max}$\} \cite{Fonseca:2021wxt}, $\Lambda_{1.4}$ \cite{Abbott:2018exr}, $(M,R)_{\rm J0030+0451}$ and \cite{Miller:2019cac}) in the leftmost column and with an additional (yet fictitious) NICER radius measurement for PSR J0740+6620 of $R=11$, 12 or 13 km with an estimated standard deviation of $\sigma_R=0.5$ km in the other three columns.
The highlighted most probable M-R sequences (2nd row), EoS (3rd row) and $\Lambda_1-\Lambda_2$
(4th row) relationships correspond to the parameter sets with at least 75\% 
of the maximum probability as shown in the first row.
\label{fig:BA-fict-set1}}
\end{figure*}

%%%%%%%%%%%%%%%  Figure 18
\begin{figure*}[!ht]
\begin{center}
%\begin{array}
\begin{tabular}{l|c|c|c}
\hline
&&&\\[-2mm]
set 2 &set 2 + $R=11$ km & set 2 + $R=12$ km & set 2 + $R=13$ km\\%[-2mm]
&&&\\[-2mm]
\hline
&&&\\[-2mm]
\includegraphics[width=0.2\textwidth]{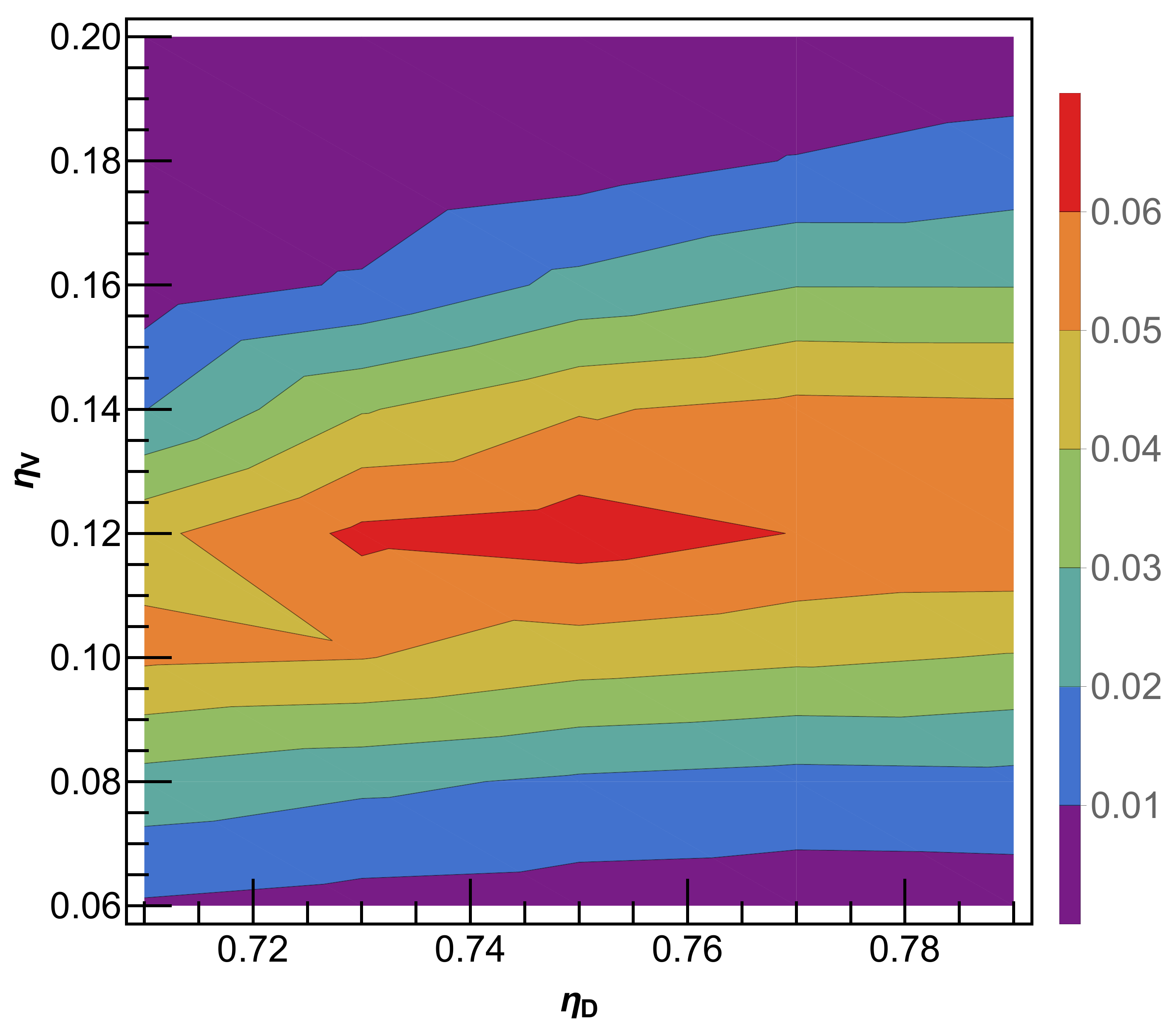} & 
\includegraphics[width=0.2\textwidth]{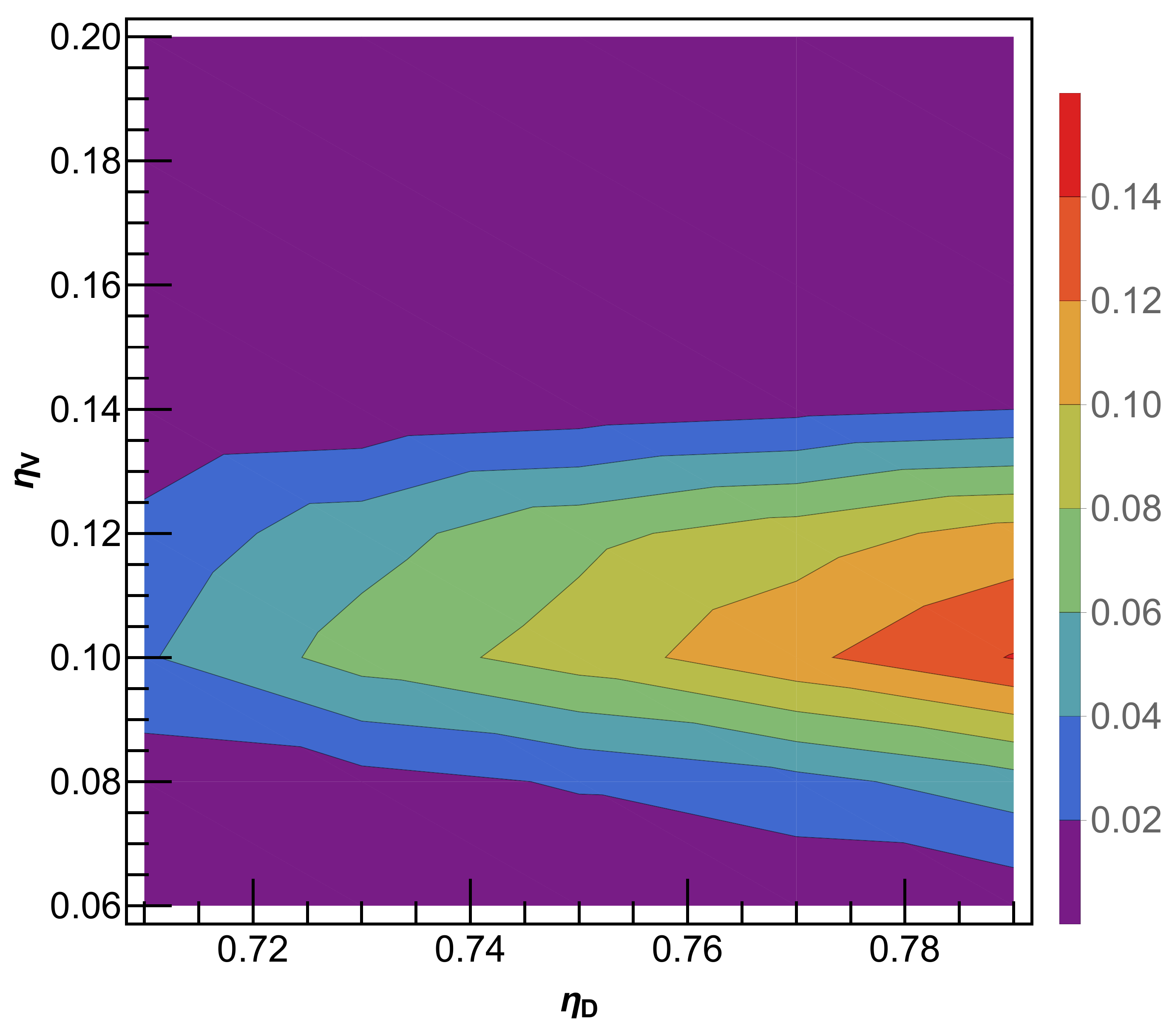} & 
\includegraphics[width=0.2\textwidth]{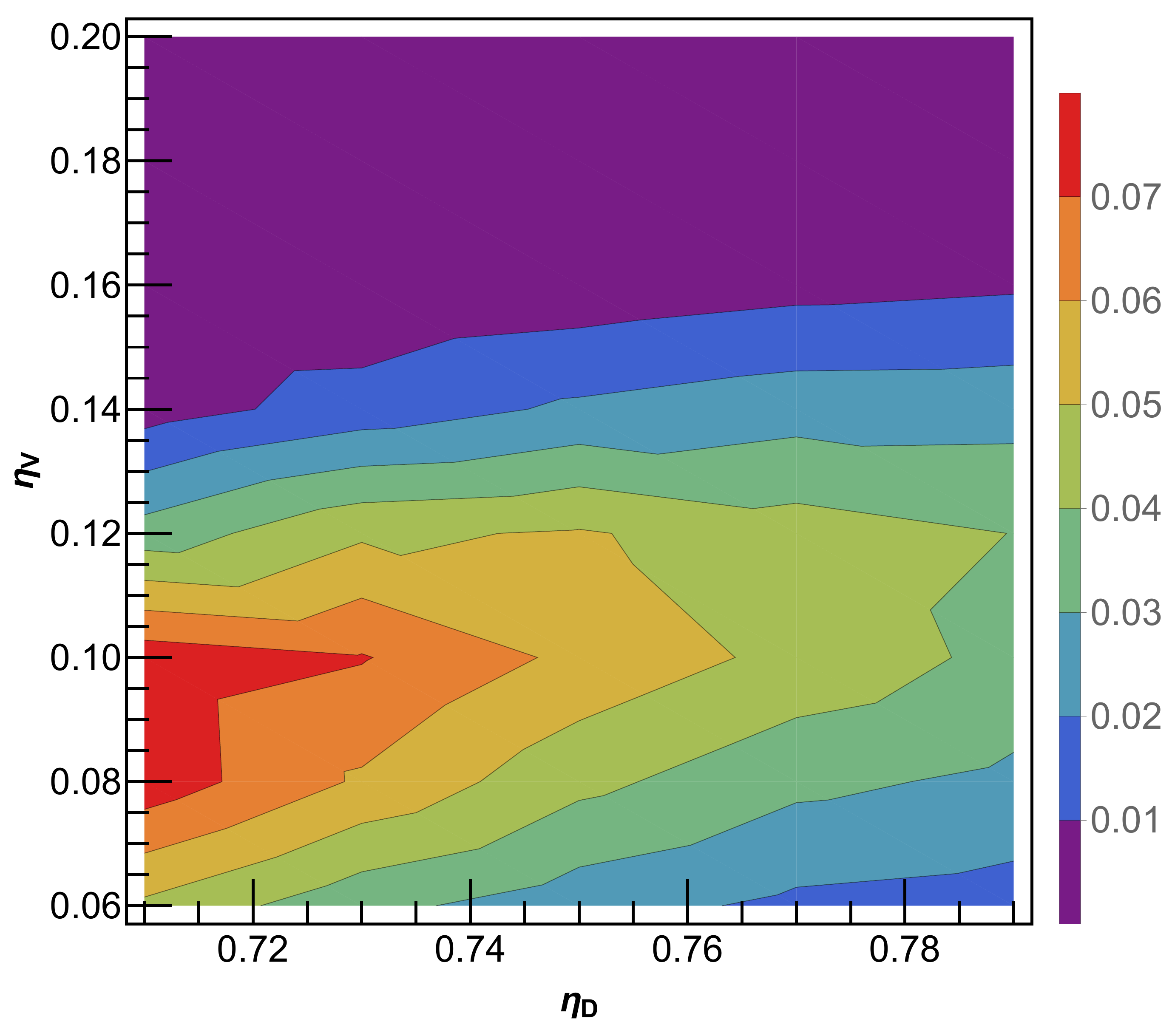} & 
\includegraphics[width=0.2\textwidth]{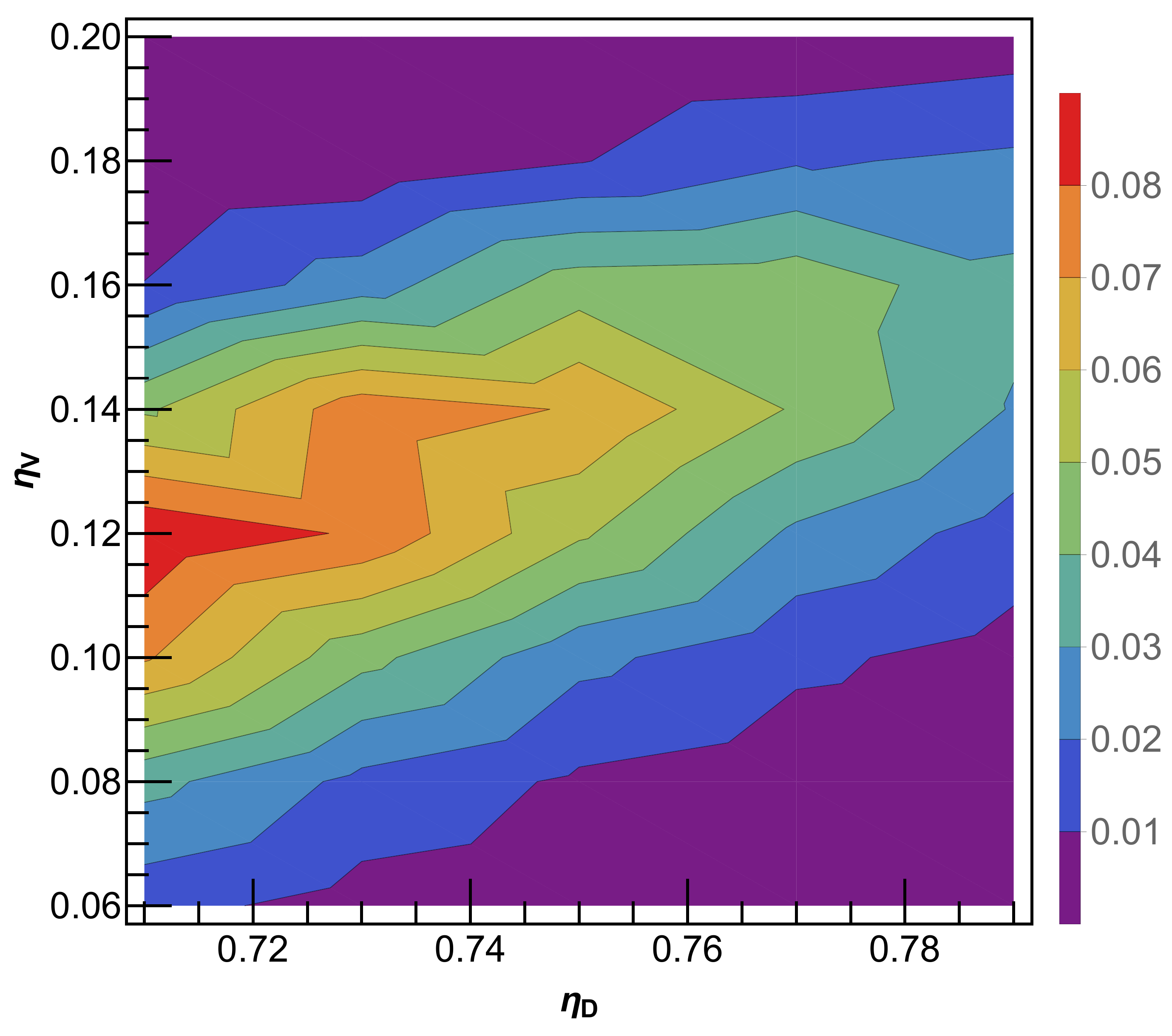} 
\\
%&&&\\[-5mm]
\hline
&&&\\[-2mm]
\includegraphics[width=0.2\textwidth]{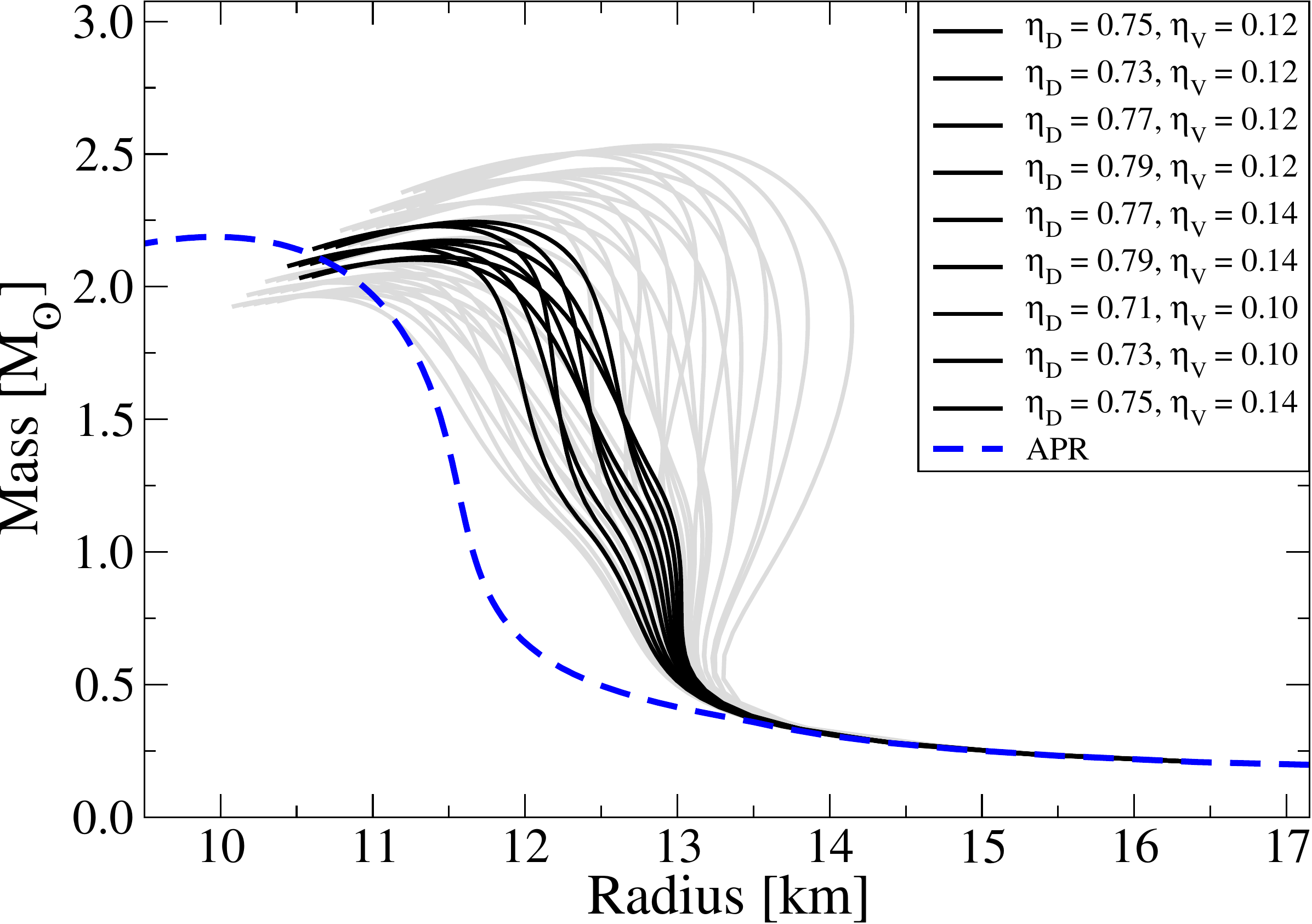} & 
\includegraphics[width=0.2\textwidth]{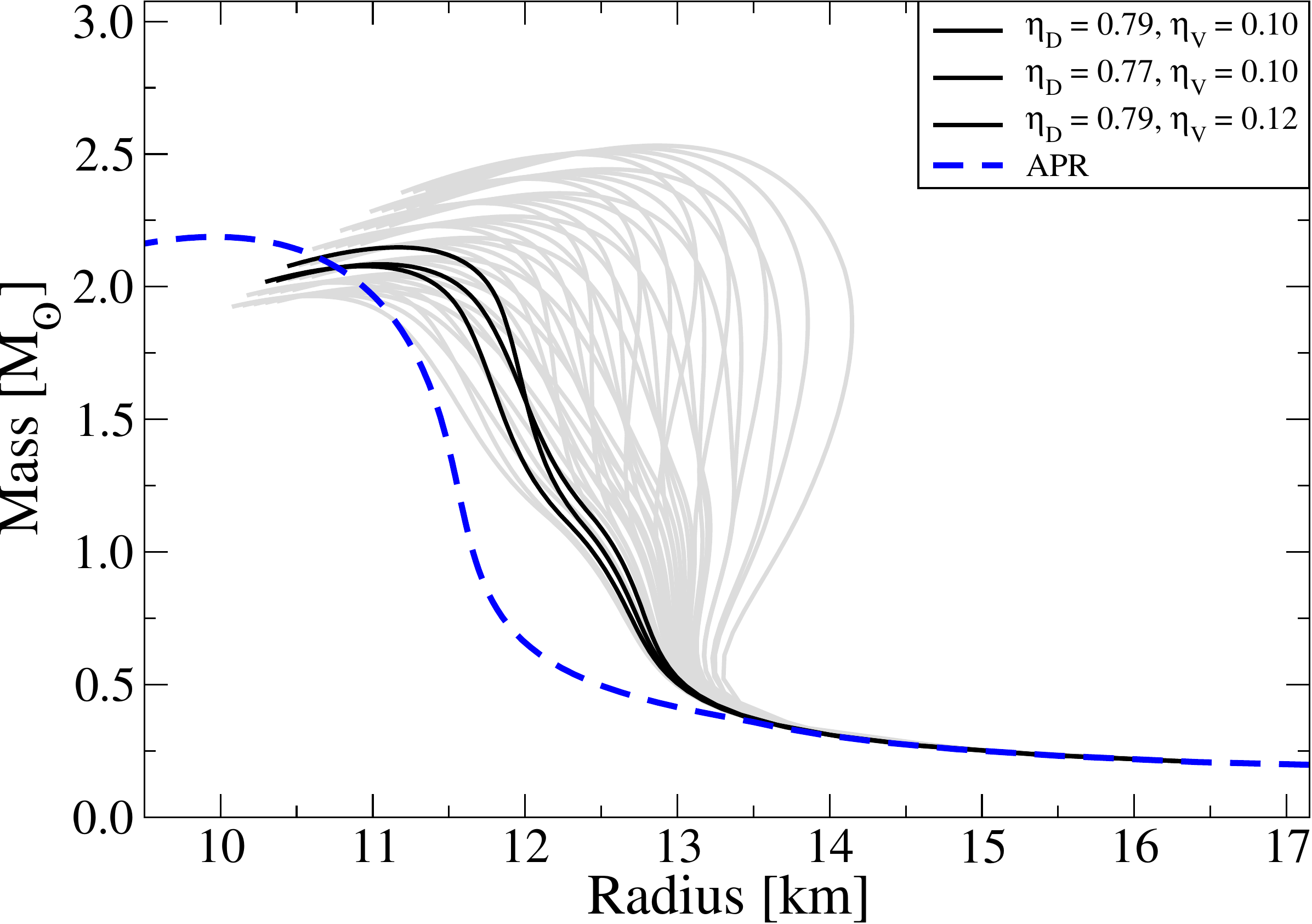}& 
\includegraphics[width=0.2\textwidth]{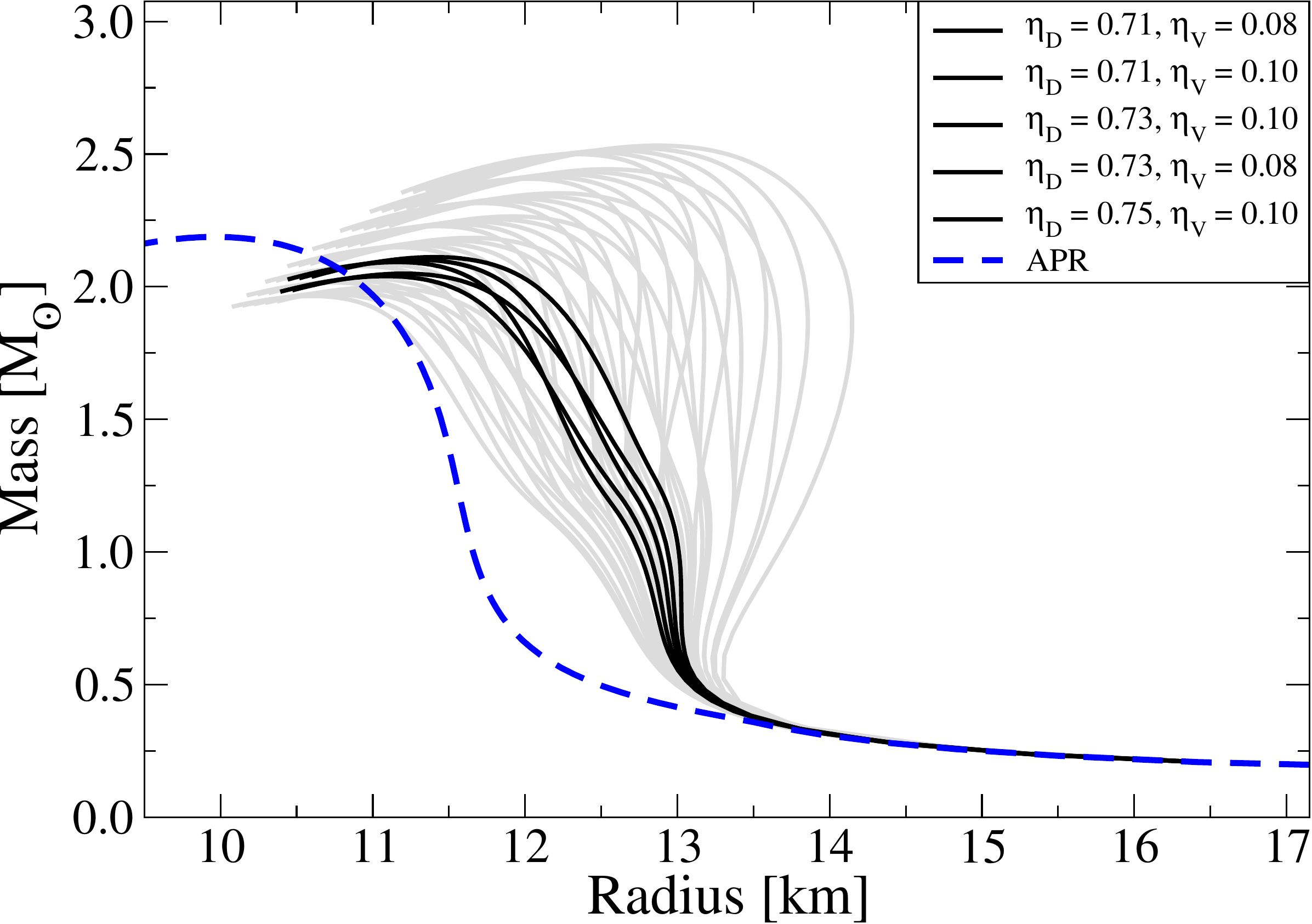}& 
\includegraphics[width=0.2\textwidth]{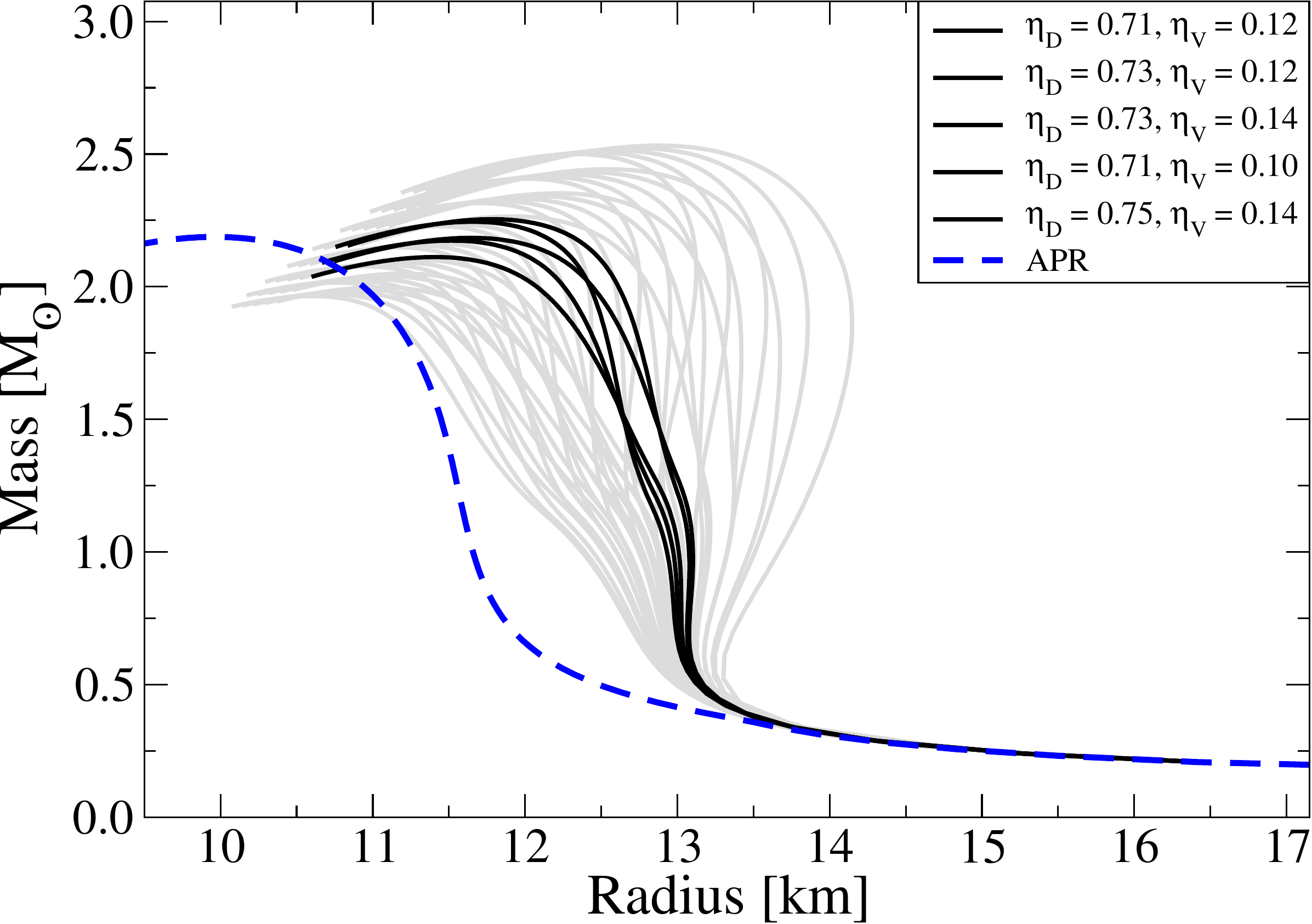}
\\
%&&&\\[-5mm]
\hline
&&&\\[-2mm]
\includegraphics[width=0.2\textwidth]{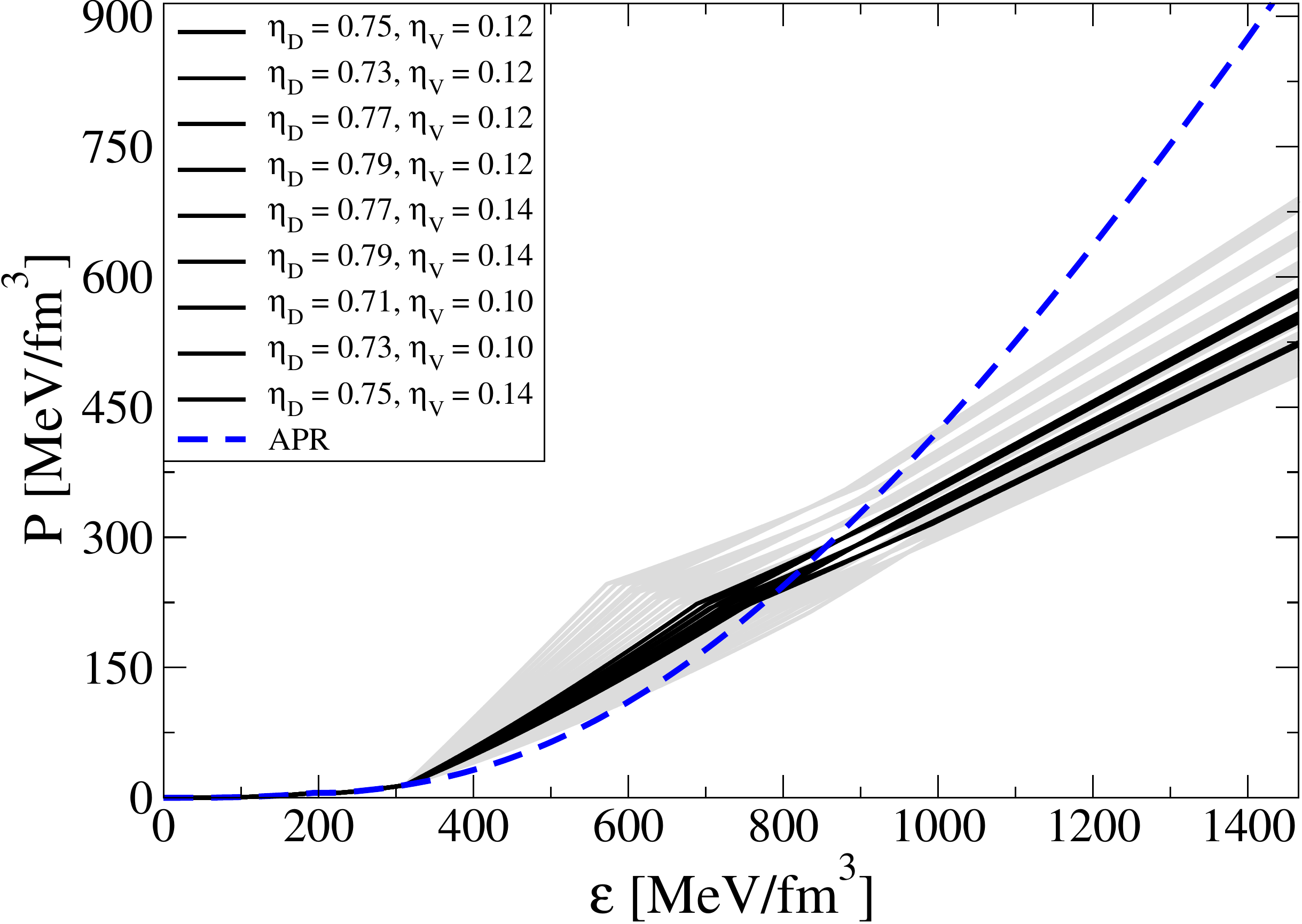} & 
\includegraphics[width=0.2\textwidth]{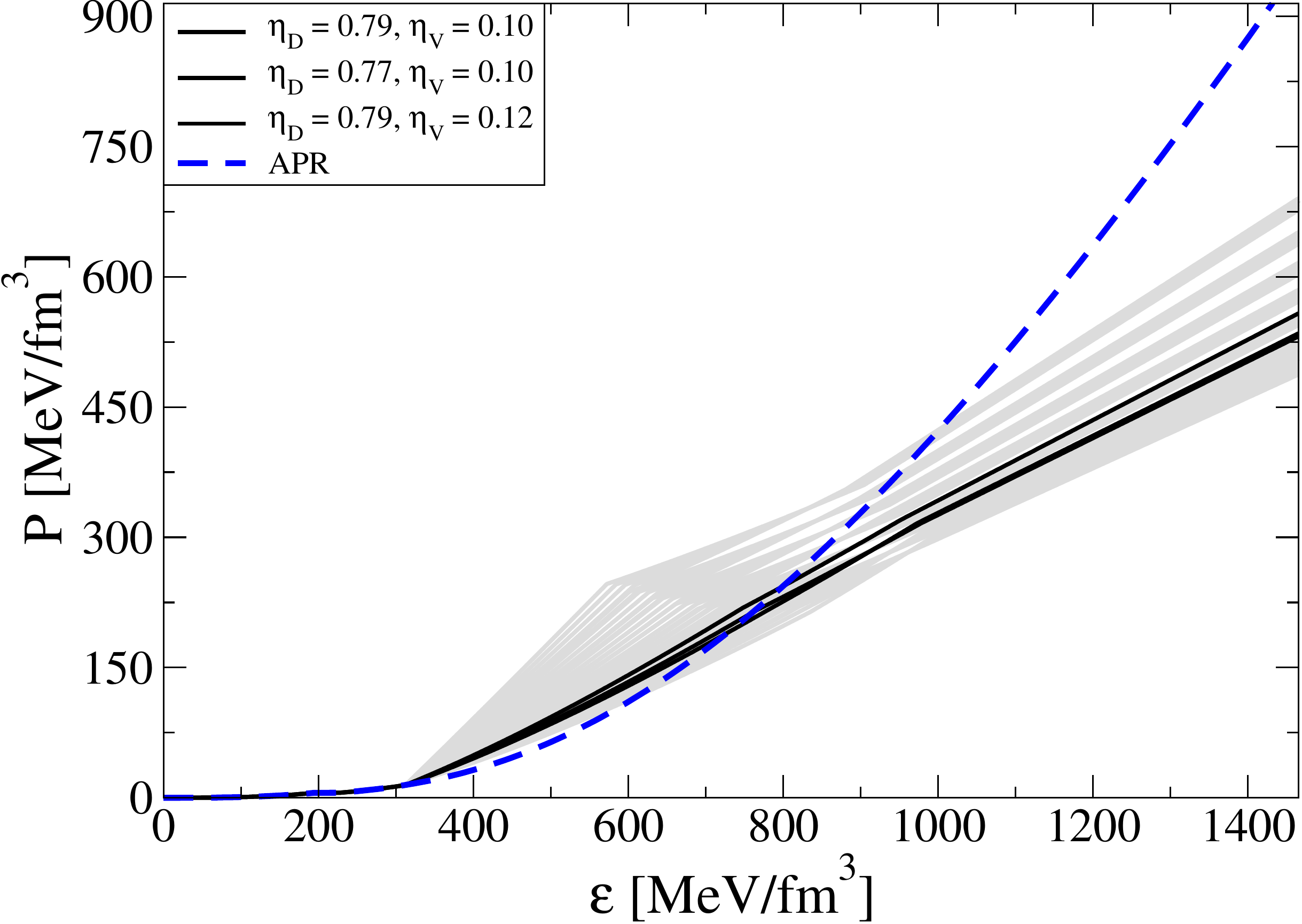}& 
\includegraphics[width=0.2\textwidth]{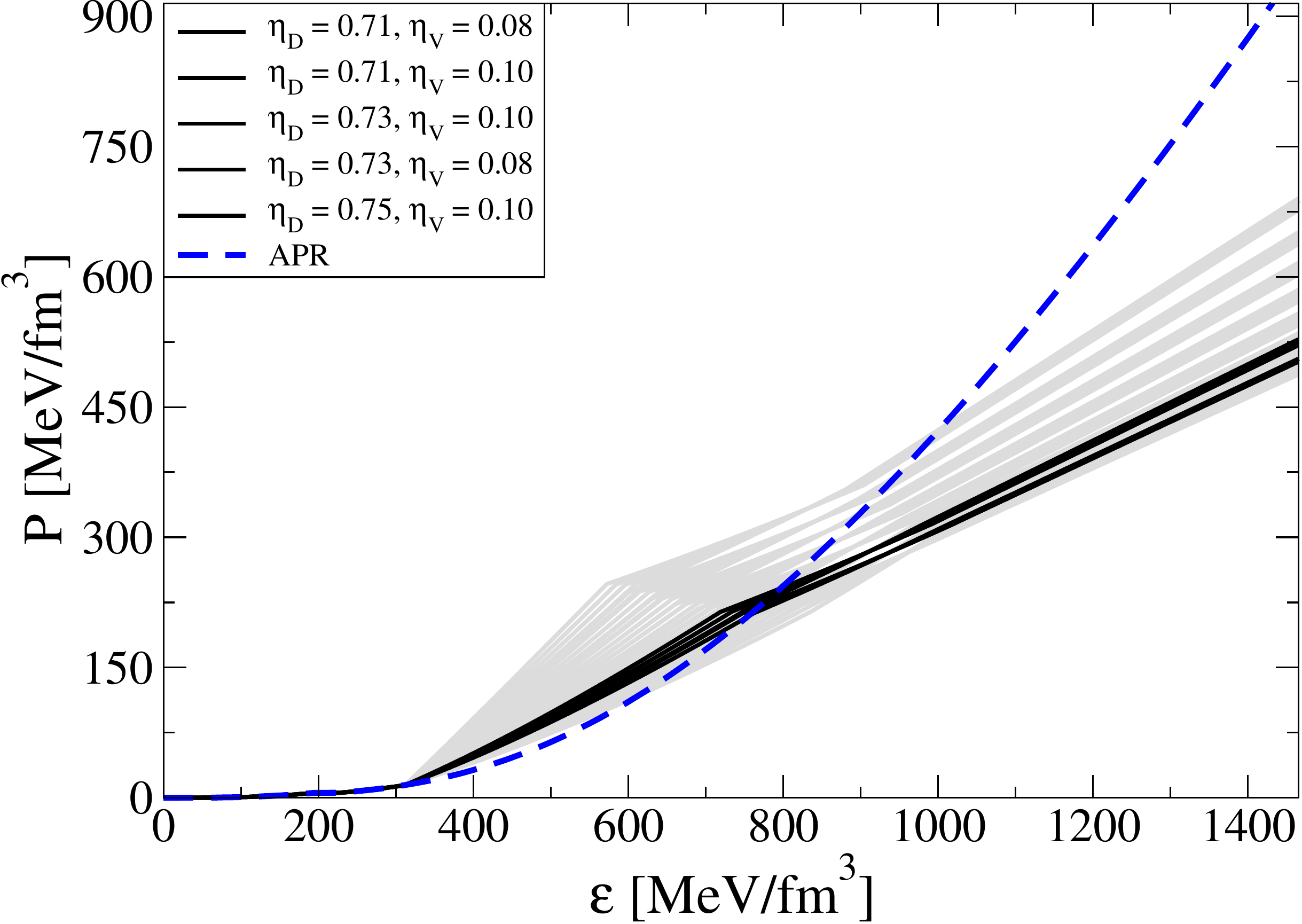}& 
\includegraphics[width=0.2\textwidth]{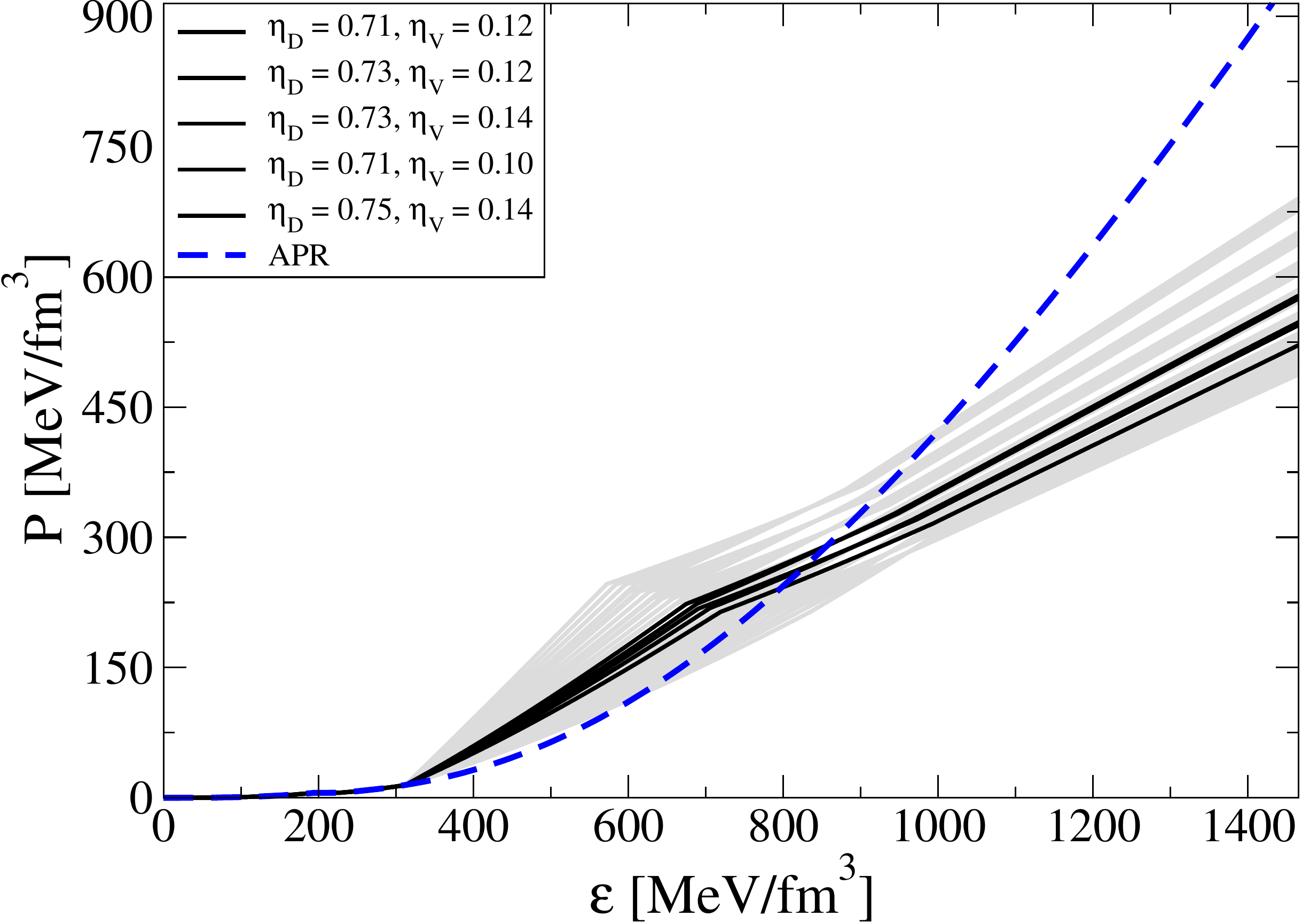}
\\
%&&&\\[-5mm]
\hline
&&&\\[-2mm]
\includegraphics[width=0.2\textwidth]{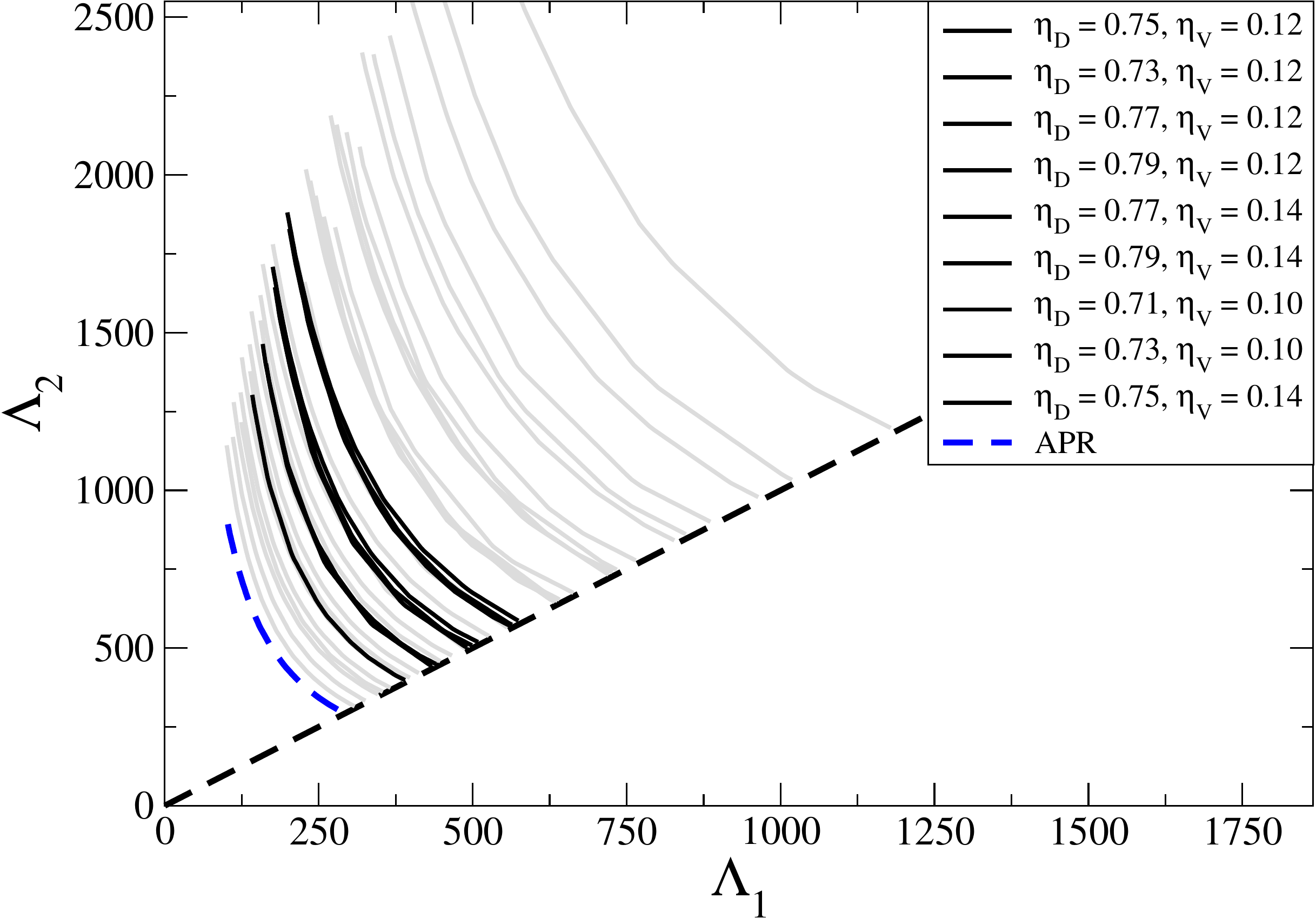} & 
\includegraphics[width=0.2\textwidth]{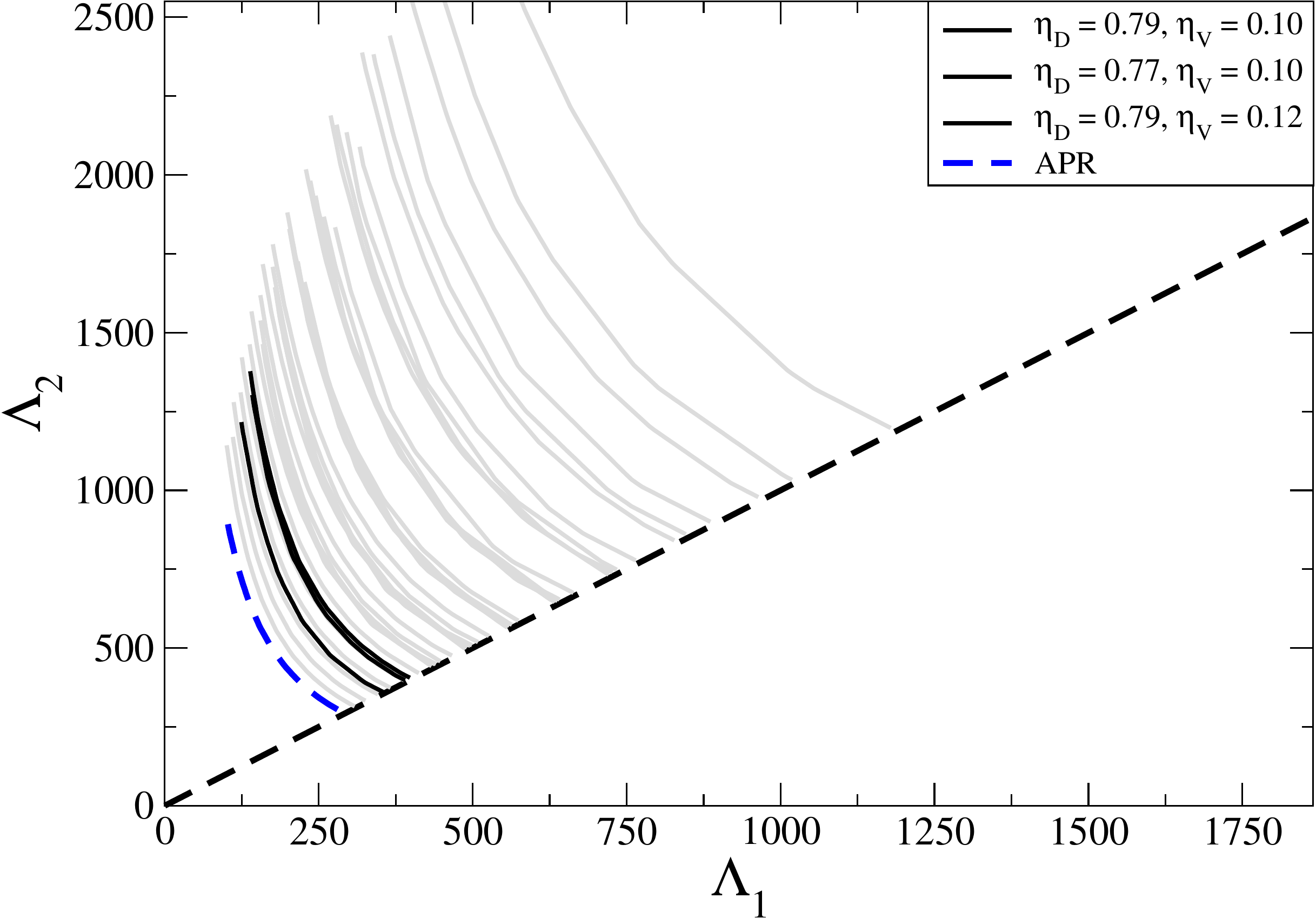}& 
\includegraphics[width=0.2\textwidth]{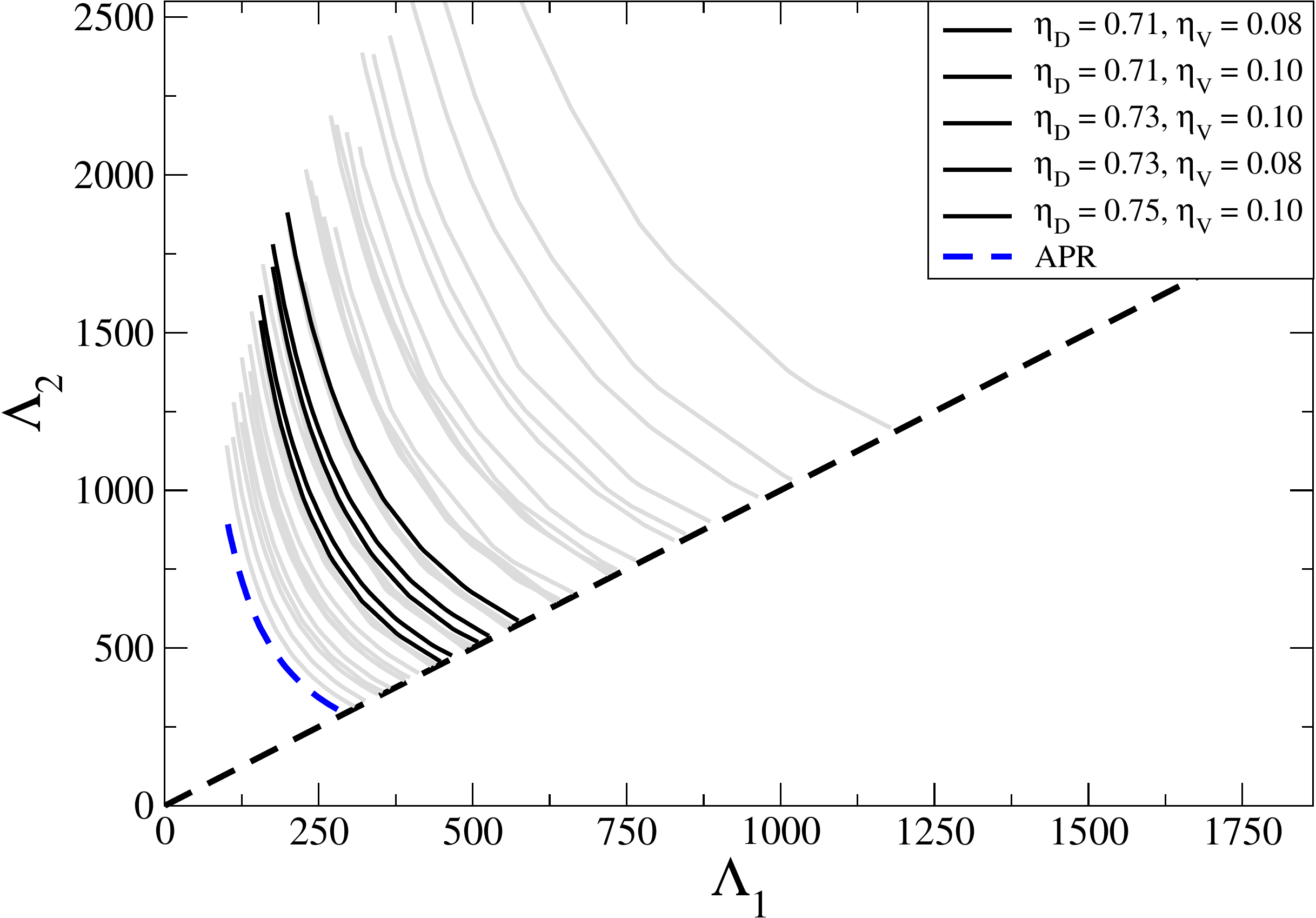}& 
\includegraphics[width=0.2\textwidth]{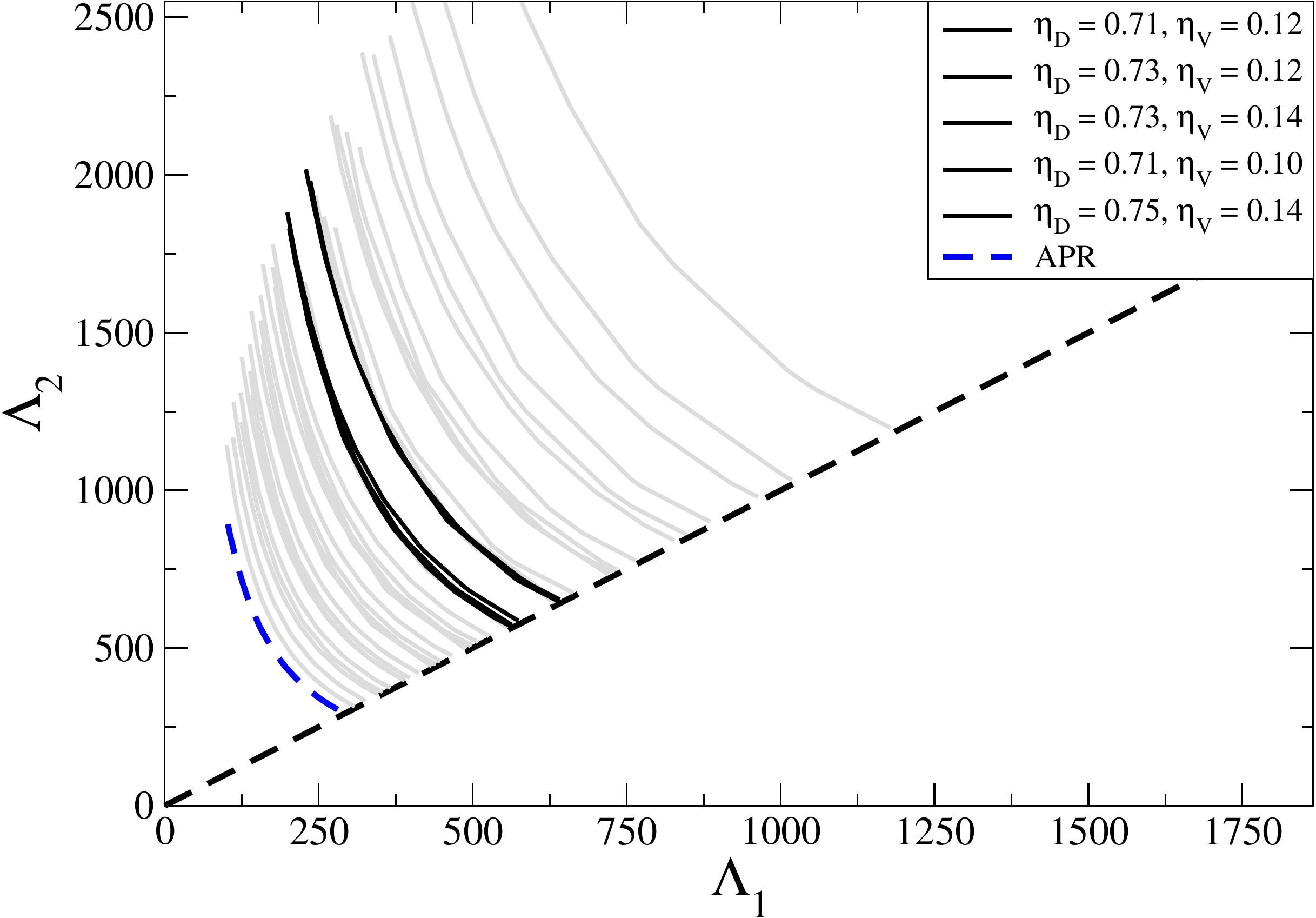}
\\
\hline 
%\end{array}$
\end{tabular}
\end{center}
\caption{Results of the BA for set 2 which includes the constraints (inf\{$M_{\rm max}$\} \cite{Fonseca:2021wxt}, $\Lambda_{1.4}$ \cite{Abbott:2018exr},
$(M,R)_{\rm J0030+0451}$ \cite{Miller:2019cac}, {sup}\{$M_{\rm max}$\} \cite{Rezzolla:2017aly}) in the leftmost column and with an additional (yet fictitious) NICER radius measurement for PSR J0740+6620 of $R=11$, 12 or 13 km with an estimated standard deviation of $\sigma_R=0.5$ km in the other three columns.
The highlighted most probable M-R sequences (2nd row), EoS (3rd row) and $\Lambda_1-\Lambda_2$ (4th row) relationships correspond to the parameter sets with at least 75\% 
of the maximum probability as shown in the first row.
\label{fig:BA-fict-set2}}
\end{figure*}

%%%%%% Figure 19 %%%%%%
\begin{figure*}[!ht]
\begin{center}
%\begin{array}
\begin{tabular}{l|c|c|c}
\hline
&&&\\[-2mm]
set 3&set 3 + $R=11$ km & set 3 + $R=12$ km& set 3 + $R=13$ km\\%[-2mm]
&&&\\[-2mm]
\hline
&&&\\[-2mm]
\includegraphics[width=0.2\textwidth]{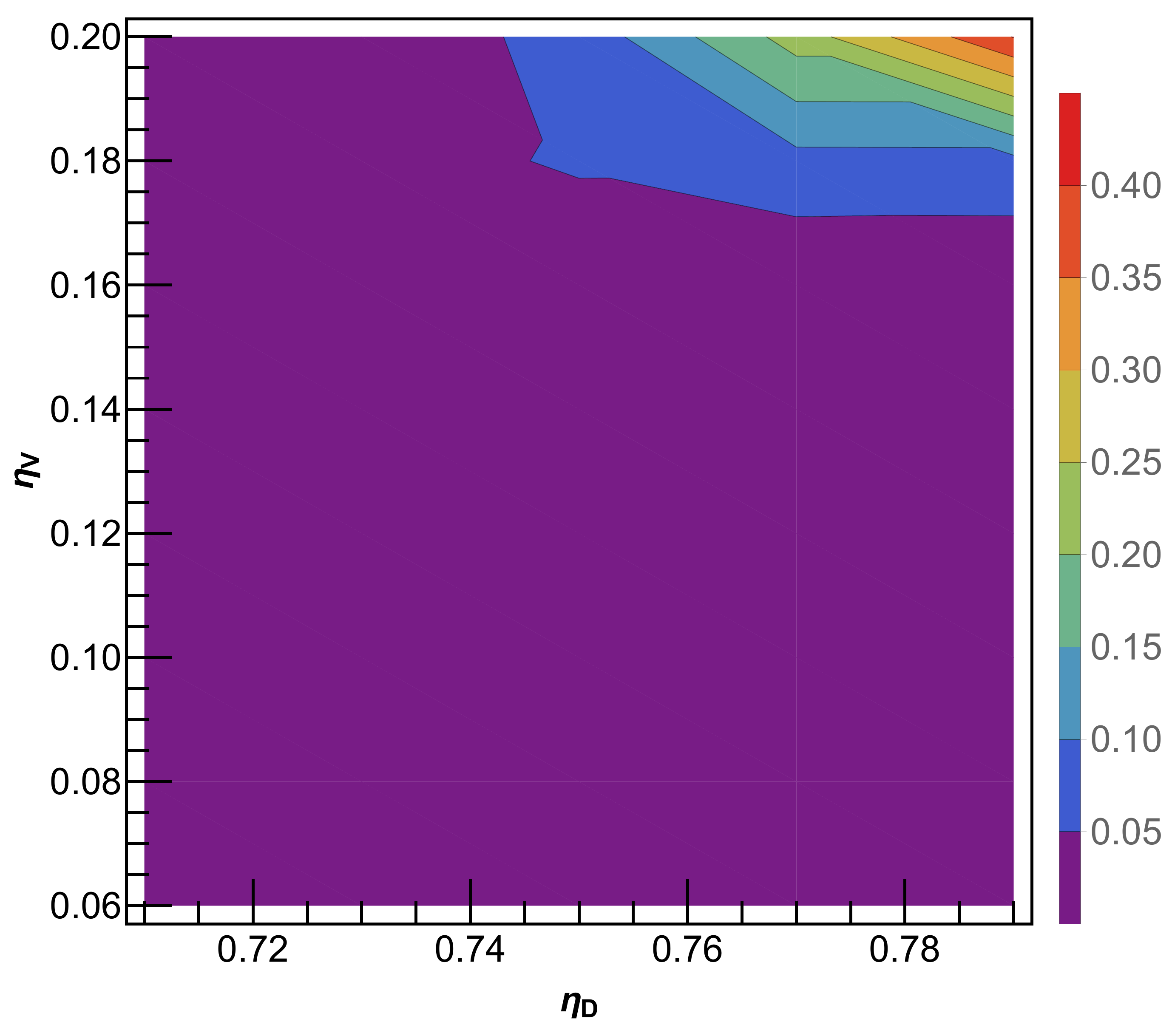} & 
\includegraphics[width=0.2\textwidth]{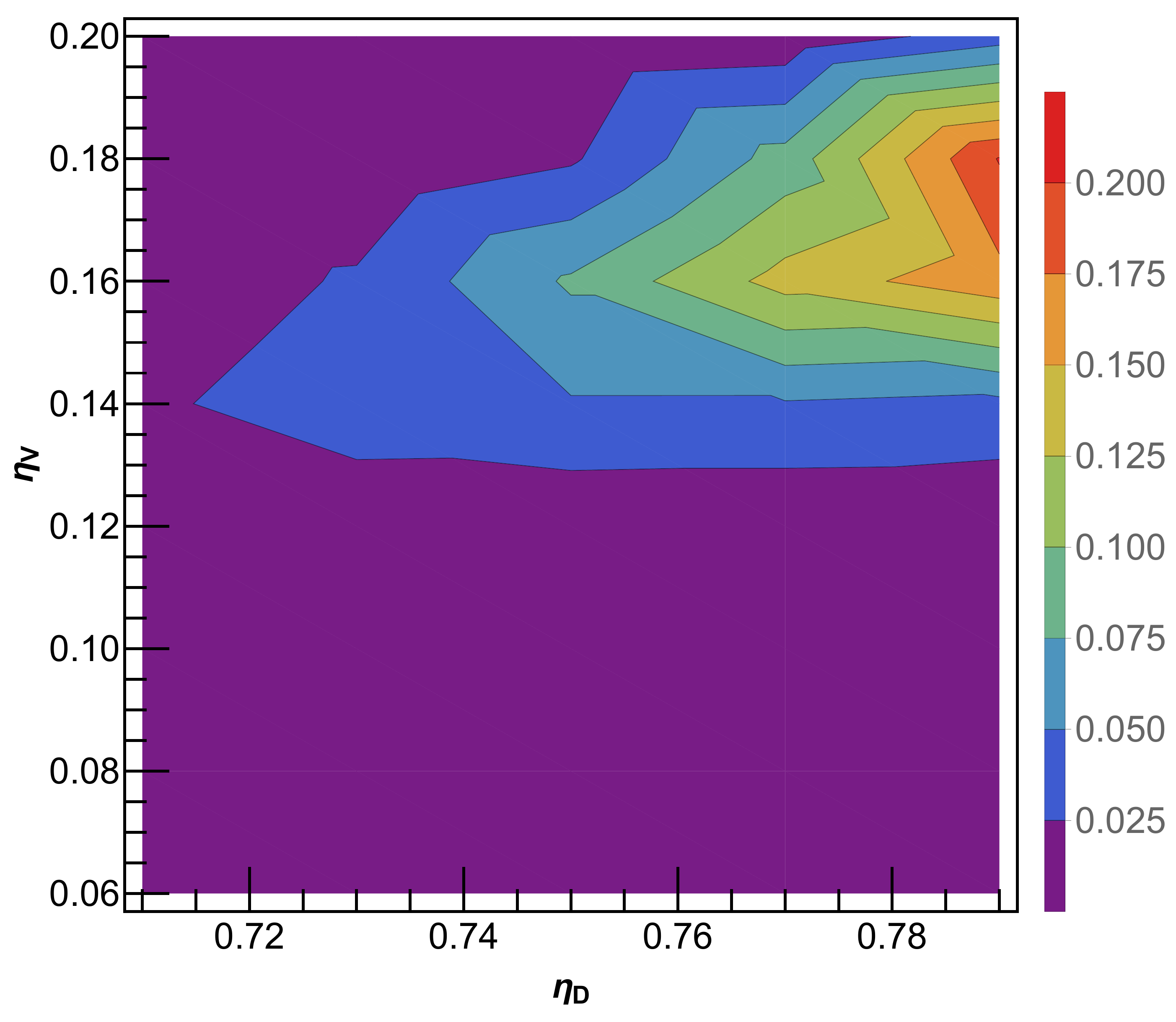} & 
\includegraphics[width=0.2\textwidth]{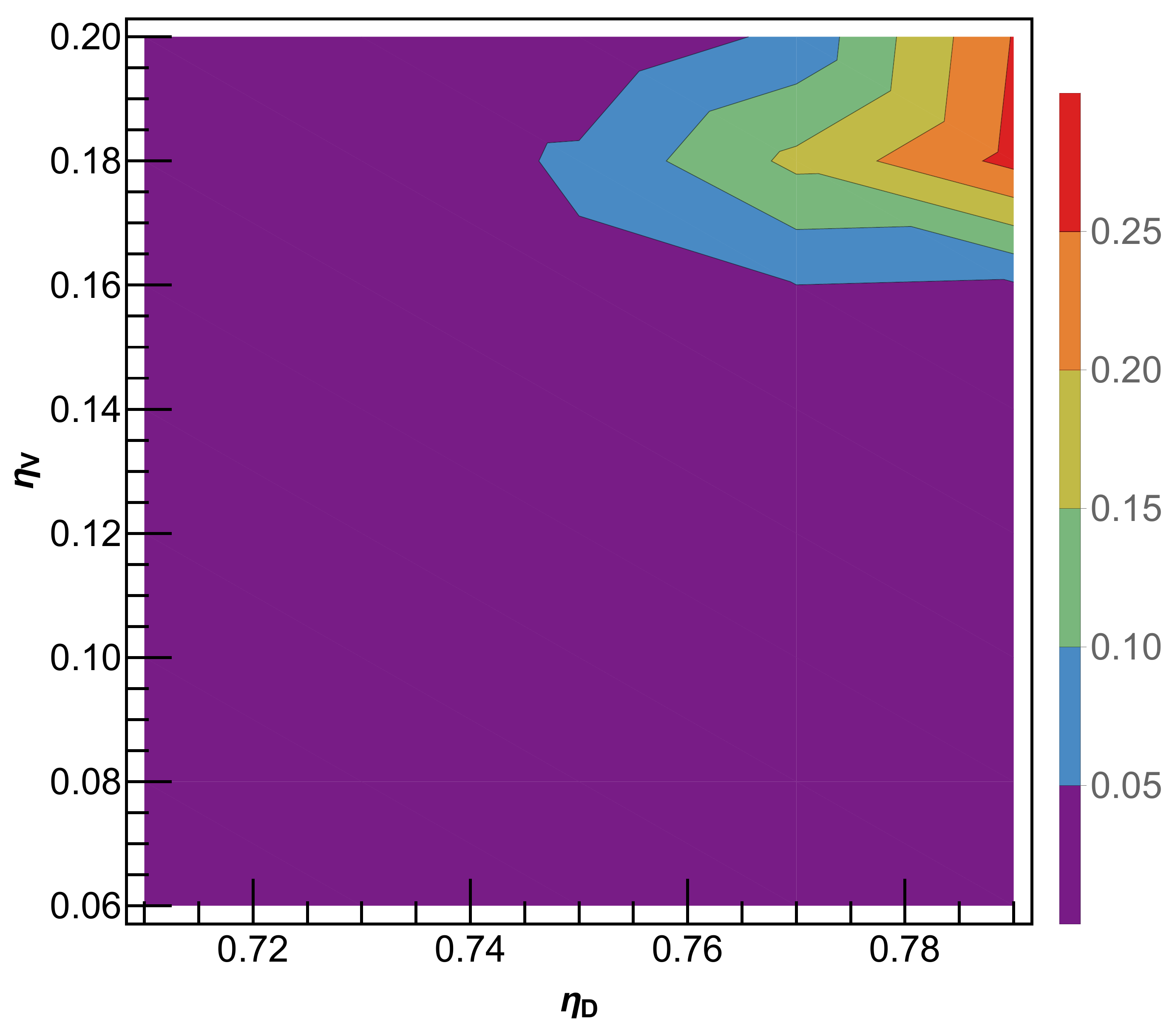} & 
\includegraphics[width=0.2\textwidth]{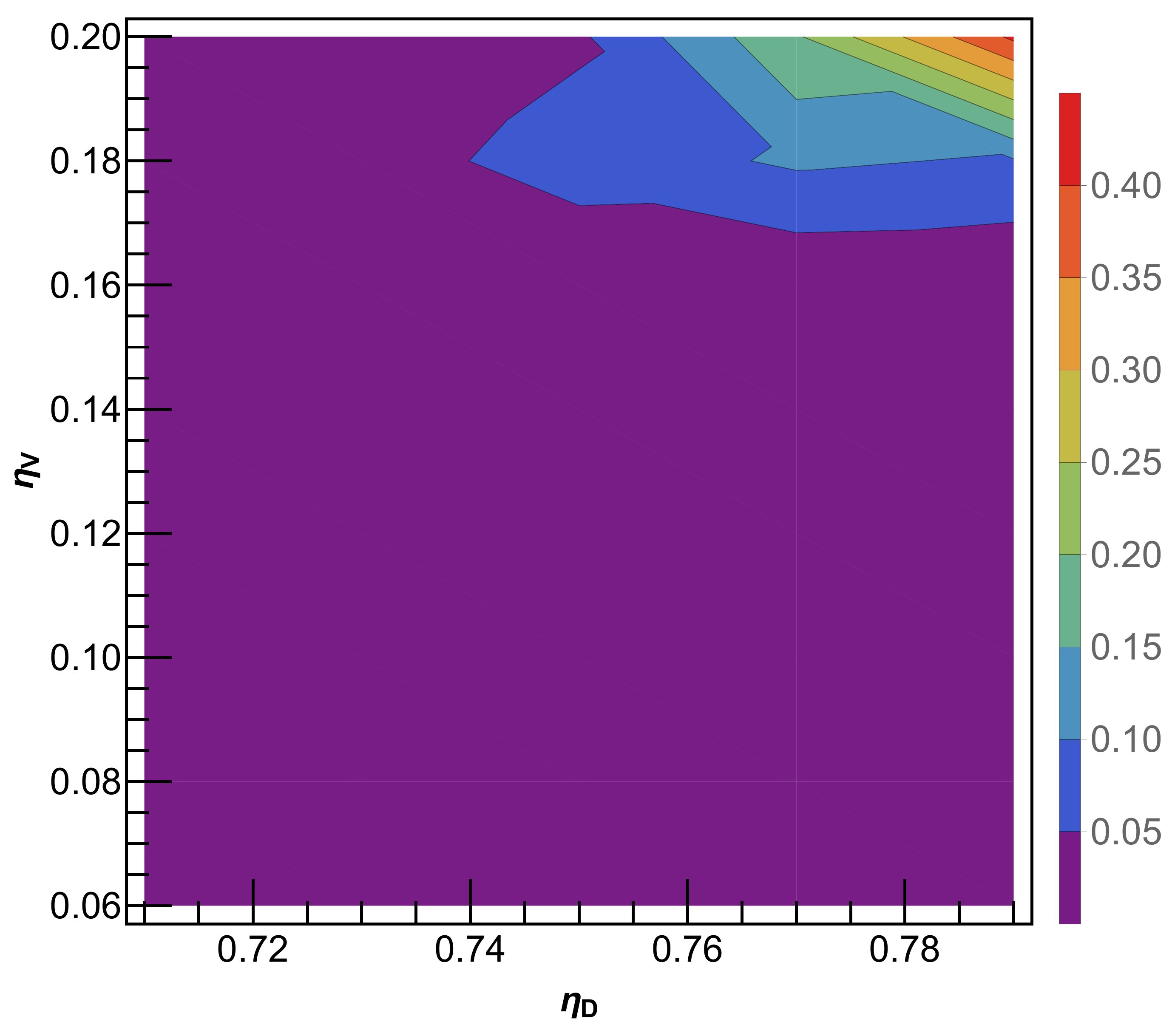} 
\\
%&&&\\[-5mm]
\hline
&&&\\[-2mm]
\includegraphics[width=0.2\textwidth]{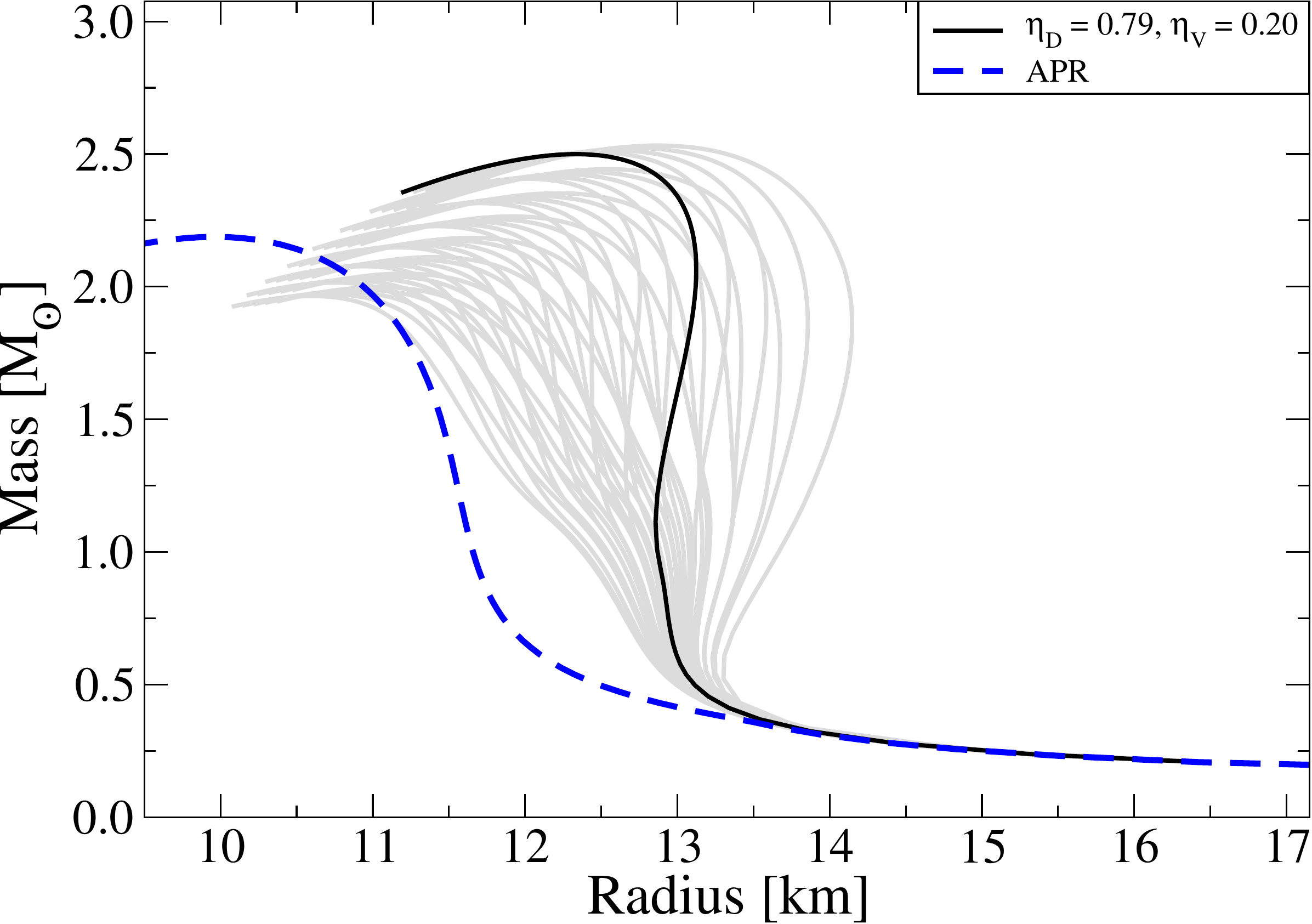} & 
\includegraphics[width=0.2\textwidth]{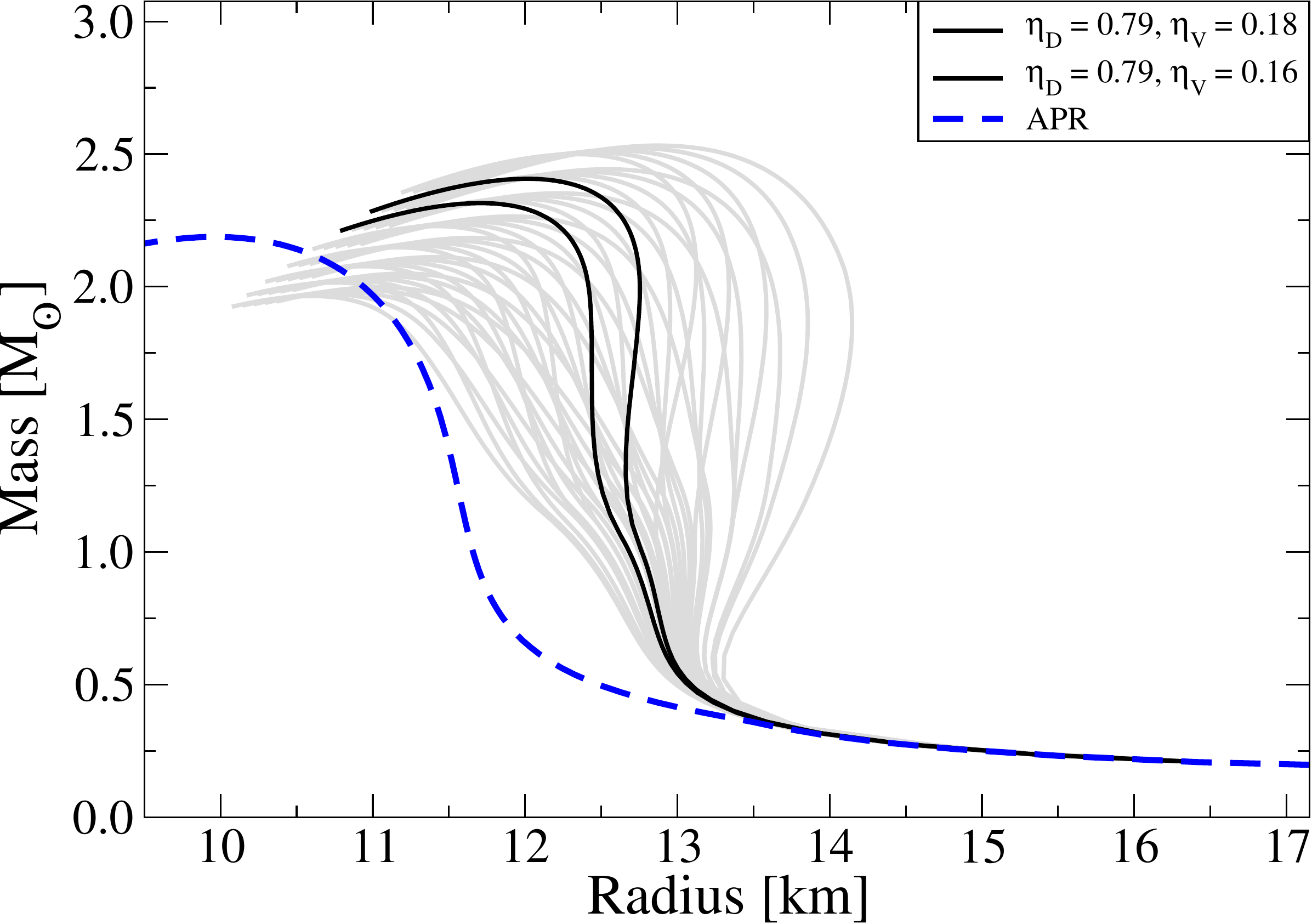}& 
\includegraphics[width=0.2\textwidth]{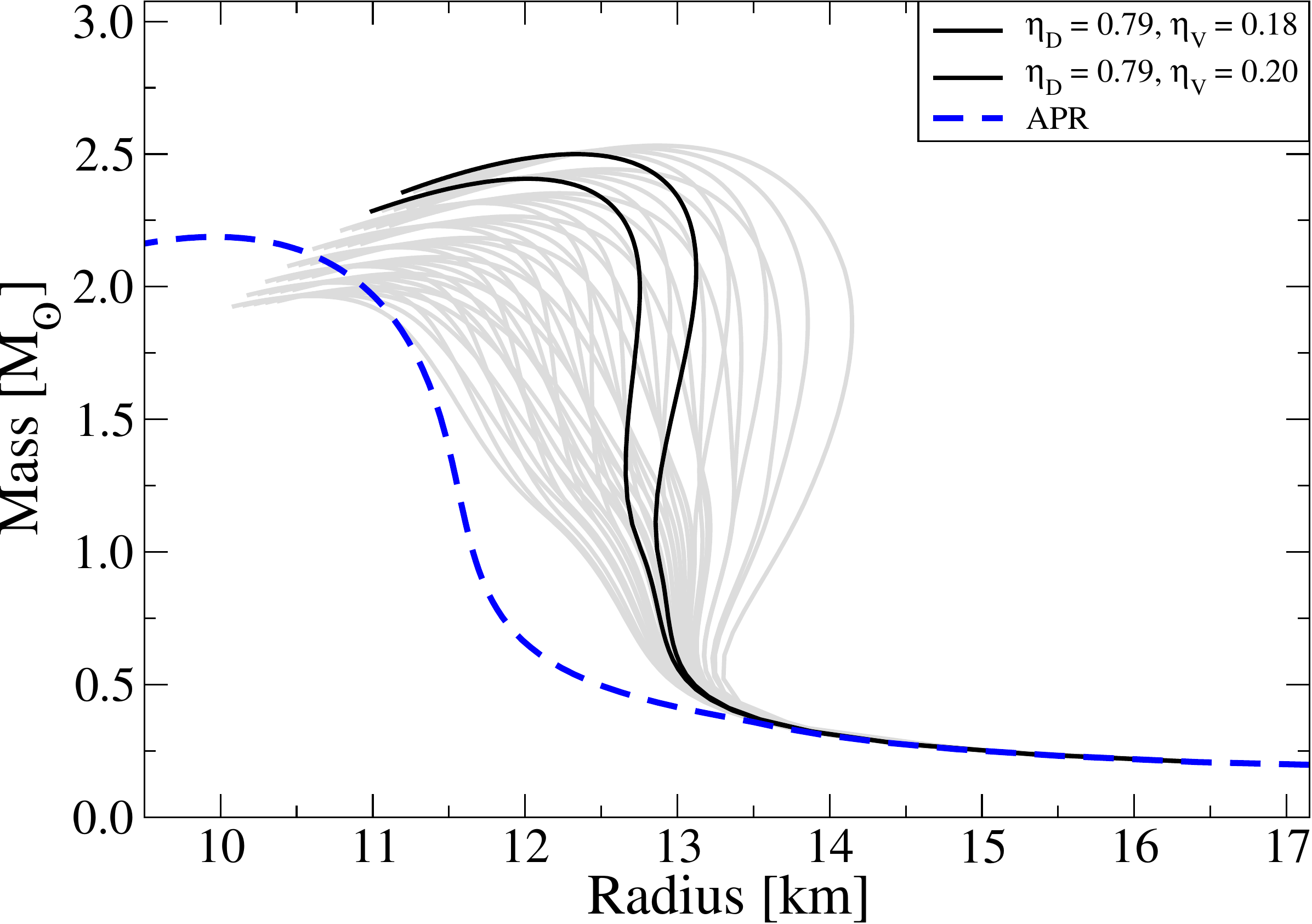}& 
\includegraphics[width=0.2\textwidth]{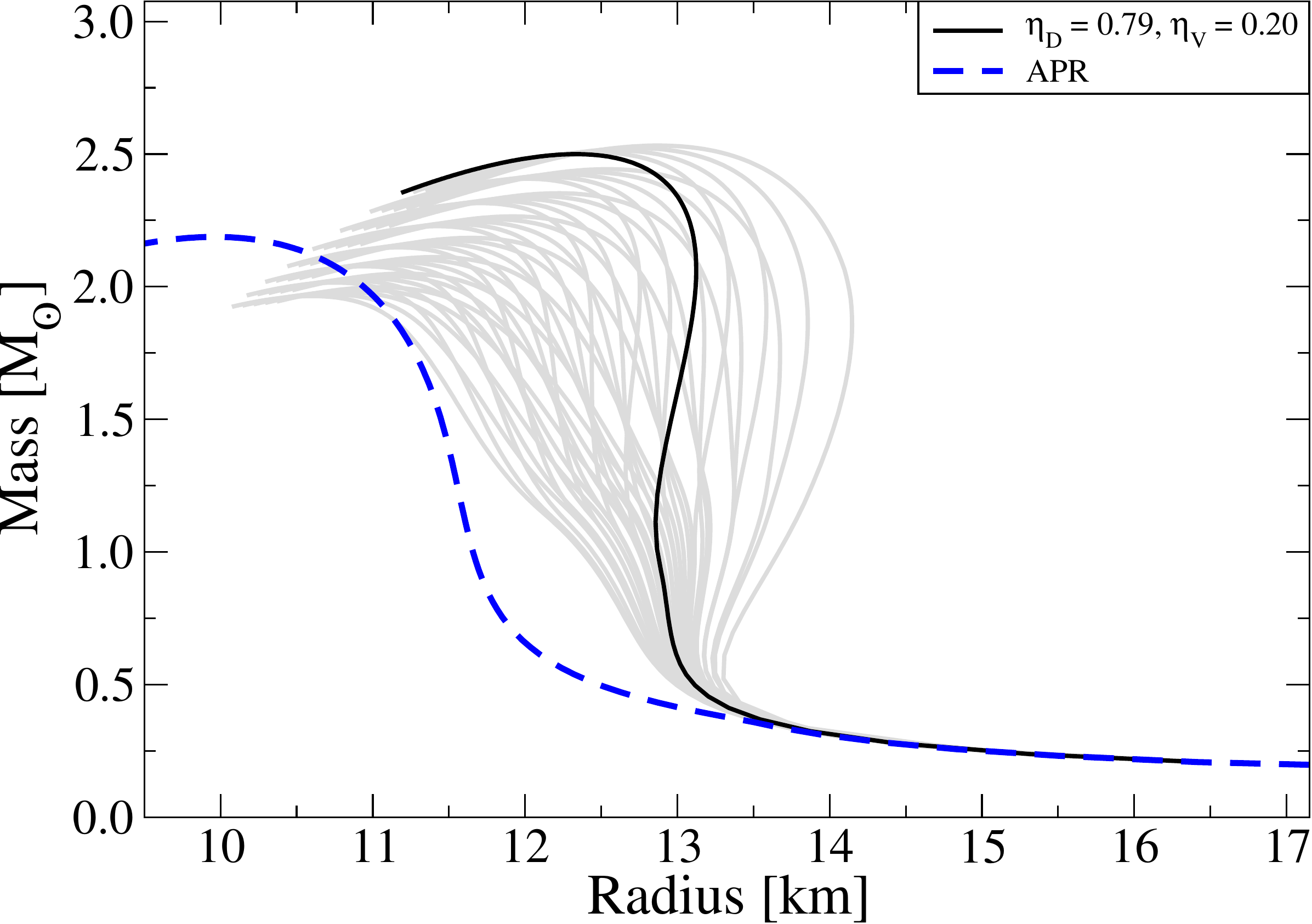}
\\
%&&&\\[-5mm]
\hline
&&&\\[-2mm]
\includegraphics[width=0.2\textwidth]{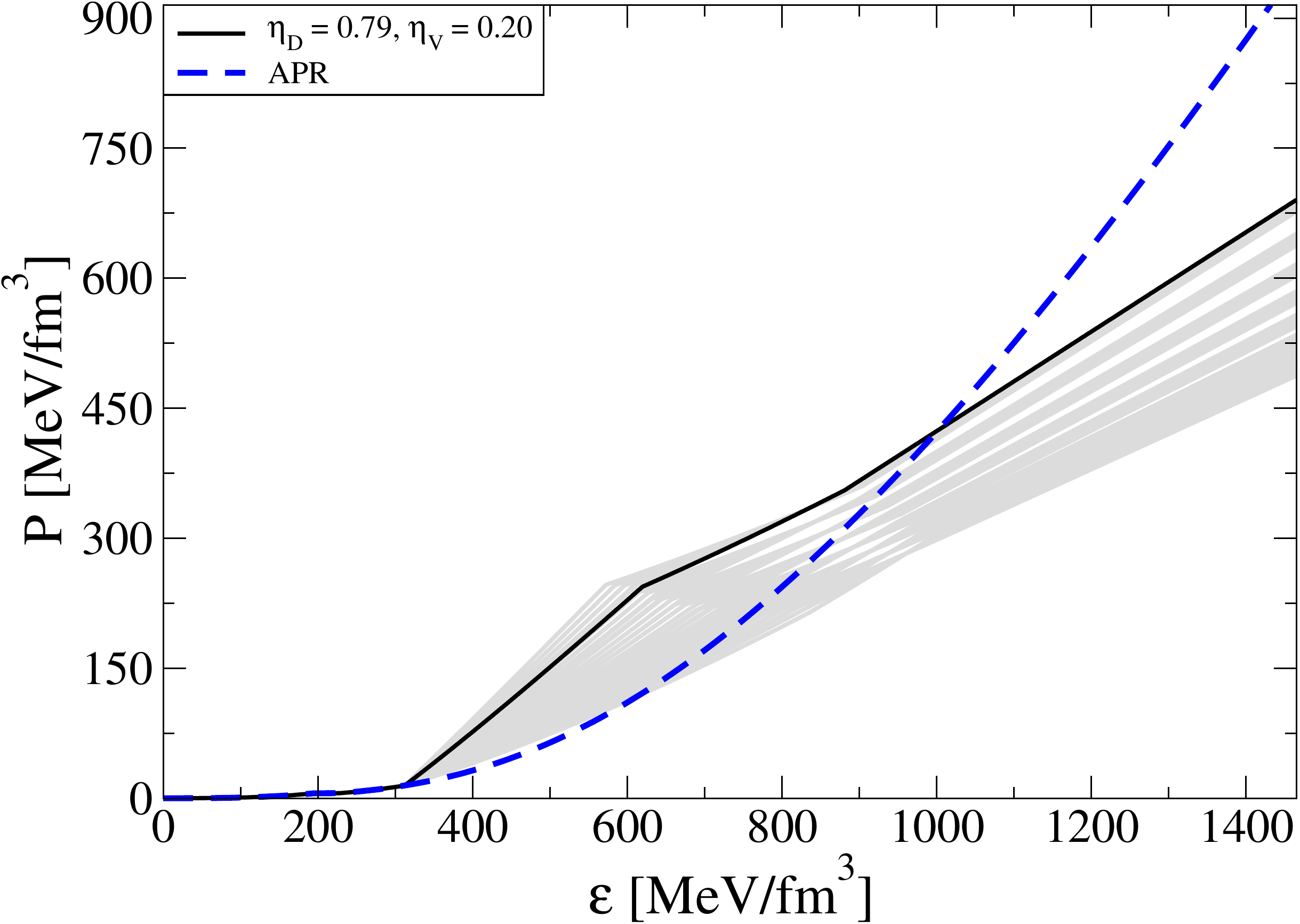} & 
\includegraphics[width=0.2\textwidth]{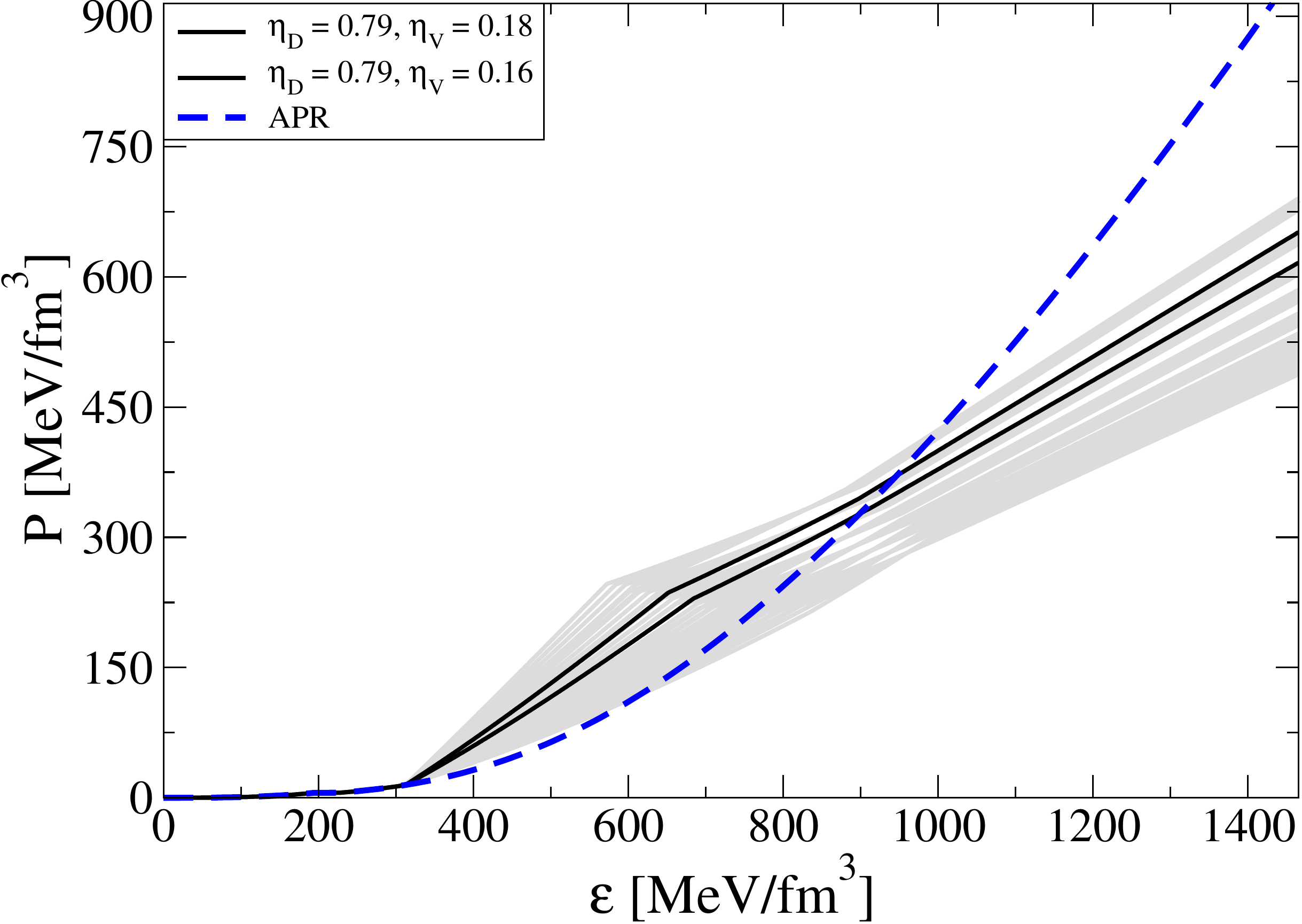}& 
\includegraphics[width=0.2\textwidth]{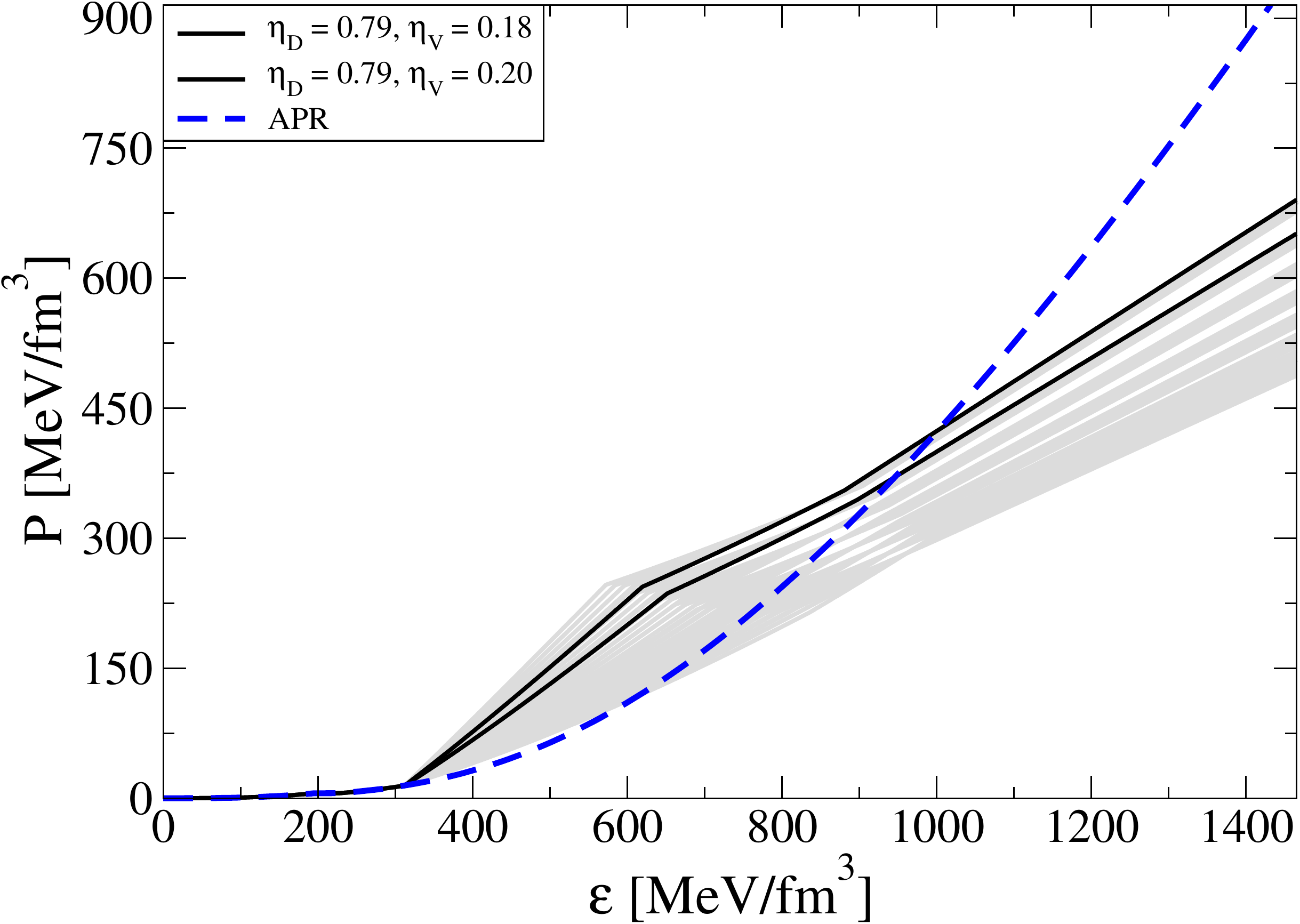}& 
\includegraphics[width=0.2\textwidth]{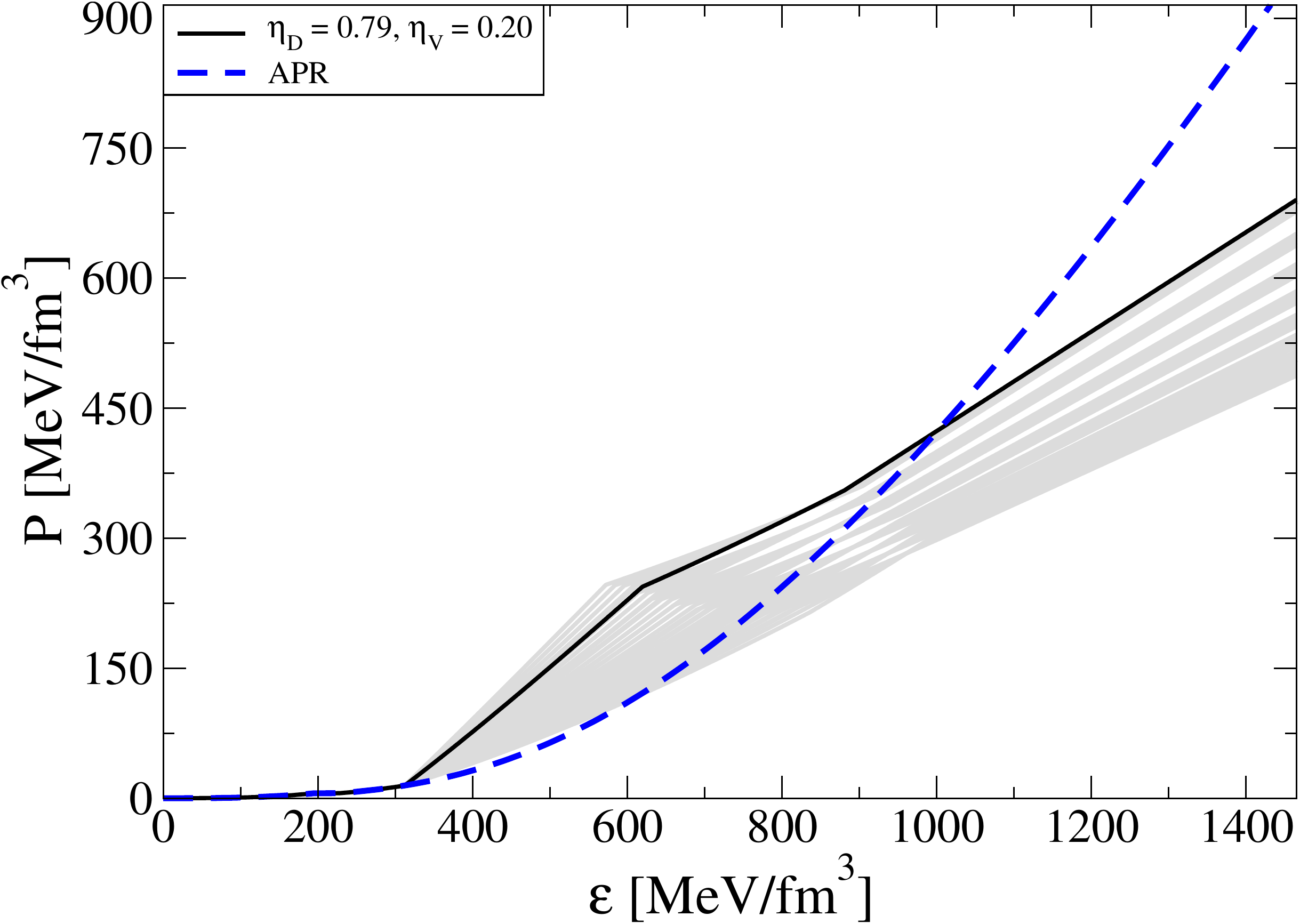}
\\
%&&&\\[-5mm]
\hline
&&&\\[-2mm]
\includegraphics[width=0.2\textwidth]{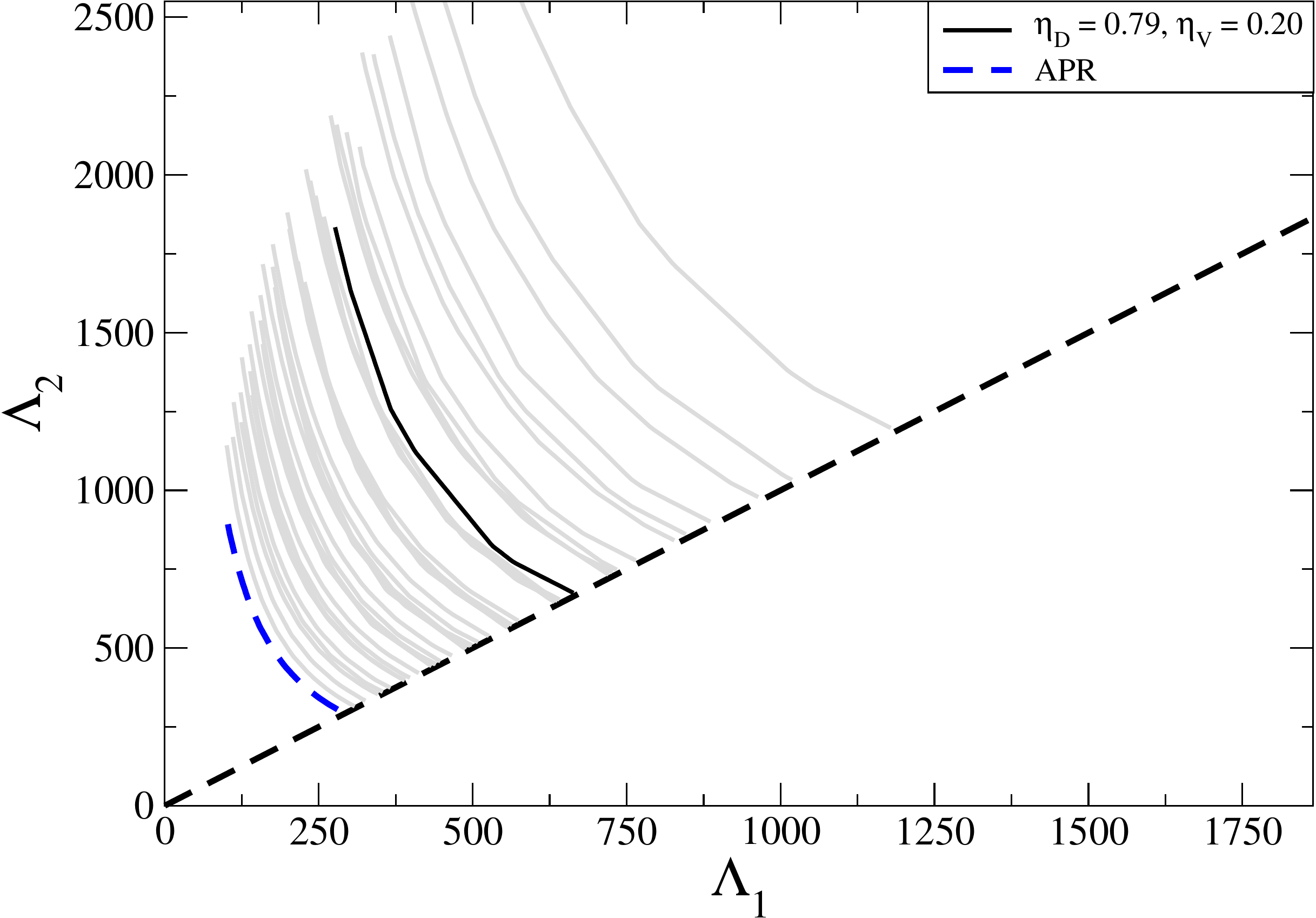} & 
\includegraphics[width=0.2\textwidth]{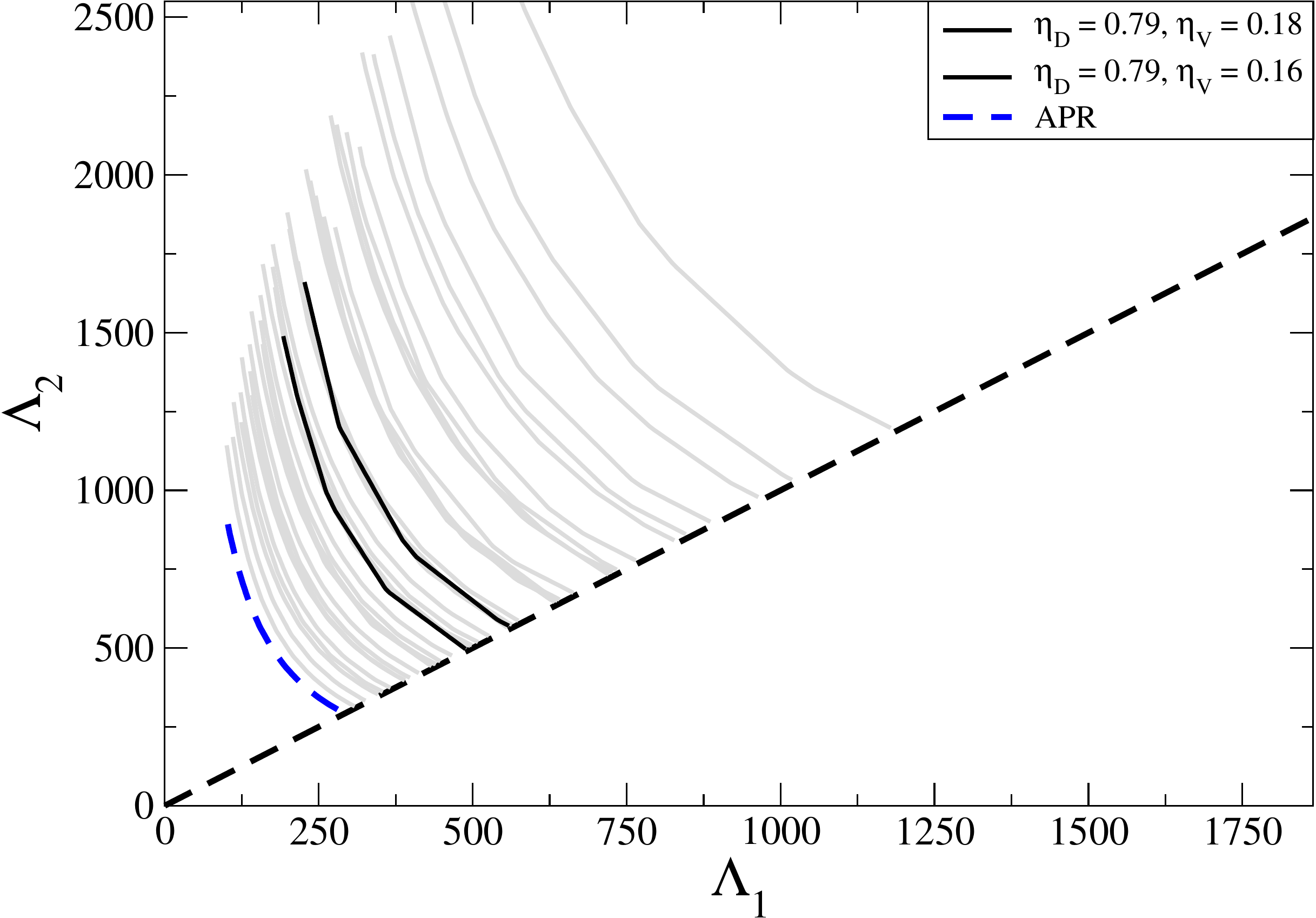}& 
\includegraphics[width=0.2\textwidth]{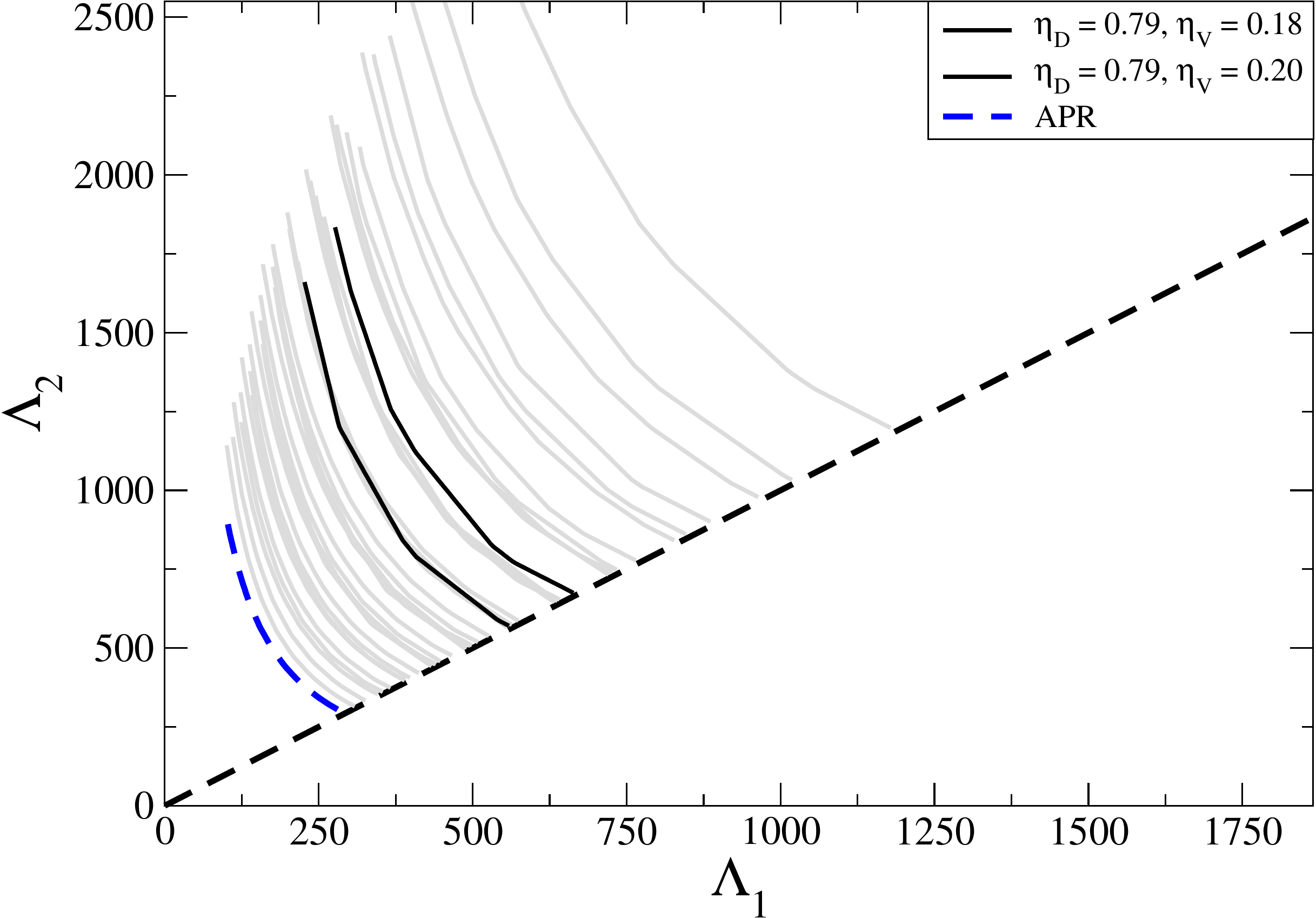}& 
\includegraphics[width=0.2\textwidth]{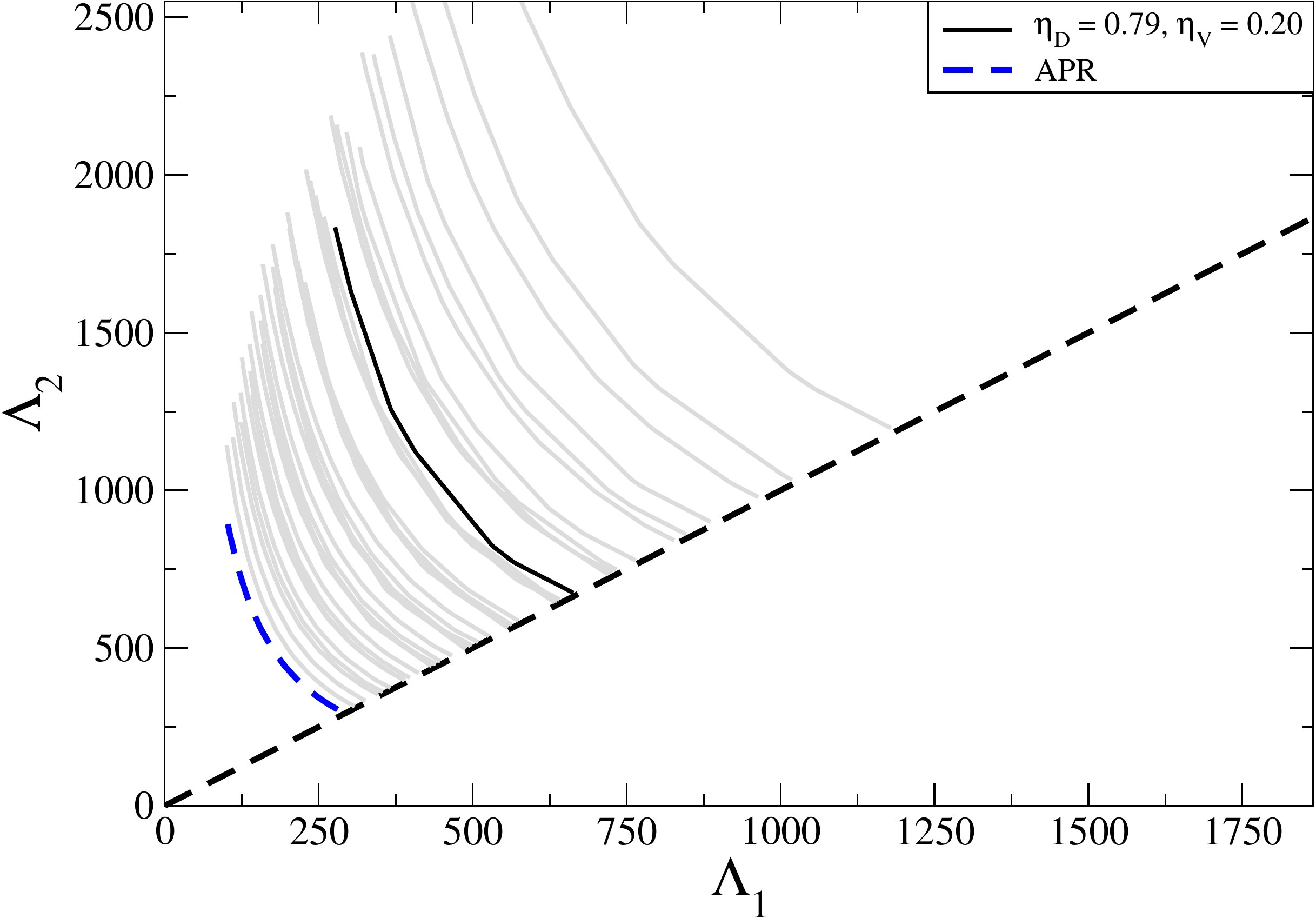}
\\
\hline 
%\end{array}$
\end{tabular}
\end{center}
\caption{Results of the BA for set 3 of  constraints (inf\{$M_{\rm max}$\} \cite{Abbott:2020khf}, $\Lambda_{1.4}$ \cite{Abbott:2018exr},
$(M,R)_{\rm J0030+0451}$ \cite{Miller:2019cac}) 
in the leftmost column and with an additional (yet fictitious) NICER radius measurement for PSR J0740+6620 of $R=11$, 12 or 13 km with an estimated standard deviation of $\sigma_R=0.5$ km in the other three columns.
The highlighted most probable M-R sequences (2nd row), EoS (3rd row) and $\Lambda_1-\Lambda_2$ (4th row) relationships correspond to the parameter sets with at least 75\% 
of the maximum probability as shown in the first row.
\label{fig:BA-fict-set3}}
\end{figure*}

The situation becomes more interesting when we consider a new, fictitious mass and radius measurement which anticipates the outcome of the NICER observation of the heavy pulsar PSR J0740+6620. In figure~\ref{fig:BA-fict} each column corresponds to the results for such a radius measurement with a value of $R= 11$ km, $R= 12$ km or $R= 13$ km, whereas the rows represent the same three aforementioned cases for the constraints. Just like before, case 3) displays a more selective effect on the $\eta_V$ parameter, however the inclusion of this new radius observation favours the highest $\eta_D$ values for most of the plots. 

\begin{figure*}[!ht]
	\centering
	\includegraphics[width=0.14\linewidth]{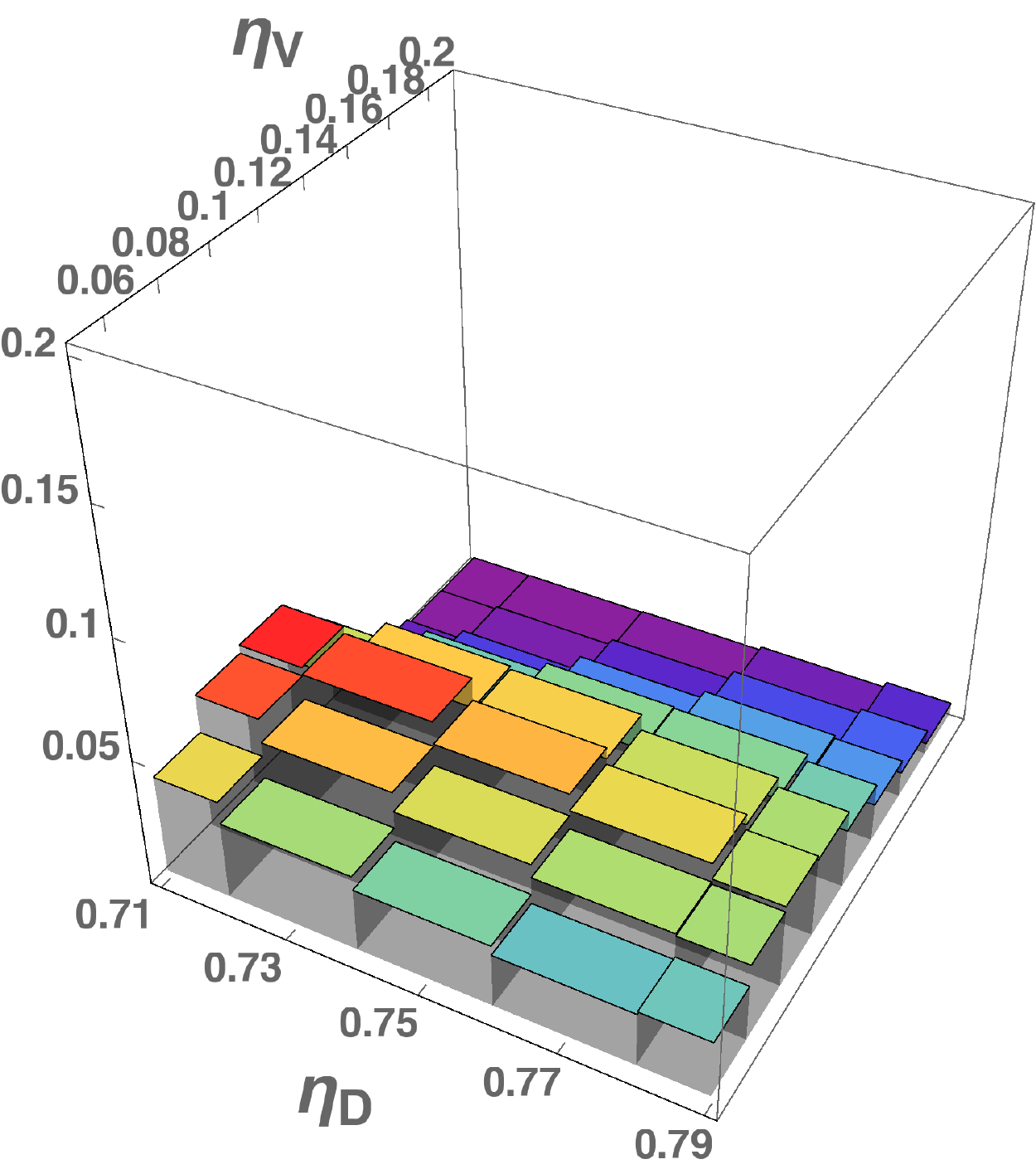}
	\includegraphics[width=0.16\linewidth]{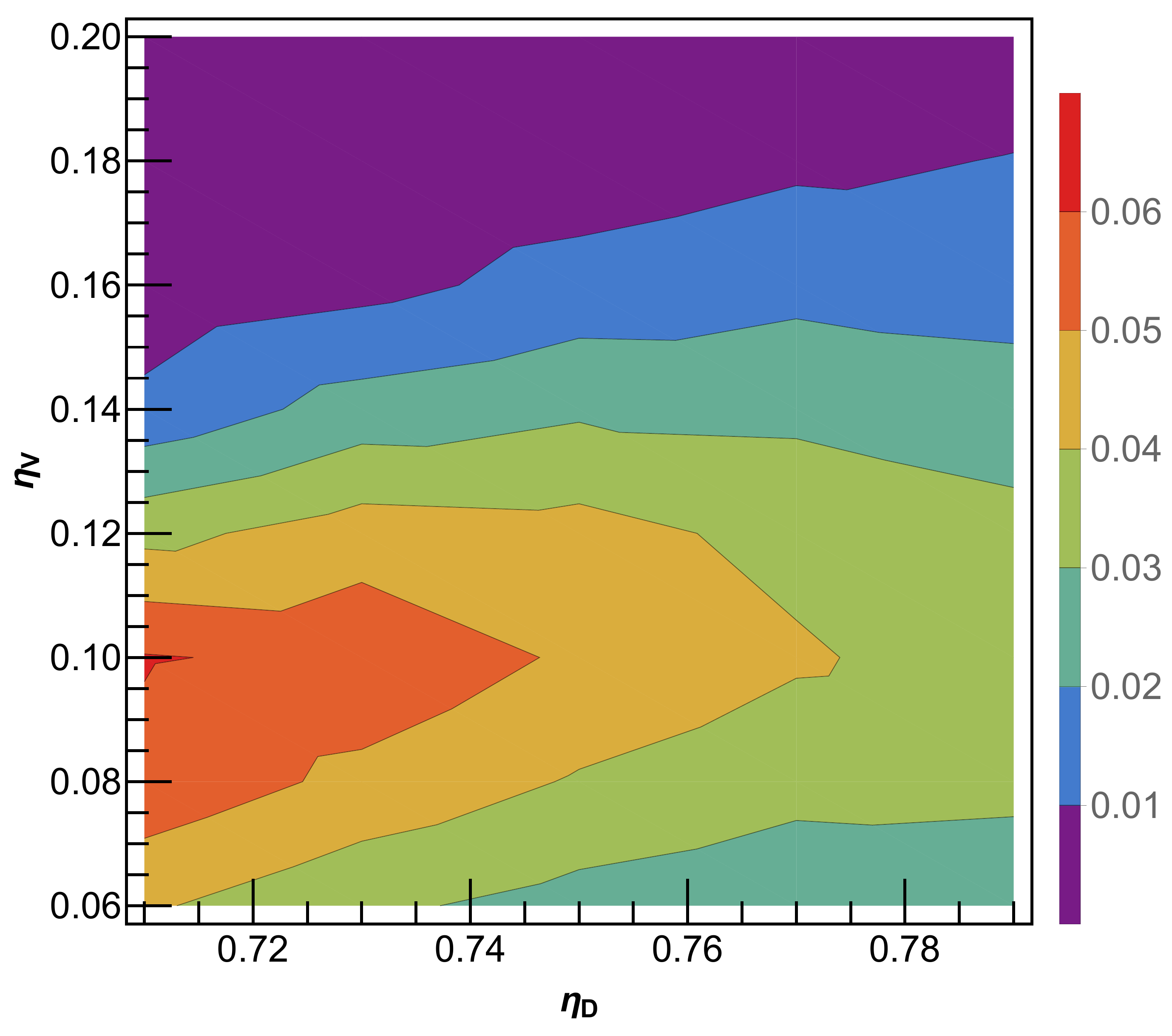}
	\includegraphics[width=0.2\linewidth]{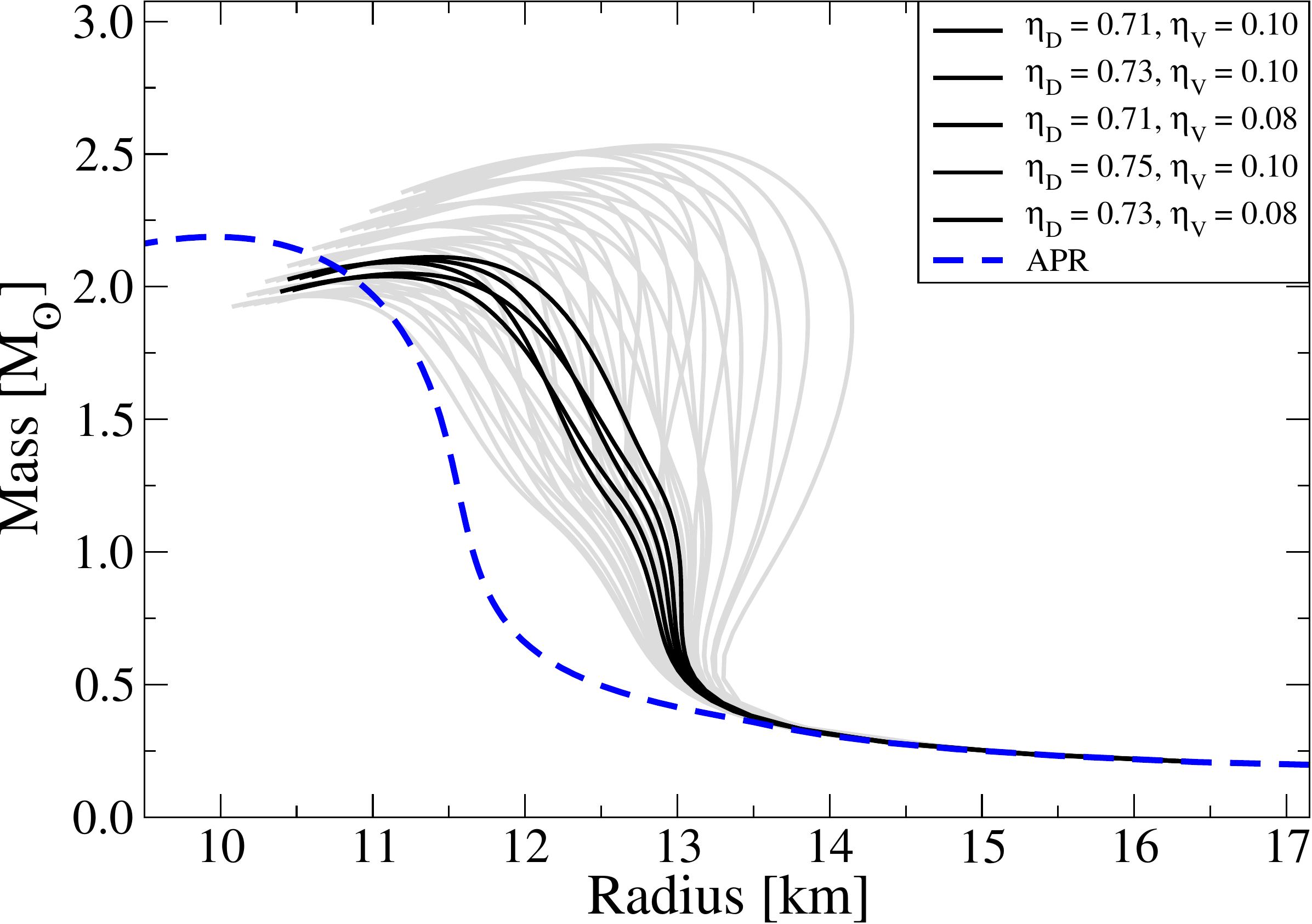}
	\includegraphics[width=0.2\linewidth]{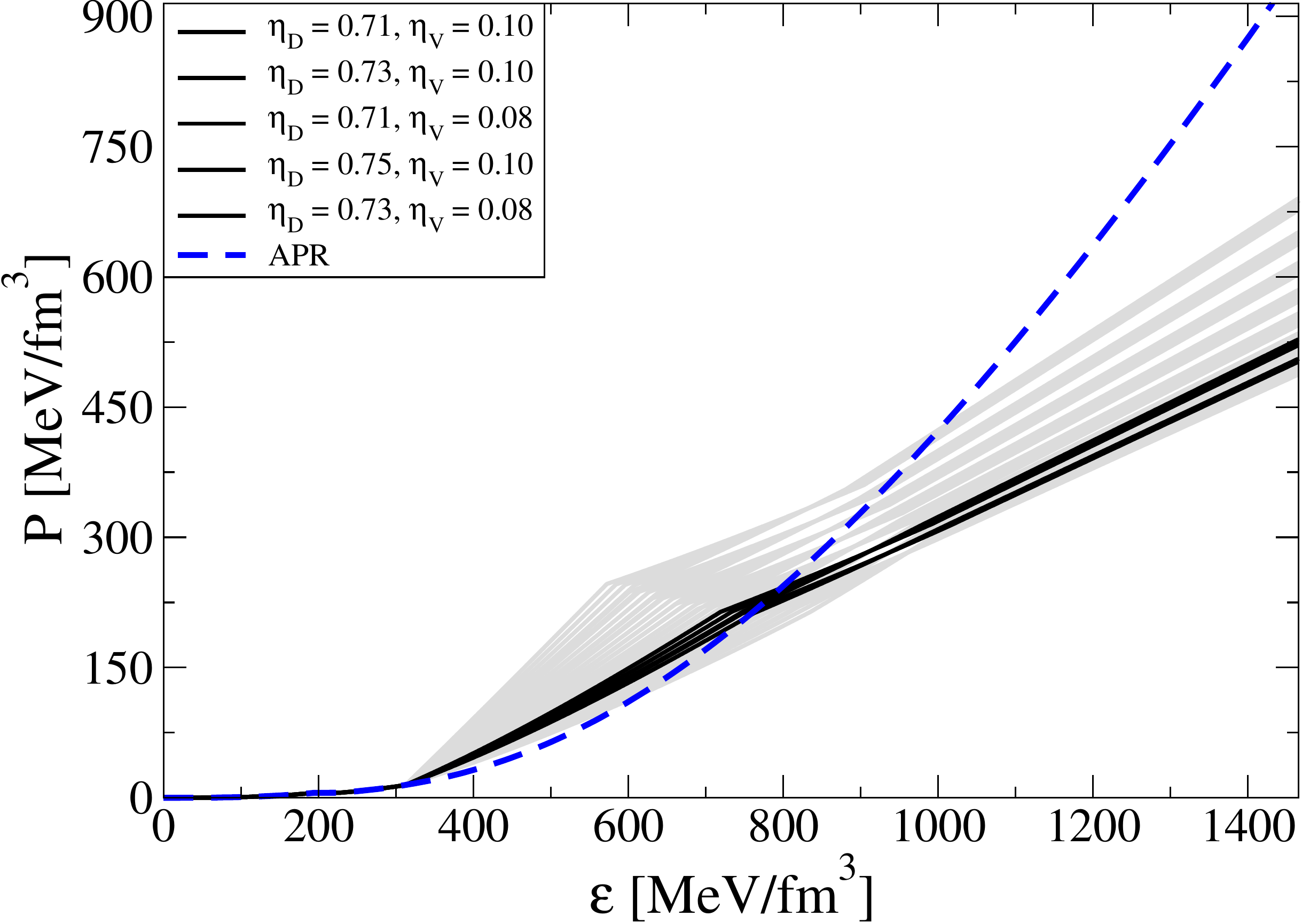}
	\includegraphics[width=0.2\linewidth]{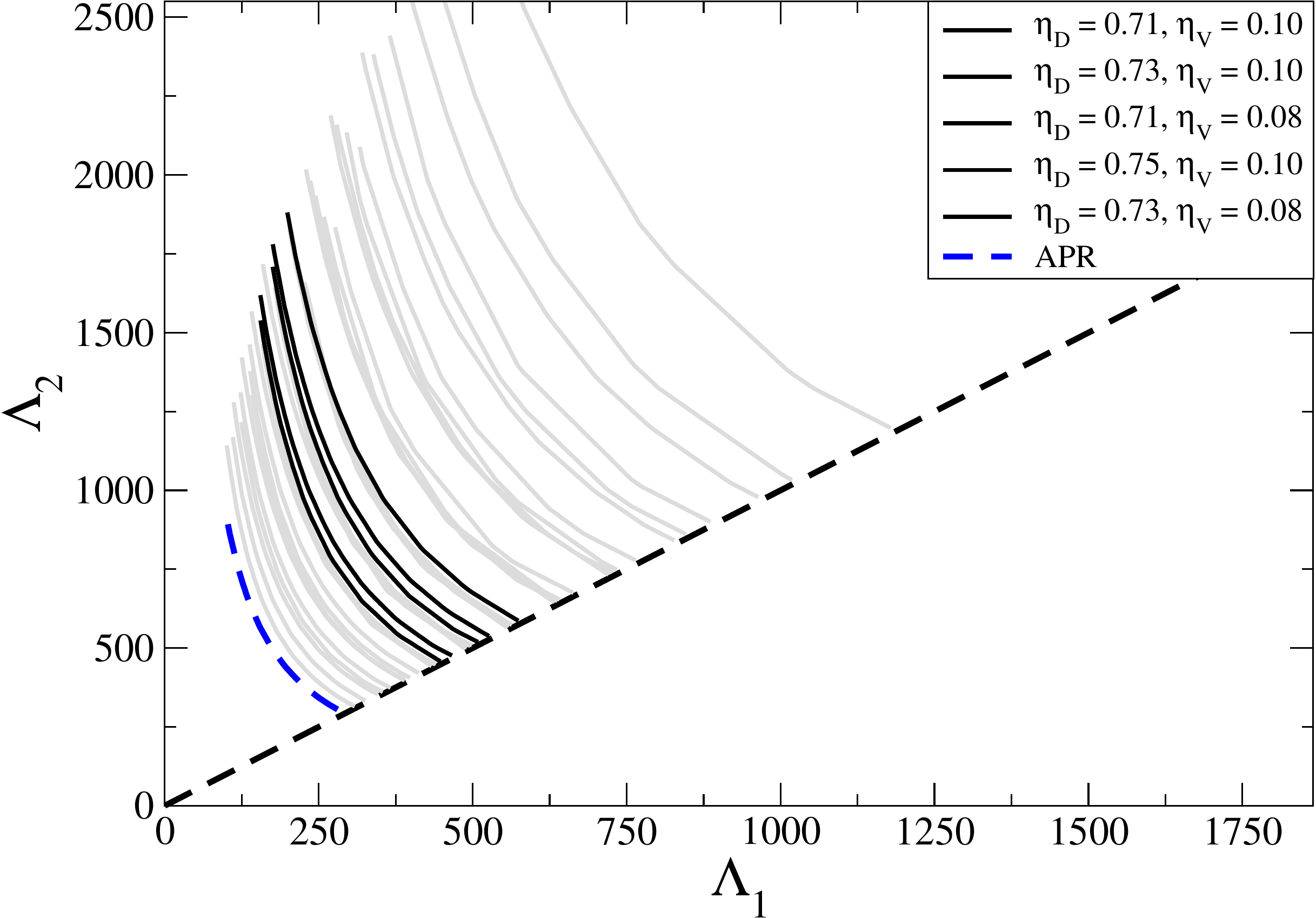}
	\caption{Bayesian analysis considering  $\Lambda_1-\Lambda_2$ from GW170817, M-R from J0030+0451 and M-R from J0740+6620 measurements.}
	\label{fig:BA_real}
\end{figure*}

In addition, figures~ \ref{fig:BA-fict-set1}, \ref{fig:BA-fict-set2} and \ref{fig:BA-fict-set3} present the equivalent contour plot constraints for the sets 1, 2 and set 3, respectively, as well as the Bayesian most probable EoS together with their mass-radius curves and tidal deformabilities for the GW170817 case.  While this paper was under revision in the refereeing process, the NICER experiment has published its results for the $M-R$ measurement \cite{Miller:2021qha}, shown in Fig.~\ref{fig:M-R_final}.
We have performed a BA using the NICER $M-R$ measurements for
PSR J0740+6620 \cite{Miller:2021qha} and PSR J0030+0451 \cite{Miller:2019cac} as well as the LVC tidal deformability constraint from GW170817 \cite{Abbott:2018exr}.
The result is shown in Fig.~\ref{fig:BA_real}.
A comparison with the three columns for fictitious radii in Fig.~\ref{fig:BA-fict-set1} shows that the actual result of the NICER measurement is well compatible with the $R=12$ km case. 
Therefore, the method of fictitious radii anticipation can be considered a reliable tool in predicting the implications of future $M-R$ measurements, e.g., by the NICER collaboration.
Moreover, for the most probable parameter set, 
the corresponding EoS lead to the $M-R$ sequences highlighted in Fig.~\ref{fig:BA_real} which fulfill the $2~M_\odot$ mass constraint but do not reach masses above $2.5~M_\odot$ which would be required if the lighter object in GW190814 was a (hybrid) neutron star.

\section{Conclusions}
\label{sec:conclusion}
In the present work we have applied Bayesian analysis methods to investigate the most likely quark-hadron hybrid EoS among a family of models that are agnostic about the detailed microphysical scenario of the hadron-to-quark matter transition. 
The model uses the APR EoS for densities below $2.0~n_0$ and above $n=4.0~n_0$ a set of nonlocal, color superconducting chiral quark model parametrizations of the NJL type with a covariant Gaussian formfactor. The vector meson coupling $\eta_V$ and the scalar diquark coupling $\eta_D$ are varied as free parameters that determine the stiffness of high-density quark matter.
The transition between both regimes is constructed by a new two-zone interpolation that realizes a smooth crossover behaviour, due to the assumption that the nature of the transition is a mixing of phases (crossover transition). 
Thus, in performing this interpolation we exclude the possibility of a first-order transition associated with a jump in the density, setting the corresponding parameter $\Delta n=0$. %However, the construction itself allows such a possibility.
In future work, we plan to systematically investigate also the extension of the new two-zone interpolation construction to the cases with $\Delta n\neq 0$.
We note, however, that the interpolation construction may be viewed as a shortcut that replaces three microphysical effects:
1) stiffening of the nuclear EoS due to quark substructure effects (quark Pauli blocking modeled, e.g., by a baryon excluded volume), 2) softening of the quark matter EoS at low densities due to confining effects (modeled, e.g., by a medium-dependent bag pressure) and 3) mixed-phase effects due to the occurrence of finite-size structures (pasta phases).
This possible microphysical underpinning of the interpolation construction, in particular within the two-zone version suggested in this paper, has been discussed and illustrated in the Introduction, but a detailed exploration is deferred to future work.

In our Bayesian study we have applied standard constraints for mass and radius measurements in set 1 and demonstrated their effect of narrowing the viable range of parameter values. 
We obtained the result that the most probable EoS lie along a line of proportionality between $\eta_V$ and $\eta_D$, whereby the higher values of $\eta_V$ are favorable for obtaining larger maximum masses of hybrid neutron stars. This finding confirms similar results of earlier studies in \cite{Klahn:2006iw,Klahn:2013kga} and the recent work \cite{Baym:2019iky} which do not employ Bayesian methods.

In the sets 2 and 3 we explored the nonstandard constraints of an upper limit on the maximum mass and the high mass of the lighter companion object of GW190814 as a lower limit on the maximum mass, respectively.
While for the set 2 the narrowing of the parameter range due to the upper limit on the maximum mass leads to an exclusion of the higher values of the vector meson coupling, the high value of the lower limit on the maximum mass for set 3 
allows only the highest possible vector couplings resulting in stiffer EoS and thus larger maximum masses and radii. 
It is a remarkable fact that within the present interpolation approach the  $2.6~M_\odot$ companion object in GW190814 could be a hybrid star with quark matter core. 
This possibility has been pointed out before in several works, among them~\cite{Blaschke:2020vuy,Dexheimer:2020rlp}. 

Finally, we have used the Bayesian approach to explore the consequences that the radius measurements on the $2~M_\odot$ pulsar PSR J0740+6620 by the NICER experiment may have for 
neutron star phenomenology. 
%}
%An important observation is that in order to reach very massive compact stars the equation of state should feature a steep rise of the constant speed of sound reaching high values near the causality limit~\cite{Tan:2020ics,Blaschke:2020vuy}. 

Before the NICER radius measurements on J0740+6620 
\cite{Miller:2021qha,Riley:2021pdl} were released,  
we have performed Bayesian analyses with fictitious results for that radius measurement, anticipating 11 km, 12 km and 13 km as a possible outcome.
For a radius value as small as 11 km or less, the hybrid star scenario for the $2.6~M_\odot$ object in GW190814 could be excluded and this event was a binary black hole merger. 
In that case the present two-zone interpolation approach with continuous crossover is not suitable since it does not produce stable hybrid stars with small enough radii above 
$2~M_\odot$. 

%As discussed, e.g., in \cite{Baym:2017whm} such a two-zone interpolation could be used to parametrize a first-order phase transition resulting in a density jump in the inner core and sufficiently compact high-mass hybrid star configurations.

%It has been demonstrated recently in Ref.~\cite{Blaschke:2020vuy} that for a NICER radius measurement on J0740+6620 resulting in 10 km or less, the only possibility to explain the star structure is a hybrid star scenario with a large quark matter core, since there is no realistic hadronic EoS model that could explain the smallness of neutron stars with a mass exceeding $2~M_\odot$.

For a radius of J0740+6620 as large as 13 km or more, the likely interpretation of GW190814 is that of a neutron star - black hole merger where the neutron star possibly had a 
color superconducting quark matter core.  
%However, a purely hadronic interior could not be excluded in that case.

We have performed a Bayesian analysis with the actual results 
for the NICER radius measurement on PSR J0740+6620 as reported by the Maryland-Illinois team \cite{Miller:2021qha} and compared the outcome for the probability contours in the model parameter space with those for the fictitious radius measurements. A striking similarity was obtained for $R=12$ km,
which proves the usefulness of the fictitious radius method for estimating the outcome of future NICER radius measurement campaigns and its impact for the dense matter EoS and neutron star phenomenology. 

The outcome of the Bayesian analysis in our setting favors the scenario that GW190814 was a binary black hole merger and not a neutron star - black hole merger.
The favored EoS which are highlighted in Fig.~\ref{fig:BA_real} favor $M-R$ sequences with a maximum mass that lies well below $2.5~M_\odot$, as shown also in that figure. 

In future extensions of the Bayesian approach to neutron star phenomenology it is desirable to widen the class of EoS by extending the present direct smooth interpolation approach 
to a first-order phase transition scenario \cite{Blaschke:2020qqj} where in the present two-zone interpolation method a nonzero density jump parameter at the matching point would be chosen, and to augment this with the possibility of a subsequent smoothing of the transition by a pasta phase construction or its emulation.

In order to compare the results of two Bayesian analyzes obtained on the same astrophysical data, it is necessary to combine the factors of the normalization of the Bayesian formula~\eqref{eq:bayes}. This allows to introduce the relative posterior probability of each analysis. The set of EoS models, whose analysis gives a greater value of the relative posterior probability can be considered to be more successful in describing the observational data. 
A comparison of the result presented in this paper with the previous result from Ref.~\cite{Blaschke:2020qqj}, where constraints corresponding to set 1 have been employed, shows that the likelihood for models considered in~\cite{Blaschke:2020qqj} exceeds that of the present work by more than 5 times.
Such a direct comparison is possible only for the case of set 1, because the other two sets have not been considered in \cite{Blaschke:2020qqj}.

The question has been raised whether one should generalize the Bayesian analysis by sytematically varying also the hadronic EoS at supersaturation densities like, e.g., within an excluded volume approach (see Ref.~\cite{Alvarez-Castillo:2016oln}).
In the present setting of the two-zone interpolation model, with a choice of the onset of the interpolated part of the EoS at the saturation density ($n_H=n_0$), there is no room for such a variation. 
Uncertainties in the knowledge of the nuclear EoS at supersaturation densities are covered by the variation of the parameters in the hadron-like zone of the interpolation. 

%Only such a more complete EoS basis for the Bayesian study will allow to draw conclusions for the key question whether neutron stars in the observable mass range can harbor deconfined quark matter in their cores. 

\subsection*{Acknowledgements}
We acknowledge the partial support by the COST Action CA16214 "PHAROS" for our international networking activities in preparing this article.
This work received support from the 
Russian Fund for Basic Research under grant no. 18-02-40137.
The work by D.B. on the new class of quark-hadron hybrid EoS 
in Section 4 was supported by the Polish National Science Centre under grant number UMO 2019/33/B/ST9/03059.
D.E.A-C. and H.G. are grateful for support from the programme for exchange between JINR Dubna and Polish Institutes (Bogoliubov-Infeld programme). 
%D.E.A-\-C. and S.T. received support  form the Heisenberg-Landau programme for scientist exchange between JINR Dubna and German Institutes.
A. G. G. would like to acknowledge to CONICET for financial support under Grant No. PIP17-700.

%\bibliography{epja}

\end{document}